\documentclass[a4paper,11pt]{article}
\pdfoutput=1 

\usepackage{jheppub} 
\usepackage{amsmath,amssymb}
\usepackage{datetime,mathdots}

\def\be{\begin{equation}}
\def\ee{\end{equation}}
\def\bea{\begin{eqnarray}}
\def\eea{\end{eqnarray}}
\def\bal{\begin{align}}
\def\eal{\end{align}}

\newcommand\bR{\mathbb{R}}

\newcommand\bZ{\mathbb{Z}}
\newcommand\bC{\mathbb{C}}

\newcommand\cN{\mathcal{N}}
\newcommand\cL{\mathcal{L}}
\newcommand\cO{\mathcal{O}}
\newcommand\cA{\mathcal{A}}
\newcommand\cB{\mathcal{B}}
\newcommand\cD{\mathcal{D}}
\newcommand\cC{\mathcal{C}}
\newcommand\cF{\mathcal{F}}
\newcommand\cH{\mathcal{H}}
\newcommand\cJ{\mathcal{J}}
\newcommand\cR{\mathcal{R}}
\newcommand\cM{\mathcal{M}}

\newcommand\cP{\mathcal{P}}
\newcommand\cS{\mathcal{S}}
\newcommand\cU{\mathcal{U}}
\newcommand\cV{\mathcal{V}}
\newcommand\cX{\mathcal{X}}

\newcommand\bfA{\mathbf{A}}
\newcommand\bfB{\mathbf{B}}
\newcommand\bfC{\mathbf{C}}
\newcommand\bfCp{\mathbf{C'}}
\newcommand\bfD{\mathbf{D}}

\newcommand\bfI{\mathbf{I}}

\newcommand\bfM{\mathbf{M}}

\newcommand\bfS{\mathbf{S}}
\newcommand\bfT{\mathbf{T}}
\newcommand\bfU{\mathbf{U}}

\newcommand\bfa{\mathbf{a}}
\newcommand\bfb{\mathbf{b}}
\newcommand\bfc{\mathbf{c}}
\newcommand\bfv{\mathbf{v}}
\newcommand\bfx{\mathbf{x}}

\newcommand\dd{\mathrm{d}}
\newcommand\ex{\mathrm{e}}
\newcommand\ii{\mathrm{i}}
\newcommand\id{\mathbf{I}}

\newcommand\qqq{\qquad\qquad}
\newcommand{\nn}{\nonumber \\ {} }
\newcommand{\nnm}{\nonumber }

\newcommand{\ta}{\mathsf{a}}   
\newcommand{\tb}{\mathsf{b}}

\newcommand{\tx}{\mathsf{x}}

\newcommand{\del}{\partial}

\newcommand{\cgn}{\mathcal{C}_{g,n}}
\newcommand{\pants}{\mathcal{C}_{0,3}}
\newcommand{\torus}{\mathcal{C}_{1,1}}
\newcommand{\ra}{\rightarrow}
\newcommand{\rev}{\text{reverse}}
\newcommand{\cyc}{\text{cyclic}}
\newcommand{\tid}{\text{id}}

\newcommand{\Hol}{\mathrm{Hol}}
\newcommand{\tr}{\mathrm{tr}}
\newcommand{\ts}[1]{\textsuperscript{\tiny #1}}

\newcommand\sgn{\text{sgn}}

\newcommand\bfs{\mathbf{s}}

\newcommand\Rep{\text{Rep}}
\newcommand\End{\text{End}}

\title{Line operators in theories of class $\mathcal{S}$, quantized moduli space of flat connections,
and Toda field theory}

\author[a,d]{Ioana Coman}
\author[a,b,c]{Maxime Gabella}
\author[a]{J\"org Teschner}

\affiliation[a]{DESY, Theory Group, Notkestrasse 85, Building 2a, 22607 Hamburg, Germany}
\affiliation[b]{Department of Mathematics, University of Hamburg, Bundesstrasse 55, 20146 Hamburg, Germany}
\affiliation[c]{Institute for Advanced Study, Einstein Drive, Princeton, NJ 08540, USA}
\affiliation[d]{Department of Theoretical Physics, NIPNE,
Str. Reactorului 30, 077125, Magurele, Romania}

\emailAdd{ioana.coman@desy.de}
\emailAdd{gabella@ias.edu}
\emailAdd{joerg.teschner@desy.de}

\abstract{
Non-perturbative aspects of $\mathcal{N}=2$ supersymmetric gauge
theories of class~$\mathcal{S}$ are deeply encoded in the algebra of functions on the moduli space $\mathcal{M}_\text{flat}$ of flat $SL(N)$-connections on Riemann surfaces.
Expectation values of Wilson and 't Hooft line operators are related to holonomies of flat connections,
and expectation values of line operators in the low-energy effective theory
are related to Fock-Goncharov coordinates on $\mathcal{M}_\text{flat}$.
Via the decomposition of UV line operators into IR line operators, we
determine their noncommutative algebra from the quantization of
Fock-Goncharov Laurent polynomials,
and find that it coincides with the skein algebra studied in the context of Chern-Simons theory.
Another realization of the skein algebra 
is generated by Verlinde network operators in Toda field theory. 
Comparing the spectra of these two realizations provides non-trivial support for their equivalence.
Our results can be viewed as evidence for the generalization of the AGT correspondence to higher-rank class~$\mathcal{S}$ theories.
}

\begin{document}

\maketitle
\flushbottom

\section{Introduction}

There has been a lot of recent progress in the study of $\mathcal{N}=2$ supersymmetric
field theories in four dimensions (see~\cite{Teschner:2014oja} for a survey).
Highlights include exact results on the expectation values of certain observables
like Wilson and 't Hooft loop operators, and powerful algorithms for the computation of the spectra of BPS states.
Many of these results are deeply related to mathematical structures on the moduli spaces of vacua.

A rich and interesting class of $\mathcal{N}=2$ theories, often referred to as class $\mathcal{S}$, arises from the twisted compactification of the six-dimensional $(2,0)$ theory with Lie algebra $\mathfrak{g}$ on a Riemann surface $\cgn$ of genus $g$ with $n$ punctures~\cite{Gaiotto:2009we}\cite{Gaiotto:2009hg}.
Class $\mathcal{S}$ theories of type $\mathfrak{g}=A_1$ admit weakly-coupled Lagrangian descriptions specified by pair of pants decompositions of $\cgn$.
Each of the $3g-3+n$ cutting curves defining the pants decomposition corresponds to an $SU(2)$ 
gauge group, while each of the $n$ punctures corresponds to an $SU(2)$ flavor group.
Alday, Gaiotto, and Tachikawa (AGT) made the striking observation 
that the four-sphere partition functions of $A_1$ theories can be expressed in terms of the correlation functions 
of Liouville conformal field theory on $\cgn$~\cite{Alday:2009aq}.

An important goal is the generalization of the AGT correspondence to higher rank.
The origin of class $\cS$ theories from the six-dimensional $(2,0)$ theory
suggests that the relations to two-dimensional conformal field theory may have interesting generalizations 
for $N>2$.
In particular, $A_{N-1}$ theories are expected to be related to 
$SU(N)$ Toda field theory~\cite{Wyllard:2009hg}, and a lot of evidence has been accumulated for 
this conjecture, including
~\cite{Passerini2010pr,Kozcaz2010af,Kanno2010kj,Gomis2010kv,Wyllard2010rp,Wyllard2010vi,Drukker2010vg,Tachikawa:2011dz,Fateev:2011hq,Estienne:2011qk,Zhang:2011au,Shiba:2011ya,Kanno:2013aha,Aganagic:2014oia,Gomis:2014eya}. 
It has been proven  in \cite{Fateev:2011hq}
 for a particular subclass of the $A_{N-1}$ theories in 
class $\cS$ which may be represented as quiver gauge theories with linear or circular quiver diagrams.
However, neither side of this higher-rank AGT correspondence is well-understood.
Usual methods do not apply because class $\cS$ theories of type $\mathfrak{g}=A_{N-1}$ with $N>2$ do not have Lagrangian descriptions in general.
It is therefore unclear what should play the role of instanton partition functions in these non-Lagrangian theories.
Similarly, the Toda conformal blocks seem to be poorly understood at the time of writing. 

Parallel developments came from the insight of Gaiotto, Moore, and Neitzke (GMN) that the BPS spectrum of a class $\mathcal{S}$ theory
is encoded in geometrical structures on
the moduli space $\cM_\text{vac}$ of vacua of the theory on $\mathbb{R}^3\times S^1$ \cite{Gaiotto:2008cd,Gaiotto:2009hg,Gaiotto:2010be,Gaiotto:2012db}. The six-dimensional description of class $\cS$ theories implies that  
$\cM_\text{vac}$ is isomorphic to Hitchin's moduli space 
$\cM_{\rm H}$
of solutions to the self-duality equations on $\cgn$ (see~\cite{Neitzke:2014cja} for a review). This space has a hyperk\"ahler structure, 
and in one of its complex structures it can be identified with the moduli space $\cM_\text{flat}$
of complex flat connections. This leads to the key relation
\bea
\cM_\text{vac} (\mathbb{R}^3\times S^1) = \cM_\text{flat} (\cgn)~.
\eea
An important manifestation of this relation is that vacuum expectation values of BPS line operators in class $\cS$ theories can be expressed as holonomies of flat connections on $\cgn$.

\begin{figure}[t]
\centering
\includegraphics[width=0.84\textwidth]{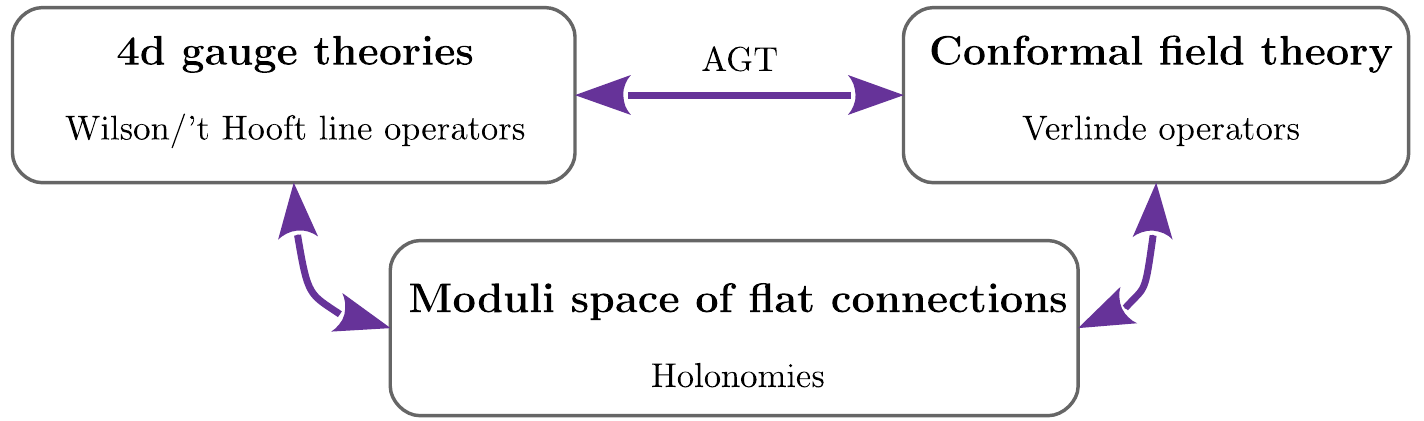}
\caption{Triangle of relations between 4d $\cN=2$ supersymmetric gauge theories labeled by Riemann surfaces $\cgn$ (\emph{left}), conformal field theory (Liouville or Toda) on $\cgn$ (\emph{right}), and the moduli space of flat $SL(N,\bC)$-connections on $\cgn$ (\emph{bottom}).
We also indicate the interpretation of line defects in each description.}
\label{AGTtriangle}
\end{figure}

It has subsequently been observed in~\cite{TV13} that the origin of the AGT correspondence can be understood from the relation to $\cM_\text{flat}$ (see~\cite{Teschner:2014nja}).
Thus there appears to be a triangle of relations between $\cN=2$ supersymmetric gauge theories, conformal field theories, and moduli spaces of flat connections, as
depicted in figure~\ref{AGTtriangle}.

In this paper, we make the first steps towards a generalization of this triangle of relations to higher rank. Given the difficulties in tackling directly higher-rank gauge theories and Toda field theory, we focus first on flat connections.
We will describe in detail the quantum algebra $\cA^\text{flat}$ of functions on the moduli space of flat $SL(N,\bC)$-connections, which the arguments of GMN relate to the algebra $\cA^\text{line}$ of quantized line operators in class $\cS$ theories of type $A_{N-1}$. We will moreover show that $\cA^\text{flat}$ can be identified with the algebra $\cA^\text{Ver}$ of Verlinde loop and network operators in $SU(N)$ Toda field theory.
Given the central role played by these algebras in the approach of~\cite{TV13},
we may regard the relations 
\bea 
\cA^\text{line} \simeq \cA^\text{flat} \simeq 
\cA^\text{Ver}
\eea 
as  support for a higher-rank AGT correspondence.

Before summarizing our results in more detail, we now give some background about line operators and their role in the AGT correspondence.

%

\subsection*{Line operators and framed BPS states}

BPS line operators in class $\cS$ theories are supersymmetric generalizations of Wilson and 't Hooft line observables, describing the effect of inserting heavy dyonic probe particles labeled by electric and magnetic charge vectors.\footnote
{
Note that allowed sets of ``mutually local'' line operators are specified by a certain topological data, which impose some restrictions on the representations~\cite{Gaiotto:2010be,Aharony:2013hda,Tachikawa:2013hya}. This subtlety will not affect our conclusions in an essential way, as noted for the $A_1$ case in~\cite{Teschner:2014nja}.
}
Such a line operator can be viewed as descending from a surface operator in the six-dimensional $(2,0)$ theory, which is labeled by a representation $\cR$ of $\mathfrak{g}$ and supported on $\Sigma =  S^1\times \wp$, with $\wp$ a path in $\cgn$.
This leads to a relation between the vacuum expectation value (vev) of the line operator $L (\cR;\wp)$ on $S^1$ and 
the classical holonomy of a flat connection $\cA$ along the path $\wp$ on $\cgn$:
\begin{equation}\label{holo}
\big\langle\,  L (\cR ;\wp)\, \big\rangle = \tr_\cR \text{Hol}_\wp \cA~.
\end{equation}
BPS line operators thus provide natural coordinate functions on $\cM_\text{flat}$.

As argued in \cite{Gaiotto:2010be,Cordova:2013bza}, the vev of a UV line operator $L$
can be represented in the IR in terms of a set of vevs of line operators $\cX_\gamma$ with charge $\gamma$ defined using the low-energy abelian gauge fields:
\bea\label{UVIRlineRel}
 L  \rightsquigarrow \sum_\gamma \overline{\underline{\Omega}}(L, \gamma) \cX_\gamma~.
\eea
The coefficients $\overline{\underline{\Omega}}(L, \gamma)$ are integers which count the BPS states supported by the line operator~$L$, called \emph{framed BPS states}.
The IR line operators $\cX_\gamma$ are Darboux coordinates on $\cM_{\rm flat}$ that are closely related to the coordinates constructed by Fock and Goncharov in their study of higher Teichm\"uller spaces~\cite{FG}.

\begin{figure}[t]
\centering
\includegraphics[width=\textwidth]{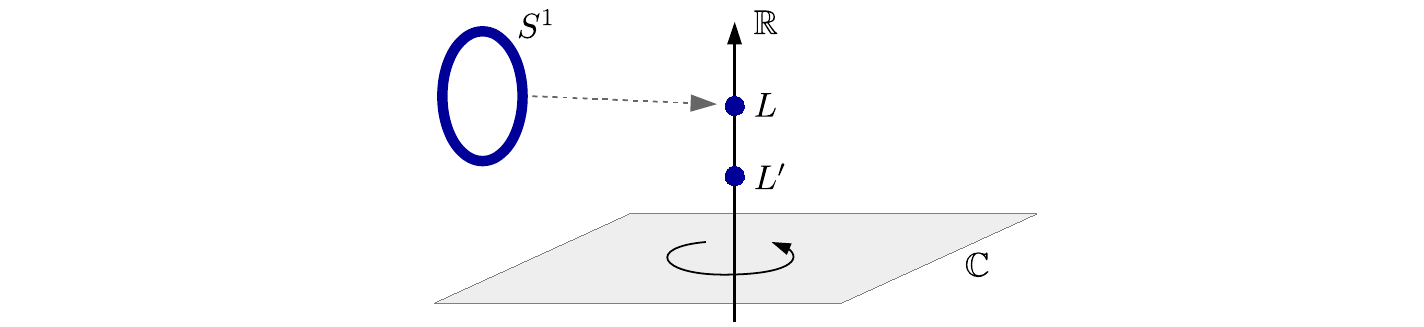}
\caption{The algebra of line operators can be quantized by twisting $\bR^3 \times S^1$ such that a plane $\bC \subset \bR^3$ rotates as one moves along the $S^1$. BPS line operators wrap $S^1$ and align along the axis $\bR$ at the origin of $\bC$.}
\label{GMNtwist}
\end{figure}
The line operators can be quantized by twisting $\bR^3\times S^1$ into the fibered product $\bR\times \bC \times_q S^1$ such that a coordinate $z$ on $ \bC$ rotates as $z \to qz $ after going around $S^1$~\cite{Gukov:2006jk,Gaiotto:2010be,Ito:2011ea}.
BPS conditions then constrain line operators on $S^1$ to be located 
at points along the axis $\bR$ and at the origin of $\bC$ (see figure~\ref{GMNtwist}).  
The relation  \eqref{UVIRlineRel} between UV and IR line operators becomes 
\bea\label{linePSC}
\hat L \rightsquigarrow  \sum_\gamma \overline{\underline{\Omega}}(L, \gamma;q) \hat \cX_\gamma~,
\eea
where the noncommutative variables $\hat\cX_\gamma$ satisfy the relation $\hat \cX_\gamma \hat \cX_{\gamma'} = q^{\frac12 \langle \gamma, \gamma'\rangle}\hat  \cX_{\gamma+\gamma'}$ appearing in the quantization of the algebra of functions on higher Teichm\"uller spaces.
The coefficients $\overline{\underline{\Omega}}(L, \gamma;q)$ are the \emph{framed protected spin characters} defined in~\cite{Gaiotto:2010be} as
\bea\label{PSCbarOmegay}
\overline{\underline{\Omega}}(L, \gamma;q) = \tr_{\cH_{L}^\text{BPS}} q^{J_3} (-q^\frac12 )^{2I_3}~,
\eea
where $\cH_{L}^\text{BPS}$ is the Hilbert space of framed BPS states, and $J_3$ and $I_3$ are generators of the $SO(3)$ and $SU(2)_R$ symmetries.
GMN conjectured that framed BPS states have $I_3=0$ (``no exotics conjecture''), 
which implies in particular that the $\overline{\underline{\Omega}}(L, \gamma;q)$ are linear combinations of $su(2)$ characters with positive integral coefficients.
The coefficients of $q^m X_\gamma$ in the decomposition~\eqref{linePSC} of a line operator are dimensions of Hilbert spaces graded by the IR electromagnetic charges $\gamma$ and the $so(3)$ spins $m$.

The noncommutative algebra of IR line operators $\hat\cX_\gamma$ determines via~\eqref{linePSC} the 
algebra of UV line operators, 
which may be represented by relations of the form
\bea\label{ncProductLLp}
\hat L * \hat L' = \sum_{L''} c(L,L',L''; q) \hat L''~.
\eea
The order in which we multiply operators corresponds to their ordering along the axis $\bR$.
The algebra generated by the operators $\hat L$ can be viewed as a 
noncommutative deformation of the algebra of functions on $\cM_\text{flat}$, with $q$ the deformation parameter.
For $A_1$ theories, it was argued in \cite{Gaiotto:2010be} that this algebra is
isomorphic to the algebra of quantum geodesic length 
operators in quantum Teichm\"uller theory.

\subsection*{The role of line operators in the AGT correspondence}

Expectation values of line operators in $A_1$ theories on the four-ellipsoid
${E}_{\epsilon_1,\epsilon_2}^4=\{ x_0^2+
\epsilon_1^2(x_1^2+x_2^2)+\epsilon_2^2(x_3^2+x_4^2)=1\}$ 
can be calculated by localization~\cite{Pestun:2007rz,Gomis:2011pf,Hama:2012bg} and take the schematic form
\begin{equation}\label{funSchroe-loc}
\big\langle \,
L\,\big\rangle_{{\epsilon_1,\epsilon_2}}
\,=\,\int \dd a\;(\Psi(a))^*\,\mathsf{L}\,  \Psi(a)~.
\end{equation}
The integration is performed over variables $a=(a_1,\dots,a_h)$ representing the zero modes
of the $h=3g-3+n$ scalar fields in the vector multiplets. 
$\Psi(a)$ represents the contribution of the path integral 
over the lower half-ellipsoid with $x_0<0$, and 
$\mathsf{L}$ is a finite 
difference operator acting on the variables $a$. It is natural to interpret the right hand side of 
(\ref{funSchroe-loc}) as an expectation value in an effective zero-mode quantum mechanics.
By localization, this quantum mechanics in finite 
volume can be shown to represent the exact result for  $\langle L \rangle_{{\epsilon_1,\epsilon_2}}$.
The functions $\Psi(a)$ can be identified with the instanton partition functions (see~\cite{Tachikawa:2014dja} for a review and references) that were found to be related to Liouville conformal blocks by AGT.

The approach proposed in \cite{TV13} establishes the relation 
between the wave-functions $\Psi(a)$ in 
\eqref{funSchroe-loc}
and the Liouville conformal blocks without using the relation to the instanton partition functions observed in 
\cite{Pestun:2007rz}. 
It is based on the observation that the effective 
zero-mode quantum mechanics in which line operators take the 
form (\ref{funSchroe-loc}) coincides with the quantum-mechanical system obtained by 
quantizing a real slice $\cM_\text{flat}^{\mathbb{R}}$ in $\cM_\text{flat}$.  This follows from the
fact that the algebra $\mathcal{A}_{\epsilon_1\epsilon_2}^{\rm line}$ 
generated by the supersymmetric 
line operators on  ${E}_{\epsilon_1,\epsilon_2}^4$ factorizes 
as $\cA_{\epsilon_1\epsilon_2}^\text{line}\simeq
\cA_{\epsilon_1/\epsilon_2}^\text{flat}\times\cA_{\epsilon_2/\epsilon_1}^\text{flat}$ 
into two copies of the
noncommutative algebra $\cA_{\hbar}^\text{flat}$ obtained in the quantization of the algebra
of coordinate functions on $\cM_\text{flat}$, as argued for instance in~\cite{Nekrasov:2010ka}. 
The two copies correspond to line operators supported on $x_0=x_1=x_2=0$ and $x_0=x_3=x_4=0$,
respectively.
The same conclusion can be reached from the 
observation made in~\cite{Ito:2011ea} that the algebra of line operators supported on $x_0=x_1=x_2=0$, 
for example, is isomorphic as 
a noncommutative algebra to the algebra of line operators in $\mathbb{R}^3\times S^1$ defined via 
\eqref{ncProductLLp} with $\hbar=\epsilon_1/\epsilon_2$.
The twisting of $\mathbb{R}^3\times S^1$ inducing the noncommutativity models the 
residual effect of the curvature near the support of the line operators on 
${E}_{\epsilon_1,\epsilon_2}^4$.

Duality invariance of the expectation 
values (\ref{funSchroe-loc}) 
may then be combined with the representation theory of $\mathcal{A}_{\epsilon_1\epsilon_2}^\text{line}$  
to obtain a precise mathematical characterization of the 
wave
functions $\Psi(a)\equiv\Psi_\tau(a)$, now considered as multivalued analytic functions of the gauge 
coupling constants $\tau=(\tau_1, \dots,\tau_h)$ \cite{TV13}. 
It was furthermore shown in 
\cite{TV13} that the Virasoro conformal blocks represent the same mathematical objects.
Within
conformal field theory one may, in particular,  define a natural family of operators 
called the Verlinde loop operators representing the action 
of the quantized algebra of functions on $\cM_\text{flat}$ on spaces of
conformal blocks. 

Having established the relation between wave-functions $\Psi_\tau(a)$ and the Liouville conformal blocks
it remains to notice that the functions $\Psi(a)$ in 
\eqref{funSchroe-loc} must coincide with the instanton partition functions  
defined in \cite{Nekrasov:2002qd}. 
Different arguments in favour of this identification 
can be found in \cite{Pestun:2007rz} and in \cite{Nekrasov:2010ka}. This line of argument establishes the  validity of the relations between conformal blocks
and instanton partition functions conjectured in~\cite{Alday:2009aq} for all $A_1$ theories 
of class $\cS$.\footnote{A somewhat similar approach to the case $\epsilon_2=0$ had previously been outlined in 
\cite{Nekrasov:2011bc}.} A  crucial role is played by the relation of the
algebra $\cA_{\hbar}^\text{flat}$ to the algebra $\cA^\text{Ver}$ generated by the Verlinde line operators.
This relation is generalized to theories of higher rank in our paper, thereby supporting 
the natural generalization of the AGT correspondence to class $\cS$ theories of type $A_{N-1}$.

\subsection*{Overview}

In this paper, we start a program to generalize the AGT correspondence to higher rank based on the central role of line operators and the moduli space of flat connections in the approach of~\cite{TV13}.

The six-dimensional origin of class $\cS$ theories of type $A_{N-1}$ suggests that there should 
exist a family of line operators that correspond to coordinate functions 
on the moduli space $\cM_{g,n}^N \equiv \cM_{\rm flat}^{SL(N,\bC)} (\cgn) $ of flat $SL(N,\bC)$-connections.
While for $A_1$ theories traces of holonomies along simple closed loops as in~\eqref{holo} were enough to parameterize $\cM_{g,n}^2$, for higher rank we consider in addition
some functions associated with \emph{networks} on $\cgn$.
These network functions are constructed from a collection of holonomies along open paths that are contracted at junctions with $SL(N)$-invariant tensors.
Such networks arise naturally from products of simple curves by applying $SL(N)$ skein relations at their intersections.
The algebra $\cA_{g,n}^N$ of functions on $\cM_{g,n}^N$ can be described in terms of a set of generators (loop and network functions) and the relations that they satisfy (section~\ref{secAlgebra}).
A standard way to quantize $\cA_{g,n}^N$ into a noncommutative algebra is to use quantum skein relations to resolve intersections in a product of operators
(this deformation can be defined using the Reshetikhin-Turaev construction of knot invariants in terms of quantum group theory). 
This gives a product of the form~\eqref{ncProductLLp}, where the operators $\hat L$ may  
also be network operators.

In section~\ref{secExamples}, we give several explicit examples of algebras $\cA_{g,n}^N$ and their quantizations for basic surfaces: three-puncture sphere $\pants$, one-punctured torus $\torus$, four-punctured sphere $\cC_{0,4}$. Using pants decompositions of surfaces $\cgn$,
we can build the algebras $\cA_{g,n}^N$ from these building blocks, in the spirit of the ``tinkertoys" approach \cite{Chacaltana:2010ks}.

In section~\ref{secFG}, we describe an explicit representation of the algebra $\cA_{g,n}^N$ in terms of Fock-Goncharov coordinates~\cite{FG}.
Loop and network functions are expressed in terms of positive Laurent polynomials, as in the relation~\eqref{UVIRlineRel} between UV and IR line operators.
The natural quantization of Fock-Goncharov coordinates then determines uniquely the quantization of $\cA_{g,n}^N$.
For all the cases that we compared, we find that the resulting quantum relations coincide with the ones obtained from skein quantization.
Furthermore, the existence of such quantum relations turns out to lead to a unique quantization of the Fock-Goncharov polynomials, as in~\eqref{linePSC}.
This allows us obtain many examples of framed protected spin characters~\eqref{PSCbarOmegay} in higher-rank theories.

In section~\ref{secCFT}, we define Verlinde network operators, which are natural generalizations of Verlinde loop operators~\cite{Alday:2009fs,Drukker:2009id} acting on spaces of conformal blocks in Toda field theory.
We show that the algebra $\mathcal{A}^\text{Ver}$ generated by the Verlinde network operators 
can be identified with the quantized algebra $\cA_{g,n}^N$.
To see this, we first observe that the braiding matrix in Toda field theory, from which the Verlinde network operators are built, is related via a twist to the standard R-matrix of the quantum group $\cU_q(sl_N)$. In turn, this R-matrix is used to construct the quantum skein algebra defining the quantized version of $\cA_{g,n}^N$.

As an outlook, we make a few observations in section~\ref{secSpectrum} about the comparison of the spectra of the operators representing the skein algebra in Toda field theory and in the quantum theory of the moduli space of flat connections, respectively.
The appendices collect some background about Fock-Goncharov coordinates and about quantum groups.

\newpage

\section{Algebra of loop and network operators}\label{secAlgebra}

This section describes relevant background on the algebra $\cA_{g,n}^N$ of functions on the moduli space $\cM_{g,n}^N$ of flat connections on a punctured Riemann surface $\cgn$, which can be described in terms of generators and relations.
We construct a set of generators for $\cA_{g,n}^N$ 
consisting of functions associated with simple loops and networks
naturally associated with a pair of pants decomposition of $\cgn$. 
Other functions can then be obtained by taking products of generators and resolving intersections with skein relations.
The number of generators obtained in this way typically exceeds
the dimension of the moduli space, as is reflected in the existence of 
polynomial relations between the coordinate functions.
The algebra $\cA_{g,n}^N$ has a Poisson structure, and it can be deformed into a noncommutative algebra 
$\cA_{g,n}^N(q)$ by applying the skein relations that encode the representation theory of the quantum group 
$\cU_q(sl_N)$.
We study in detail the case of pair of pants $\pants$, which is the building block for any $\cgn$.
This will illustrate the crucial role of networks at higher rank.
We also present some results for the punctured torus $\torus$ and the four-punctured sphere $\cC_{0,4}$.
Using pants decompositions one may use these results to get a set of coordinates 
allowing us to cover $\cM_{g,n}^N$ at least locally.

\subsection{Moduli space of flat connections}\label{modulispace}

The close relationship between 4d $\cN=2$ supersymmetric theories in class $\cS$ and Hitchin systems is revealed by compactifying on a circle $S^1$.
The moduli space $\cM_\text{vac}$ of vacua of such theories with gauge group $G$ on $\bR^3\times S^1$ can be identified with the moduli space of solutions to Hitchin's equations on $\cgn$ \cite{Gaiotto:2009hg}\cite{Gaiotto:2010be}.
These equations imply that the complex connections $\cA(\zeta)$ built out of a connection $A$ and a one-form $\varphi$, 
\bea
\cA (\zeta) = R  \zeta^{-1}\varphi + A + R \zeta \bar \varphi~,
\eea
are flat for all values of the parameter $\zeta \in \bC^*$ ($R$ is the radius of $S^1$).\footnote{We will restrict to the case $\zeta=1$ henceforth.}
$\cM_\text{vac}$ is a hyper-K\"ahler space, and with the appropriate choice of complex structure it is identified with the moduli space of flat $G_\bC$-connections on $\cgn$, with singularities at the punctures. We will only consider the cases
where the singularities at the punctures are of regular type\footnote
{Regularity means that the connection is gauge-equivalent to a meromorphic connection with simple poles at the punctures.}. 
Flat connections modulo gauge transformations are then
completely characterised by the representation of the fundamental group $\pi_1(\cgn)$ generated by the 
holonomy matrices.
The holonomy of a flat $G_\bC$-connection $\nabla=\dd + \cA$ along a closed curve $\gamma\in \pi_1(\cgn)$ is 
given by $\text{Hol}(\gamma) = \mathcal{P} \exp \int_\gamma \cA \in G_\bC$.
The moduli space $\cM_\text{vac}$ is thereby identified with the space of representations of 
$\pi_1(\cgn)$ into $G_\bC$  called the character variety:
\bea
\cM_\text{vac} \simeq \text{Hom}\big(\pi_1(\cgn), G_\bC\big)/G_\bC~.
\eea
More explicitly we have
\begin{equation*}
\cM = \bigg\{\, (\bfA_1, \ldots, \bfA_g, \bfB_1,\ldots, \bfB_g, \bfM_1, \ldots , \bfM_n ) \,|\, \prod_{i=1}^g \bfA_i\bfB_i\bfA_i^{-1}\bfB_i^{-1} = \prod_{a=1}^n {\bfM_a} \,\bigg\} \bigg/ G_\bC,
\end{equation*}
where ${\bfA_i,\bfB_i}\in G_\bC$ are holonomy matrices for based loops going around the A- and B-cycles for each of the $g$ handles, and ${\bfM_a}\in G_\bC$ are holonomy matrices for based loops going around each of the $n$ punctures.
These matrices are considered modulo the action of $G_\bC$ by simultaneous conjugation.

For 4d $\cN=2$ theories of type $A_{N-1}$, the complexified gauge group is $G_\bC=SL(N,\bC)$.
We are thus interested in the moduli space $\cM_{g,n}^N \equiv \cM_\text{flat}^{SL(N,\bC)}(\cgn)$ of flat $SL(N,\bC)$-connections on a Riemann surface $\cgn$, modulo gauge transformations.
It has a dimension given by
\bea\label{dimMgnN}
\dim[ \cM_{g,n}^N] &=& -\chi(\cgn) \dim[SL(N,\bC)]  = (2g+n-2) (N^2-1)~,
\eea
with the Euler characteristic $\chi(\cgn) = 2 - 2g -n$.
We can furthermore fix the conjugacy classes of the holonomies ${\bfM_a}$ around the punctures
(as we will see, this amounts to restricting to a symplectic leaf of the Poisson variety $\cM_{g,n}^N$).
The moduli space $\bar{\cM}_{g,n}^N$ of flat connections with $\bfM_a$ in fixed conjugacy classes has the dimension
\bea\label{dimMflatfixed}
\dim[\bar{\cM}_{g,n}^N] &=& -\chi(\cgn) \dim[SL(N,\bC)]  -n \text{rank}[SL(N,\bC)]\nn
&=&  (2g+n-2) (N^2-1) - n(N-1) ~.
\eea

\subsection{Trace functions} \label{SECalgfun}

The algebra $\cA_{g,n}^N\equiv \text{Fun}^\text{alg} (\cM_{g,n}^N)$ of algebraic functions on $\cM_{g,n}^N$ can be described in terms of generators and relations, as we now review.
Traces of holonomy matrices provide coordinate functions for $\cM_{g,n}^N$ (see e.g.~\cite{Goldman:2009} for a review).
They can be expressed as traces of words made out of letters given by the holonomy matrices ${\bfA_i,\bfB_i,\bfM_a}$.
The relation coming from the fundamental group $\pi_1(\cgn)$ allows us to eliminate one of the holonomy matrices, say $\bfM_n$, which leaves $(1-\chi)$ independent letters (for $n>0$).
General upper bounds are known for the maximal length of words that form a generating set of $\cA_{g,n}^N$.
The generators can be taken to be traces of words with lengths up to $N(N+1)/2$ for $N\leq 4$, or up to $N^2$ for $N>4$ (see references in~\cite{Sikora:2011}).
The difference between the number of generators and the dimension of $\cM_{g,n}^N$ is then accounted for by the existence of polynomial relations $\cP_\alpha=0$, which are consequences of the Cayley-Hamilton theorem.
The algebra $\cA_{g,n}^N$ is thus described as the polynomial ring generated by trace functions quotiented by polynomial relations:
\bea
\cA_{g,n}^N = \bC\left[ \tr {\bfA_i}, \tr {\bfA_i \bfB_j}, \cdots \right] / \{\cP_\alpha \} .
\eea

Note that this algebraic structure of $\cM_{g,n}^N$ does not distinguish between different surfaces with the same number of letters. 
For example, the description of $\cA_{g,n}^N$ in terms of generators and relations is the same for the one-punctured torus $\cC_{1,1}$ and for the three-punctured sphere $\cC_{0,3}$, which both have $(1-\chi) = 2$ letters.

\paragraph{Examples for $N=2$:}
Let us first consider the particularly simple example of $SL(2,\bC)$-connections on the one-punctured torus $\torus$.
The moduli space is given by
\bea
\cM_{1,1}^2 = \{ ({\bf A, B,M}) |   {\bf ABA}^{-1}\bfB^{-1} = \bfM  \} / SL(2,\bC) ~.
\eea
The algebra of functions $\cA_{1,1}^2$ is generated by the trace functions
\bea
\tr \bfA, ~ \tr \bfB, ~ \tr \bfA\bfB~.
\eea
Since the dimension of $\cM_{1,1}^2$ is 3, there is no relation between these 3 generators.
However, the generators are related to the trace of the holonomy around the puncture via
\bea\label{C11SL2relABM}
(\tr \bfA)^2+(\tr \bfB)^2+(\tr \bfA\bfB)^2-\tr \bfA\tr \bfB\tr \bfA\bfB = -\tr \bfM + 2~,
\eea
and therefore they do satisfy a relation once we fix the conjugacy class of $\bfM$ to obtain $\bar{\cM}_{1,1}^2$.

A more typical example is the sphere $\cC_{0,4}$ with four punctures, which we label by $A$, $B$, $C$, $D$.
The moduli space is 
\bea
\cM_{0,4}^2 = \{ (\bfA, \bfB,\bfC,\bfD) |   \bfA\bfB\bfC\bfD= \id  \} / SL(2,\bC) ~.
\eea
The holonomies $\bfA$, $\bfB$, $\bfC$ can be taken to be 3 independent letters, while $\bfD= (\bfA\bfB\bfC)^{-1}$.
The algebra of functions $\cA_{0,4}^2$ is generated by traces of words with maximal length equal to 3:
\bea
\bfA,\ \bfB,\ \bfC,\  \bfA\bfB\equiv  \bfS,\   \bfB\bfC\equiv \bfT,\ \bfC\bfA\equiv \bfU,\ \bfA\bfB\bfC=\bfD^{-1},\ \bfC\bfB\bfA~.
\eea
These 8 trace functions satisfy 2 polynomial relations (we use the notation $A_1 \equiv \tr \bfA$):
\bea \label{SumProdRelationsC04}
D_1 + \tr \bfC\bfB\bfA &=&   A_1 T_1  + B_1 U_1 + C_1 S_1 - A_1 B_1 C_1~,\\
D_1  \cdot \tr \bfC\bfB\bfA &=&  S_1T_1U_1 + S_1^2 + T_1^2 + U_1^2 +  A_1^2 + B_1^2 + C_1^2  \nn
&& -  A_1 B_1 S_1  - B_1 C_1 T_1 - C_1 A_1 U_1 -4 ~. \nonumber
\eea
The first relation allows to eliminate $\tr \bfC\bfB\bfA$ since it is linear.
The algebra of functions $\cA_{0,4}^2$ is then described as the quotient of the polynomial ring
\bea
\bC[A_1,B_1,C_1,D_1,S_1,T_1,U_1]
\eea 
by the quartic polynomial
\bea\label{quarticPolynC04}
\cP_1 &=&  S_1T_1U_1 + S_1^2 + T_1^2 + U_1^2 +A_1 B_1 C_1D_1+A_1^2 + B_1^2 + C_1^2 + D_1^2  \nn
&&  - (A_1 B_1 + C_1 D_1) S_1 - (B_1 C_1 + D_1 A_1  )T_1  -(C_1 A_1+ B_1 D_1 )U_1  -4~.
\eea
This gives a 6-dimensional quartic hypersurface in $\bC^7$.

In general, the number of words in $r$ letters of length up to 3 is
\bea
r + {r \choose 2} + {r \choose 3} = \frac{r(r^2 + 5)}{6}~.
\eea
This number of generators becomes quickly much larger than the dimension $3(r-1)$ of $\cM_{g,n}^2$, which implies that there are many polynomial relations.

\paragraph{Example for $N=3$:}
The description of $\cA_{g,n}^3$ for surfaces with 2 letters is very similar to that of $\cA_{g,n}^2$ for surfaces with 3 letters (see for example~\cite{Lawton:2006a} and references therein).
The generators of $\cA_{1,1}^3$ can be taken to be the traces of the 10 following words with length up to 6 (note that the Cayley-Hamilton theorem implies that $\bfA^{-1}\sim \bfA^2$ so it counts as 2 letters):
\bea \label{SL3LawtonGenerators}
\bfA^{\pm1},  ~  \bfB^{\pm1},~   (\bfA\bfB)^{\pm1} , ~
  (\bfA \bfB^{-1})^{\pm 1} ,~     (\bfA\bfB\bfA^{-1}\bfB^{-1})^{\pm 1}  ~.
\eea
These generators satisfy 2 relations (similar to the relations~\eqref{SumProdRelationsC04} for $\cA_{0,4}^2$):
\bea\label{LawtonRelSumProd}
\tr \bfA\bfB\bfA^{-1}\bfB^{-1} + \tr \bfB\bfA\bfB^{-1}\bfA^{-1} &=& (\ldots) ~,\nn
\tr \bfA\bfB\bfA^{-1}\bfB^{-1}   \cdot  \tr \bfB\bfA\bfB^{-1}\bfA^{-1} &=& (\ldots)   ~.
\eea
Eliminating $ \tr \bfB\bfA\bfB^{-1}\bfA^{-1} $ with the first relation, we can then describe $\cA_{1,1}^3$ as an 8-dimensional sextic hypersurface in $\bC^9$.

\subsection{Poisson structure}

There is a Poisson structure on the moduli space $\cM_{g,n}^N$, see~\cite{Audin} for a review and further
references.
Note that unlike the algebraic structure described in the previous subsection, the Poisson structure does distinguish between surfaces with the same $\chi(\cgn)$.
Goldman gave a general formula for the Poisson bracket of trace functions in terms of intersections of curves~\cite{Goldman:1986}:
\bea\label{GoldmanFormula}
\{\tr \Hol(\alpha), \tr \Hol(\beta ) \} = \sum_{p\in \alpha \cap\beta } \epsilon(p;\alpha,\beta)  \left[ \tr \Hol (\alpha_p  \beta _p) -\frac1N \tr \Hol(\alpha) \tr \Hol(\beta)  \right]~,
\eea
where $\epsilon(p;\alpha,\beta)=\pm 1$ is the oriented intersection number at the point $p$, and $\alpha_p, \beta _p$ are the curves $\alpha,\beta$ based at $p$.

As an illustration we can consider $\cC_{1,1}$ with $N=2$.
Since the A- and B-cycles intersect once, Goldman's formula gives
\bea
\{ \tr \bfA, \tr \bfB\} = \tr \bfA\bfB - \frac12 \tr \bfA \tr \bfB~.
\eea
Note that the right-hand side can be written as the derivative of the relation~\eqref{C11SL2relABM} by $\tr \bfA\bfB$.
This indicates that the Poisson structure on $\cM_{1,1}^2$ is compatible with its structure as an algebraic variety.
The Poisson algebra for the traces functions~\eqref{SL3LawtonGenerators} on $\cM_{0,3}^3$ and $\cM_{1,1}^3$ has been studied in~\cite{Lawton:2007}.

\subsection{Classical skein algebra} \label{SECskein}

\begin{figure}[t]
\centering
\includegraphics[width=\textwidth]{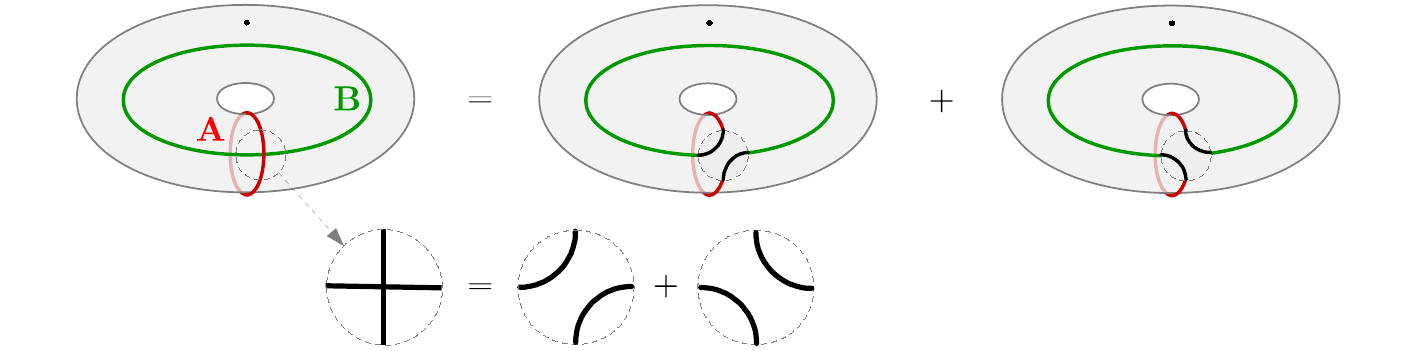}
\caption{\emph{Top}: The product of the trace functions for the A- and B-cycles on $\torus$ gives the trace functions associated with the curves obtained by resolving the intersection.
\emph{Bottom}: Locally, the crossing is replaced by two pairs of non-intersecting segments. This is the classical skein relation for $N=2$.}
\label{SkeinSL2}
\end{figure}

Relations between functions in $\cA_{g,n}^N$ have a topological origin.
Let us take again the example of $\cC_{1,1}$ with $N=2$.
The product of the traces associated with the A- and B-cycles is given by
\bea\label{skeintrAtrB}
\tr \bfA\tr \bfB = \tr \bfA\bfB + \tr \bfA\bfB^{-1}~.
\eea
Graphically, we can interpret this as resolving the intersection of the A- and B-cycles into a pair of curves that curl up around the torus in two ways (see figure~\ref{SkeinSL2}).
This procedure is reminiscent of the classical skein relations in knot theory, which are linear relations between knot diagrams (projections of knots onto a plane) that differ only locally around an intersection.
In fact, skein relations are nothing else than graphical representations of $\epsilon$-tensor identities such as $\epsilon_{ab}\epsilon^{cd} = \delta^c_a\delta^d_b - \delta^d_a \delta^c_b$, which can be used to derive the relation~\eqref{skeintrAtrB}.

The $SL(2)$ skein relation implies that $\cA_{g,n}^2$ can be described in terms of simple curves without self-intersections.
In the case of $\cC_{0,3}$ with $N=2$, the trace function $\tr \bfA \bfB^{-1}$, which corresponds to a figure-8 curve surrounding the punctures $A$ and $B$ and intersecting itself once, can be expressed in terms of non-intersecting curves as
\bea
\tr \bfA\bfB^{-1} &=& \bfA^a_d  \bfB^{-1b}_{\quad c} \delta^c_a\delta^d_b = \bfA^a_d \bfB^{-1b}_{\quad c} (\epsilon_{ab}\epsilon^{cd} +\delta^d_a \delta^c_b)  \nn 
&=&- \tr \bfA\bfB + \tr \bfA \tr \bfB~,
\eea
where we used $\bfB= - \epsilon (\bfB^\text{t})^{-1} \epsilon$ and $\tr \bfB^{-1}=\tr \bfB$.
Similarly, $\tr \bfC\bfB\bfA$ on $\cC_{0,4}$ corresponds to the Pochhammer curve with three self-intersections and can be expressed in terms of simple curves by applying the skein relation, as in~\eqref{SumProdRelationsC04}.

\begin{figure}[tb]
\centering
\includegraphics[width=\textwidth]{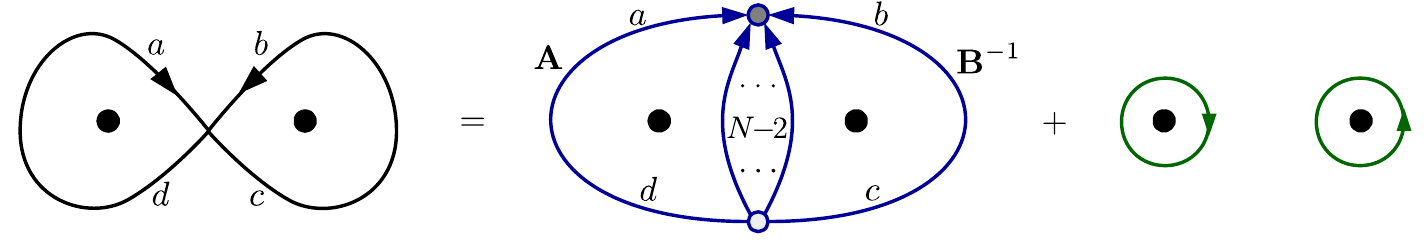}
\caption{The application of the skein relation for general $N$ to a self-intersecting curve produces a network with two $N$-valent junctions connected by an edge with multiplicity $N-2$.}
\label{netFig8Skein}
\end{figure}

An important difference for $N>2$ is that the skein relations involve $N$-valent junctions, associated with $SL(N)$-invariant $\epsilon$-tensors (see for example~\cite{Kuperberg}).
In the case $N=3$, the $\epsilon$-tensor identity
\bea
\delta_a^c\delta_b^d = \epsilon_{abm}\epsilon^{cdm}+ \delta_a^d \delta_b^c
\eea
corresponds to the skein relation expressed graphically as
\bea\label{SkeinSL3}
\includegraphics[width=0.93\textwidth]{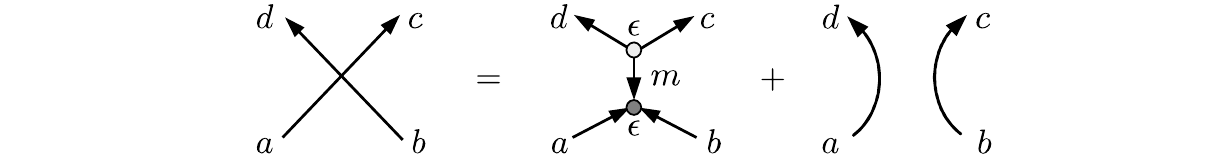}  
\eea
For general $N>2$, the resolution of a self-intersection via skein relations therefore produces a network with two junctions.
Let us consider again a figure-8 curve around two punctures:
\bea
\tr \bfA \bfB^{-1} &=&   \bfA^d_a  \bfB^{-1c}_{\quad b} \delta^a_c\delta^b_d = \bfA^d_a  \bfB^{-1c}_{\quad b} \left(\frac{1}{(N-2)!} \epsilon^{abm_1 \cdots m_{N-2}}\epsilon_{cdm_1 \cdots m_{N-2}} + \delta^a_d \delta^b_c \right)~. 
\eea
This relation can be represented graphically as in figure~\ref{netFig8Skein}.

The natural appearance of junctions in the algebra $\cA_{g,n}^N$ motivates us to adopt a set of generators for $\cA_{g,n}^N$ which does not consist exclusively of trace functions, but also includes network functions.

\subsection{Loop and network functions} \label{secLoopNetwork}

The set of generators of $\cA_{g,n}^N$ described in subsection~\ref{SECalgfun} involves traces of large words of the holonomy matrices $\bfA_i$, $\bfB_i$, $\bfM_a$, which are
generally associated with curves that have self-intersections.
In this paper, we will mostly trade such self-intersecting curves for networks with junctions using the 
skein relations.
Our generators are thus associated with simple loops (without self-intersections) and with networks.

\begin{figure}[tb]
\centering
\includegraphics[width=\textwidth]{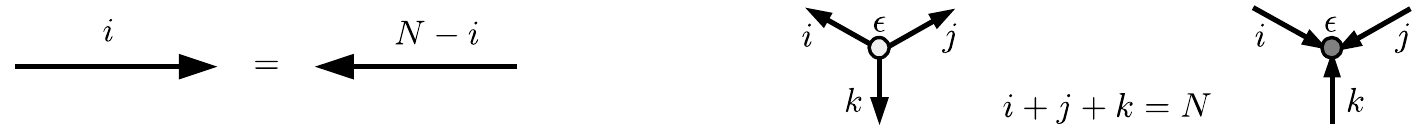}
\caption{\emph{Left}: Loop functions are depicted as oriented simple closed curves labeled by integers $i =1, \ldots , N-1$. Reversing the orientation amounts to replacing $i$ by $N-i$.
\emph{Right}: Trivalent junctions with all outgoing edges (source) or all incoming edges (sink). Edges are labeled by integers $i$, $j$, $k$ that sum to $N$.}
\label{fundAntifundArrows}
\end{figure}

\paragraph{Loops:}
Each loop on $\cgn$ gives $N-1$ trace functions $A_i$, $i=1, \ldots N-1$, which can be taken to be the coefficients of the characteristic polynomial of the associated holonomy matrix $\bfA  \in SL(N,\bC)$:
\bea\label{charpolyn}
\det (\bfA - \lambda \id) = (-\lambda)^N +(- \lambda)^{N-1}A_1+(- \lambda)^{N-2} A_2 +\cdots  -  \lambda A_{N-1} +1  ~.
\eea
The coefficients $A_i$ are sums of all principal $i\times i$ minors of $\bfA$:
\bea\label{AiSumsOfMinors}
A_i &=& \sum_{n_1, \cdots, n_{N-i}} [\bfA]_{n_1 \cdots n_{N-i}}~,
\eea
where we denote the determinant of $\bfA$ with the rows and columns $n_1$, \ldots, $n_{N-i}$ removed by
\bea\label{MinorDef}
 [\bfA]_{n_1 \cdots n_{N-i}} = \frac{1}{i!} \epsilon_{n_1 \cdots n_{N-i} m_1m_i} \bfA^{m_1}_{l_1} \cdots \bfA^{m_i}_{l_i} \epsilon^{n_1 \cdots n_{N-i} l_1 \cdots l_i}~.
\eea
We can also write the loop functions $A_i$ as traces of exterior powers of $\bfA$:
\bea\label{coeffCharPolAi}
A_1 &=& \tr \bfA~, \qquad A_2 = \frac12 [( \tr \bfA)^2 - \tr( \bfA^2)]~, \quad \ldots~, \quad 
A_i = \tr ( \wedge^i \bfA )~.
\eea
The loop function $A_i$ corresponds to the $i^\text{th}$ fundamental antisymmetric representation~$\wedge^i \square$ of $SL(N,\bC) $.
Note that if we replace $\bfA$ by its inverse $\bfA^{-1}$ in the expression for $A_i$ we obtain the complex conjugate representation $A_{N-i}$.
For example, the fundamental representation corresponds to $A_1= \tr \bfA$, while the antifundamental representation corresponds to $A_{N-1} = \tr \bfA^{-1}$.
We can thus represent loop functions graphically by an oriented loop labeled by an integer $i =1,\ldots, \lfloor{N/2}  \rfloor$, where $\lfloor{N/2}  \rfloor$ is the integral part of $N/2$.
Reversing the orientation corresponds to replacing $\bfA$ by $\bfA^{-1}$, the fundamental $A_1$ by the antifundamental $A_{N-1}$, and so on (see figure~\ref{fundAntifundArrows} left).
For even $N$, the $(N/2)^\text{th}$ representation is self-adjoint, and so the corresponding loop does not need an orientation.

\begin{figure}[h]
\centering
\includegraphics[width=\textwidth]{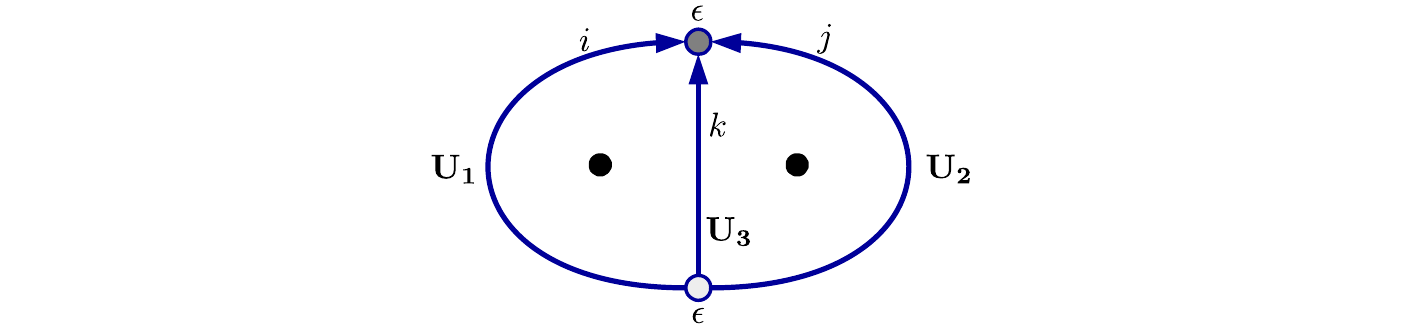}
\caption{Network with two trivalent junctions and three edges surrounding two punctures.
The edges carry the $i$\ts{th}, $j$\ts{th}, and $k$\ts{th} antisymmetric representations of $SL(N,\bC)$, with $i+j+k=N$. The network function $N_{ijk}$ is constructed by contracting $i$ copies of the holonomy matrix $\bf{U_1}$, $j$ copies of $\bf{U_2}$, and $k$ copies of $\bf{U_3}$ with the two $\epsilon$-tensors at the junctions.}
\label{networkExample}
\end{figure}

\paragraph{Networks:}
We also construct functions associated with networks.
By a \emph{network}\footnote{Closely related objects go under the names of spin networks, trace diagrams, tensor diagrams, birdtracks, webs, etc.} we mean a closed directed graph whose edges carry antisymmetric representations and whose vertices carry invariant tensors.\footnote{We prefer this definition to the one given by Samuel Johnson in his \emph{Dictionary of the English Language} (1755):
``Network: Any thing reticulated or decussated, at equal distances, with interstices between the intersections.''}
The vertices, or junctions, are in principle $N$-valent but they can always be resolved into trivalent junctions, on which we therefore focus.
The three edges that meet at a junction (either all outgoing or all incoming) are labeled by positive integers $i$, $j$, $k$ satisfying $i+j+k=N$, which indicate that they carry respectively the $i$\ts{th}, $j$\ts{th}, $k$\ts{th} antisymmetric representations (see figure~\ref{fundAntifundArrows} right).
Junctions do not have labels since there is only one invariant tensor $\epsilon_{m_1 \cdots m_N}$ in $\wedge^i \square \otimes \wedge^j\square \otimes \wedge^k \square$.
A network function is defined by contracting the holonomy matrices along the edges with the $\epsilon$-tensors at the junctions. The resulting function is invariant under simultaneous conjugation of the holonomy matrices.
For example, a network consisting of two junctions connected by three oriented edges as in figure~\ref{networkExample} gives $(N-1)(N-2)/2$ network functions $N_{ijk}$ of the form
\bea\label{networkdefined}
N_{ijk} = \frac1{i!j!k!} \epsilon_{m_1 \cdots m_N}  \bfU^{m_1}_{{\bf1}\ n_1}\cdots\bfU^{m_i}_{{\bf1}\ n_i}   \bfU^{m_{i+1}}_{{\bf2}\ n_{i+1}} 
\cdots \bfU^{m_{i+j}}_{{\bf2}\ n_{i+j} } \bfU^{m_{i+j+1}}_{{\bf3}\ n_{i+j+1}}\cdots \bfU^{m_N}_{{\bf3}\ n_N}  \epsilon^{n_1 \cdots n_N} ~,
\eea
where ${ \bf U_1,\bfU_2,\bfU_3}$ are the holonomy matrices (in the fundamental representation) along the three edges.
If we reverse the orientation of the edges we obtain another set of $(N-1)(N-2)/2$ network functions, which we denote by $\bar N_{ijk}$.

\paragraph{From networks to self-intersecting curves:}

Network functions can also be expressed in terms of the holonomy matrices
$\bfA_i$, $\bfB_i$, $\bfM_a$ generating a representation of $\pi_1(\cgn)$. In order to do this, let us recall
that a flat $SL(N)$-connection can always be trivialized by gauge transformations in
any simply-connected domain. Covering $\cgn$ by simply-connected domains one may describe
the flat connections in terms of the constant transition functions from one domain to another.
Equivalently, one may describe flat connections using branch-cuts, a collection of curves or
arcs on $\cgn$ such that cutting along the branch-cuts produces a simply-connected domain.
The holonomy $\text{Hol}(\gamma)$ will then receive contributions only when the curve
$\gamma$ crosses branch-cuts. $\text{Hol}(\gamma)$ can therefore be represented as the product
of ``jump"-matrices associated with the branch-cuts crossed by $\gamma$, taken in the order in which the
different branch-cuts are crossed.

\begin{figure}[h]
\centering
\includegraphics[width=\textwidth]{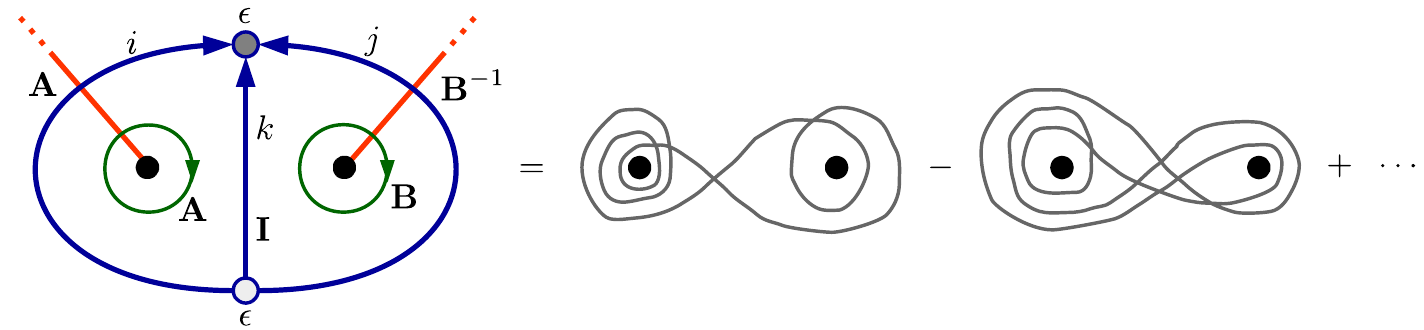}
\caption{The branch-cuts (orange lines) encode the holonomies $\bfA$ and $\bfB$ around the punctures. The edges of the network $N_{ijk}$ are associated with the holonomies of the branch-cuts that they cross: $i$ times $\bfA$, $j$ times $\bfB^{-1}$, and $k$ times the identity $\id$. The right-hand side shows two of the self-intersecting curves that appear in the expansion of the network function in terms of traces.}
\label{networkExampleCuts}
\end{figure}

Considering the network $N_{ijk}$ defined in~\eqref{networkdefined}, for example, we can use the
branch-cuts depicted in figure~\ref{networkExampleCuts}. The ``jump"-matrices associated with these
branch-cuts coincide with the holonomy matrices $\bfA$ and $\bfB$ around the two punctures
depicted in figure~\ref{networkExampleCuts}. The holonomy matrix associated with the middle arc labeled by the
letter $k$ is the identity, as no branch-cut is crossed. It follows that
we can express $N_{ijk}$ as
\bea\label{networkTraceRelation}
N_{ijk} &=& \frac1{i!j!k!}\epsilon_{m_1 \cdots m_N}  \bfA^{m_1}_{\ n_1}\cdots \bfA^{m_i}_{\ n_i} \bfB_{\quad n_{i+1}}^{-1m_{i+1}} 
\cdots \bfB_{\quad n_{i+j}}^{-1m_{i+j}} \delta^{m_{i+j+1}}_{n_{i+j+1}}\cdots \delta^{m_N}_{n_N}  \epsilon^{n_1 \cdots n_N}~, \nn
&=&\frac1{i!j!}  \delta_{m_1 \cdots  m_{i+j}  }^{n_1 \cdots  n_{i+j } }  \bfA^{m_1}_{\ n_1}\cdots \bfA^{m_i}_{\ n_i} \bfB_{\quad n_{i+1}}^{-1m_{i+1}} 
\cdots \bfB_{\quad n_{i+j}}^{-1m_{i+j}}   \nn
&=& \tr \bfA ^i \bfB ^{-j} - \tr \bfA^{i-1} \bfB^{-1} \bfA\bfB^{-j+1}  + \cdots
\eea

We thus see that this point of view relates network functions to trace functions associated with self-intersecting curves as in subsection~\ref{SECalgfun}.
As illustrated on the right of figure~\ref{networkExampleCuts}, the corresponding trace functions are typically associated with very complicated curves with many self-intersections.
This is one of the advantages of working with networks instead of self-intersecting curves. 
Different labelings of a network with a given topology account for a family of intricate curves.
Notice that the relation~\eqref{networkTraceRelation} between network and trace functions can also be seen as a consequence of skein relations, given that it involves contractions of $\epsilon$-tensors.
We have shown a simple example in figure~\ref{netFig8Skein}.

\subsection{Commuting Hamiltonians}\label{comHam}

The moduli space $\bar \cM_{g,n}^N$ with fixed conjugacy classes for the holonomies around the punctures exhibits the key features of an integrable Hamiltonian system.
Firstly, $\bar \cM_{g,n}^N$ is a symplectic manifold. 
Fixing the conjugacy classes of the holonomies around the punctures corresponds to restricting to a symplectic leaf of the Poisson manifold $\cM_{g,n}^N$ viewed as a symplectic foliation.

We can moreover find a maximal set of Hamiltonians, that is a number of Poisson-commuting functions equal to half the dimension of the moduli space $\bar \cM_{g,n}^N$.
Goldman's formula~\eqref{GoldmanFormula} implies that trace functions associated with curves that do not intersect each other automatically Poisson-commute. In the case $N=2$ we can simply consider the trace functions
associated to
a maximal number of mutually non-intersecting closed curves on $\cgn$.
Cutting $\cgn$ along such a collection of curves defines 
a decomposition of $\cgn$ into $(2g-2+n)$ pairs of pants with $(3g-3+n)$. 
The traces of holonomies associated with these curves thus provides a maximal set of 
Poisson-commuting Hamiltonians:
\bea
\# \{\text{cutting loops}\} = 3g-3+n = \frac12 \dim\bar \cM_{g,n}^2~.
\eea

\begin{figure}[t]
\centering
\includegraphics[width=\textwidth]{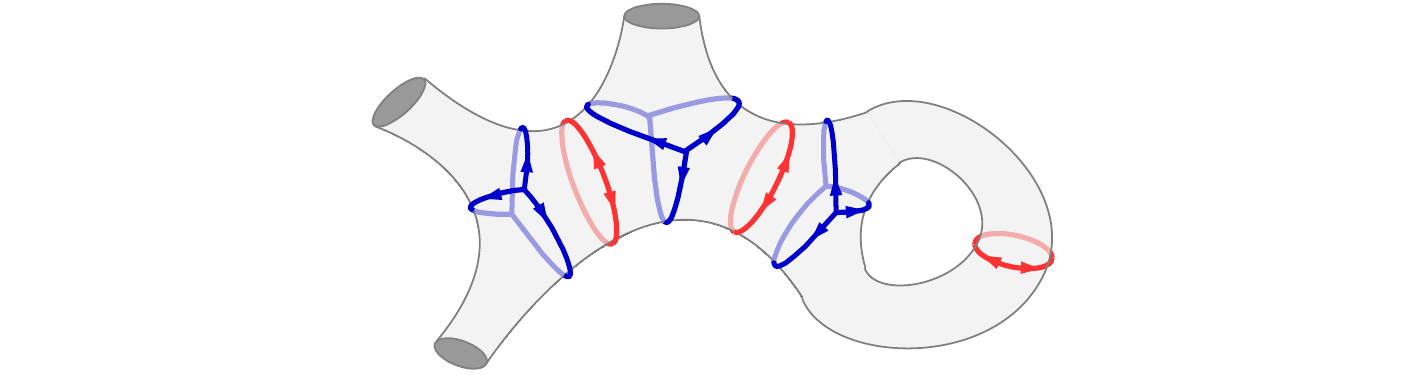}
\caption{A maximal set of commuting Hamiltonians is provided by the functions associated with cutting curves (red) and with pants networks (blue) in a pair of pants decomposition of the surface $\cgn$.}
\label{popdecomp}
\end{figure}

However, for $N>2$, the cutting curves alone do not suffice to get a maximal set of commuting Hamiltonians.
We therefore supplement them by the two-junction networks~\eqref{networkdefined} that can be put on each pair of pants (see figure~\ref{popdecomp}).
Each cutting curve provides $(N-1)$ trace functions, and each pants network provides $(N-1)(N-2)/2$ network functions.
They add up to give a number of functions which is precisely half the dimension~\eqref{dimMflatfixed} of the moduli space of flat connections with fixed holonomies around the punctures:
\bea
(3g-3+n)(N-1) + (2g-2+n) \frac{(N-1)(N-2)}{2} &=&  \frac12 \dim\bar \cM_{g,n}^N~.
\eea

The Poisson-commutativity of the trace functions associated to cutting curves among themselves  
and with the networks is obvious from the fact that they do not intersect (Goldman's formula~\eqref{GoldmanFormula} applies to self-intersecting curves, and hence to the networks as well).
On the other hand, it is not obvious that the network functions associated 
with pants networks Poisson-commute among themselves, $\{N_{ijk}, N_{lmn}\} = 0$.
Although we do not have a general proof, 
we checked that they commute up to the case $N=6$.

\subsection{Tinkertoys}\label{secTinkertoys}

In order to get a complete system of coordinates for  $\bar\cM_{g,n}^N$ we need to supplement 
the maximal set of commuting Hamiltonians described in the previous subsection by  sufficiently many additional
coordinate functions. 
A natural way to find additional  variables that do not Poisson-commute with the 
pants networks $N_{ijk}$ on $\cC_{0,3}$ is to take the pants networks $\bar N_{ijk}$ with reverse orientation. 

\begin{figure}[t]
\centering
\includegraphics[width=\textwidth]{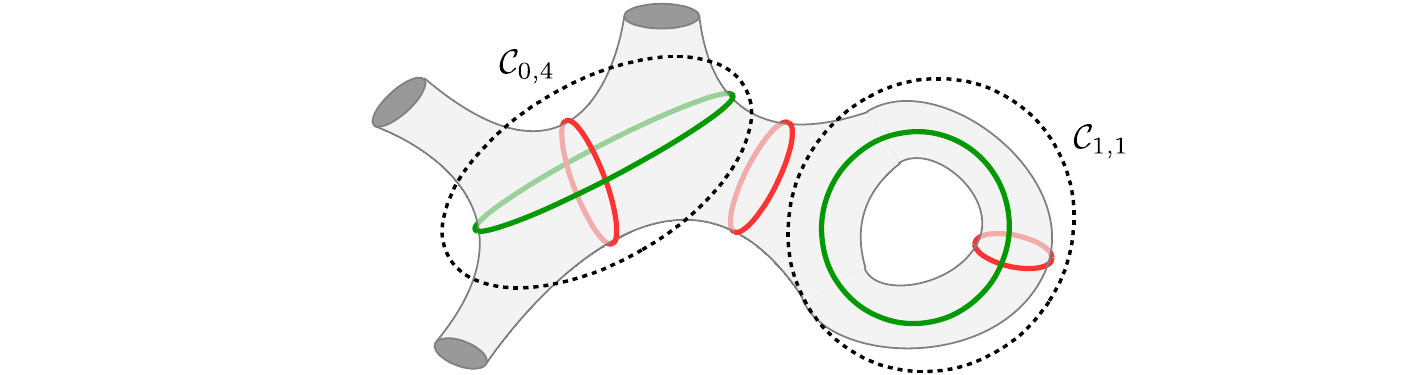}
\caption{The neighborhood of a cutting loop (red) on $\cgn$ can look either like $\cC_{0,4}$ or like $\torus$.
In each case we can find a natural conjugate loop (green).}
\label{tinkertoy}
\end{figure}

Simple additional coordinates that do not Poisson-commute with 
the trace functions associated to the cutting curves defining the pants decomposition can also be defined in a simple way. 
Each cutting curve is contained in a subsurface isomorphic either to a four-holed sphere or a one-holed torus embedded
in $\cgn$, see figure~\ref{tinkertoy} for an illustration.
In the case of $\cC_{0,4}$, natural additional coordinates are associated with curves surrounding 
another pair of holes than the cutting curve under consideration.
In the case of $\cC_{1,1}$ one may, for example consider an additional coordinate associated with the 
B-cycle if the cutting curve is the A-cycle, see figure~\ref{tinkertoy}.

Altogether, the cutting loops and their conjugate loops, together with the pants networks of both orientations provide a complete set of coordinates on $\bar\cM_{g,n}^N$ that cover this space at least locally.
We can therefore reduce the study of $\bar\cM_{g,n}^N$ for a generic Riemann surface $\cgn$ to the study of pants networks on $\pants$, and of pairs of conjugate loops on $\cC_{0,4}$ and $\torus$. 
This motivates us to focus on these three cases in the following subsections.

Note that the additional coordinates on $\bar\cM_{g,n}^N$ that we introduced 
above are not \emph{canonically} conjugate. However, it should be possible to 
define generalisations of the Fenchel-Nielsen 
coordinates, a set of Darboux coordinates for $\bar\cM_{g,n}^N$ in terms of which
one may parameterise the coordinate functions defined 
above. Such Darboux coordinates
were shown in the case $N=2$ to play a key role in the relation
to integrable systems~\cite{Nekrasov:2011bc}.

\subsection{Skein quantization} \label{SECquantization}

Motivated by the applications to supersymmetric
gauge theories, we will next discuss the  quantization of the 
moduli space $\cM_{g,n}^N$ of flat $SL(N,\bC)$-connections on a Riemann surface $\cgn$.
This means in particular to construct a family of noncommutative deformations $\cA_{g,n}^N(q)$ of
the algebra $\cA_{g,n}^N$ of functions on $\cM_{g,n}^N$ parameterized by one parameter $q\equiv \ex^\hbar$.
The loop and network functions get replaced by generators of the noncommutative algebra $\cA_{g,n}^N(q)$.
In the classical limit $q\to1$ ($\hbar\to 0$), the product $\hat A \hat B$ of two operators reduces to the commutative product $AB$ of the corresponding functions, while the commutator $[\hat A, \hat B]= \hat A \hat B - \hat B \hat A$ should reproduce the Poisson bracket $\{A, B\}$.

\begin{figure}[tb]
\centering
\includegraphics[width=\textwidth]{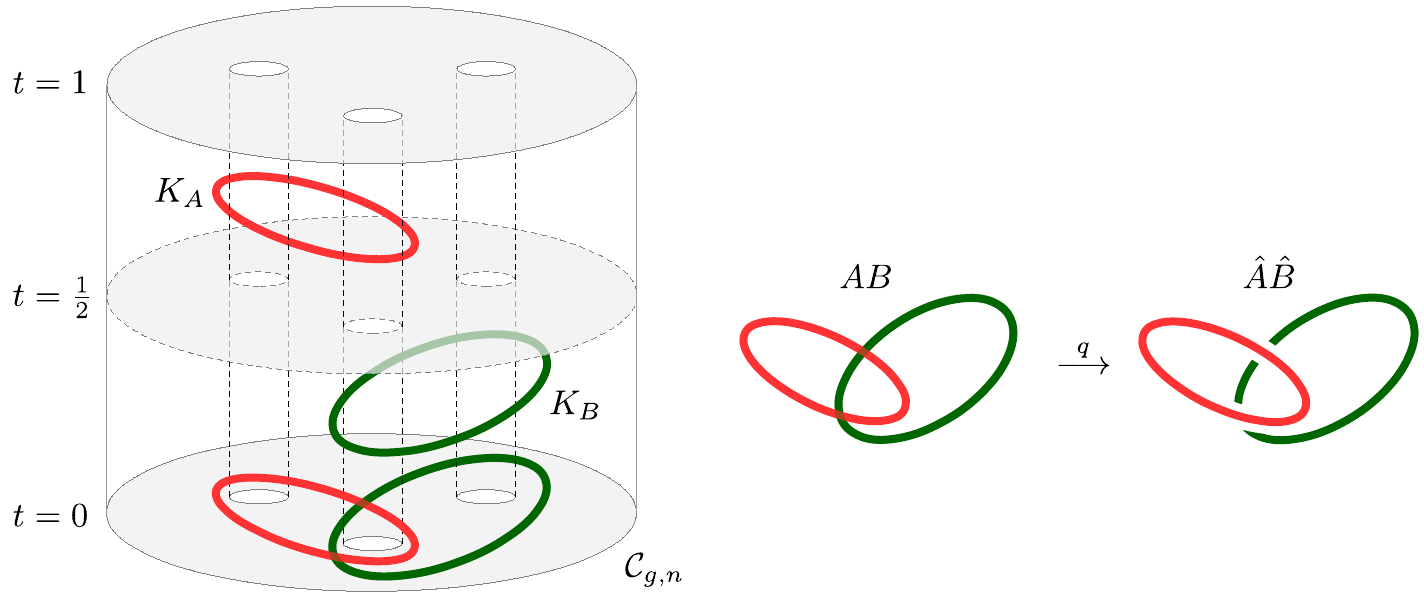}
\caption{\emph{Left}: Product of two knots $K_A$ and $K_B$ in the 3-manifold $\cgn \times [0,1]$. \emph{Right}: The corresponding loop operators that intersect classically are superposed at the quantum level.}
\label{TuraevProduct}
\end{figure}

This problem has been extensively studied in the past, starting from~\cite{Turaev:1991}, and motivated in 
particular by the relation to Chern-Simons theory\footnote{A lot of the research 
in this direction  was devoted to  Chern-Simons theories with compact gauge groups
like $SU(N)$. However, the resulting algebras $\cA_{g,n}^N(q)$ appearing in this
context turn out to be independent of the real form of the
corresponding complex group (here $SL(N)$) under consideration.}~\cite{WittenJones}\cite{Witten:1989rw}. 
Considering Chern-Simons theory on three-manifolds $M_3$ of the form $\cgn\times I$, with $I$ 
an interval with  coordinate~$t$, one may note that parameterised closed curves on $\cgn$ naturally define knots 
in $M_3$. In the context of Chern-Simons theory it is natural to 
relate the ordering of the factors in a product of generators in $\cA_{g,n}^N(q)$ to the ordering of 
observables according to 
the value of their ``time"-coordinates $t$. 
Given two knots  $K_A$ and $K_B$ one may define their formal product $K_AK_B$ to be  
the link composed of $K_A$ in $\cgn \times [1/2,1]$ and $K_B$ in $\cgn \times [0,1/2]$,
\begin{equation}\label{knotprod}
K_A K_B = \big\{ (x,t) \in \cgn \times [0,1] | (x, 2t-1) \in K_A \text{ for } t\geq \tfrac12~;  (x,2t)\in K_B 
\text{ for } t\leq \tfrac12 ~\big\}~.
\end{equation}
This operation is depicted in Figure \ref{TuraevProduct}. A natural set of relations to be imposed
on the product in $\cA_{g,n}^N(q)$ has been 
identified, severely constrained by the topological nature of Chern-Simons theory leading
to the definition of isotopy invariants of knots and links.

The first constructions of quantum $sl_N$ invariants
were provided by 
Reshetikhin and Turaev~\cite{RT1} using the representation 
theory of the quantum group $\cU_q(sl_N)$. It was later observed that
the resulting algebra can be described without the use of quantum groups in terms of generators and 
relations.
In the following we will briefly describe  the work 
of Sikora \cite{Sikora:2004} on link invariants in $\mathbb{R}^3$ which uses both points of view
(see also \cite{Kuperberg} for $SL(3)$,
and~\cite{MOY} for similar formulations).

Sikora describes in \cite{Sikora:2004} a 
construction of isotopy invariants of certain ribbon graphs called $N$-webs. The $N$-webs
are composed of oriented ribbons emanating from, or ending in, $N$-valent vertices called sources or sinks,
respectively (see \cite{Sikora:2004} for a more formal definition). We will see that the $N$-webs are closely related
to the networks considered in this paper.
This construction
can be understood as a special case of Reshetikhin and Turaev's constructions. 
It can be  described using the projections of  $N$-webs to $\mathbb{R}^2$ called 
$N$-web diagrams. The web-diagram may be  decomposed into pieces of three types: (i) crossings, 
 (ii) sinks or sources, and (iii) cups or caps 
of the form
\[
 \includegraphics[width=0.93\textwidth]{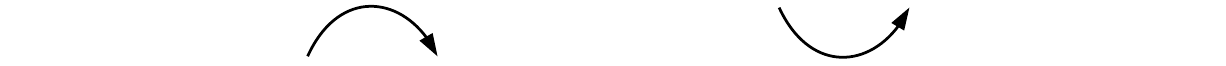} ~.
 \]
With each of these pieces one associates intertwining maps between 
tensor products of fundamental representations $\square$ of 
$\cU_q(sl_N)$. The maps associated with crossings, in particular, are in a basis 
$e_1, \ldots, e_N$ for $\mathbb{C}$ 
represented by 
\be
R(e_a \otimes e_b) = q^{\frac1{2N}}  \left\{ 
  \begin{array}{l l }
   e_b \otimes e_a  &   \qquad \text{for } a>b ~,\\
   q^{-\frac12} e_a \otimes e_b  &   \qquad \text{for } a=b~, \\
    e_b \otimes e_a   + (q^{-\frac12} - q^{\frac12}) e_a \otimes e_b &   \qquad \text{for } a<b ~,
  \end{array} \right. 
\ee
or by the inverse map $R^{-1}$, depending on which edge is on top of the other:
\bea\label{RmatCrossings}
\includegraphics[width=0.93\textwidth]{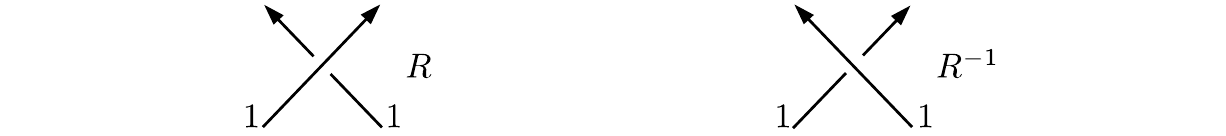} ~.
\eea
The maps associated with sinks are the unique (up to normalization) intertwining maps from the 
$N$-fold tensor product of the fundamental representation of $\cU_q(sl_N)$ to the trivial representation,
while the conjugate of this map is associated with the sources. In a similar way one associates to the caps
the unique (up to normalization) intertwining maps from the tensor products of fundamental
representations $\square$ with the anti-fundamental representations $\bar\square$ to the the trivial 
representation. The maps associated with the cups are the conjugate intertwining maps, 
respectively. Explicit formulae can be found in \cite{Sikora:2004}.
Using these building blocks one constructs the invariant associated with an $N$-web by composing 
the intertwining maps associated  to the pieces in the natural way specified by the decomposition 
of the given $N$-web diagram into pieces.

The isotopy invariants of $N$-webs defined in this way satisfy various relations that 
can be used
to calculate them explicitly.
These relations relate invariants associated with $N$-web diagrams that are identical outside of 
suitable discs $\mathbb{D}\subset\mathbb{R}^2$. A typical  example  may be graphically
represented as
\bea\label{qSkeinCrossSLN}
\includegraphics[width=0.93\textwidth]{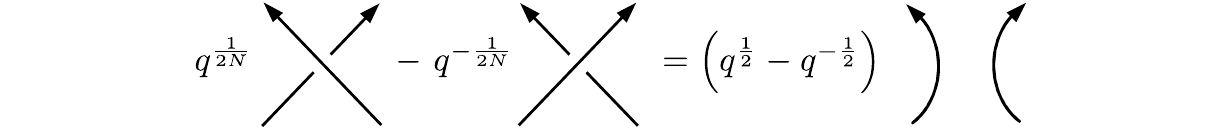}  
\eea
Such relations are quantum analogs of the skein relations discussed previously.
Before entering into a more detailed description of the skein relations, let us note that the local 
nature of the skein relations will allow us to use the same relations as defining relations for the
algebras $\cA_{g,n}^N(q)$ we are interested in.  This will be the basis for the approach 
used in the next section, as is illustrated  in  particular by Figure \ref{C11SkeinAB} below.
The three-dimensional isotopy invariance of the
$N$-web invariants ensures that the resulting algebra has a three-dimensional interpretation 
via \eqref{knotprod}. It is easy to see that the relation~\eqref{qSkeinCrossSLN} 
reproduces Goldman's bracket~\eqref{GoldmanFormula} in the limit $\hbar\rightarrow 0$.

We shall now turn to a more detailed description of the set of relations proposed in~\cite{Sikora:2004}.
The first condition in~\cite{Sikora:2004} is the crossing condition~\eqref{qSkeinCrossSLN}.\footnote
{We choose conventions that agree with the calculations in terms of Fock-Goncharov holonomies in section~\ref{secFG}.
They are related to the bracket used in~\cite{Sikora:2004} by the redefinition $q \to q^{-2}$, and then the renormalization of each junction by $\ii q^{-N(N-1)/4}$ and of each edge carrying the $i$\ts{th} antisymmetric representation by $1/[i]!$. We also introduce some signs in~\eqref{qUnknotReide1}.} 
Next, the quantum invariant for the union of two unlinked knots must be equal to the product of the quantum invariants for the knots.
There are also conditions for the contraction of a trivial knot and for the Reidemeister move of type I:
\bea\label{qUnknotReide1}
\includegraphics[width=0.93\textwidth]{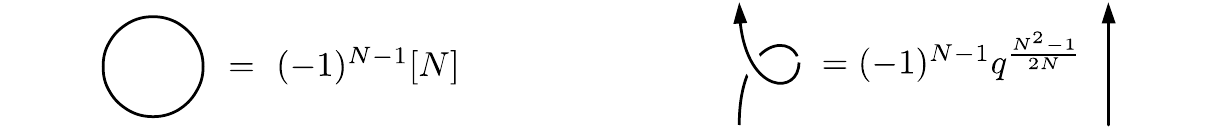}  
\eea
We have been using the notation $[n]$ defined as
\bea
[n] \equiv \frac{q^{n/2}-q^{-n/2}}{q^{1/2}-q^{-1/2}}= q^{(n-1)/2} + q^{(n-3)/2} + \cdots +q^{-(n-1)/2} ~.
\eea
Finally, there is a relation between two nearby $N$-valent junctions (a source and a sink) and a sum of positive braids labeled by permutations $\sigma$ (with lengths $l(\sigma)$):
\bea\label{SikoraSourceSink}
\includegraphics[width=0.93\textwidth]{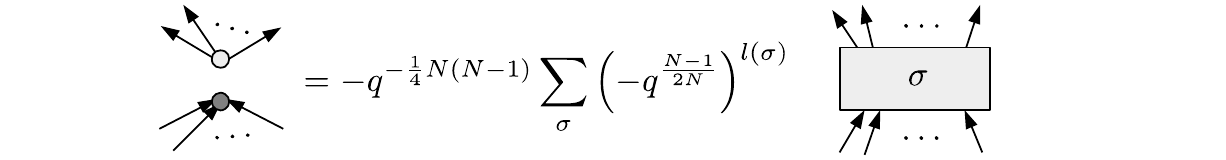}  
\eea
It was shown in~\cite{Sikora:2004} that the relations
above suffice to characterize the resulting invariant of $N$-webs uniquely.

Note that the edges do not carry labels in Sikora's formulation. For our goals it will be convenient to represent 
$i$ parallel edges between two junctions by a single edge with label $i$. This will allow us to define the quantized
counterparts of the networks introduced previously. A quantum network corresponding to the 
network  shown in figure~\ref{networkExample}, for example, may be represented by an $N$-web 
obtained by splitting the $N$ edges connecting one source and one sink
into three groups of $i$, $j$, and $k$ edges.

\begin{figure}[t]
\centering
\includegraphics[width=\textwidth]{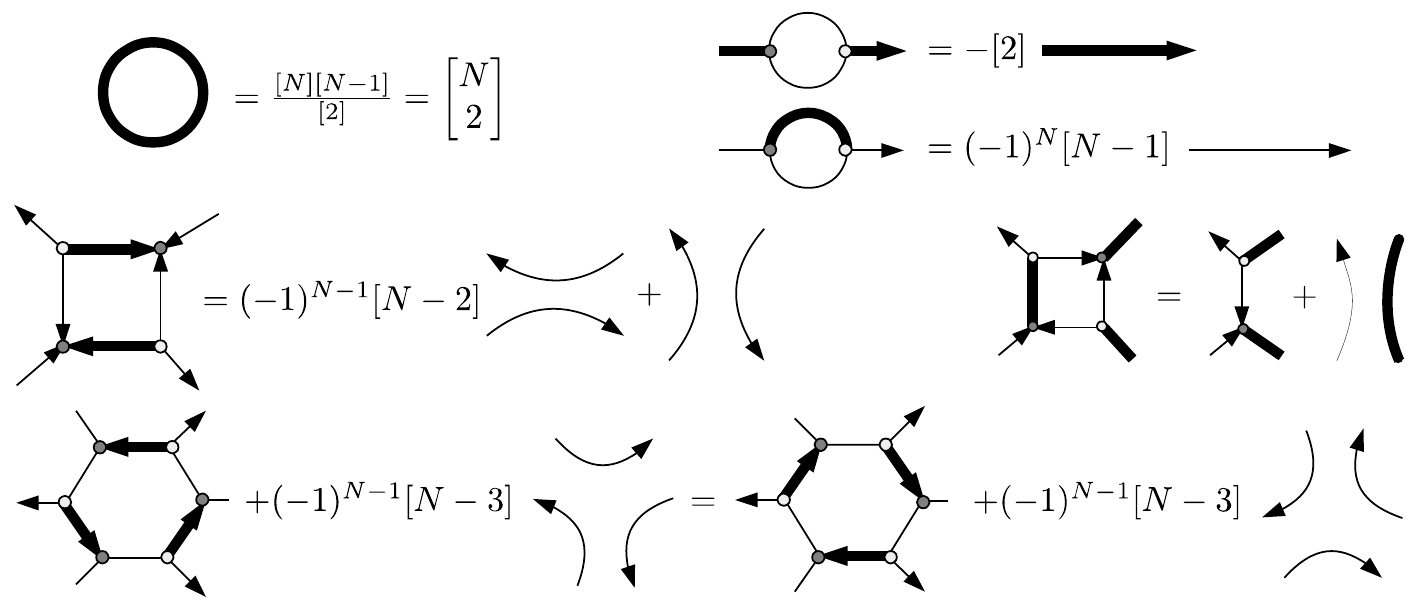}
\caption{Reduction of contractible cycles. The square reduction on the right is valid for $N=4$. Our conventions are that thin edges carry the fundamental representation, while thick edges carry the $(N-2)$\ts{th} antisymmetric representation.}
\label{ReductionSLN}
\end{figure}

The relation \eqref{SikoraSourceSink}
allows to derive skein relations for the resolution of all possible crossings in terms of $N$-web diagrams
without crossings.
Of particular interest is the following 
special case of the fundamental skein relation
obtained by contracting $(N-2)$ pairs of edges from the upper and lower parts of~\eqref{SikoraSourceSink}:
\bea\label{qSkeinSLN}
\includegraphics[width=0.93\textwidth]{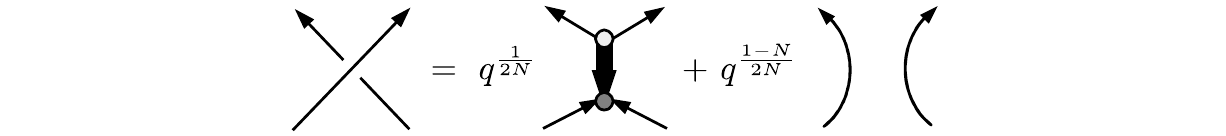}  
\eea
We indicate that the edge between the two junctions carries the label $N-2$ by drawing it thicker than the other edges associated with the fundamental representation.
The fundamental skein relation with the other ordering at the crossing has $q$ replaced by $q^{-1}$.
A large set of useful relations can be derived from the relations stated above, including 
reduction moves of contractible bubbles (digons), squares, hexagons, etc. Such relations 
were worked out in~\cite{KimGraphical,Morrison07,CKMwebs12}.
We show some examples in figure~\ref{ReductionSLN}.

Sikora's construction  allows one to recover the construction 
of quantum $sl_N$ invariants previously  given by Murakami, Ohtsuki, and Yamada in~\cite{MOY} 
(a useful summary is given in~\cite{Tachikawa:2015iba}). This construction uses
trivalent graphs with a ``flow'' built out of the following two types of vertices:
\bea\label{MOYvertices}
\includegraphics[width=0.93\textwidth]{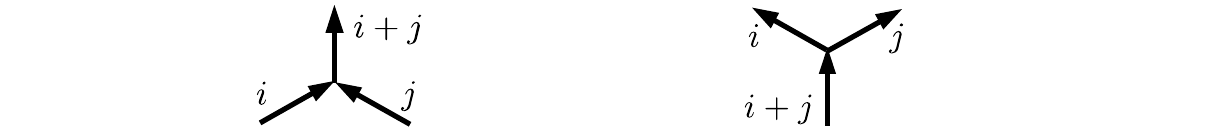}  
\eea
The edges connected at such vertices cary labels with values in 
$\{0,1,\ldots, N-1\}$. 
An edge with label 0 can be removed, and an edge with label $i$ is equivalent  
to an edge with label $N-i$ with opposite orientation, as depicted on the left of figure~\ref{fundAntifundArrows}.
The vertices \eqref{MOYvertices} can be represented in terms of pairs of the sources and sinks used in Sikora's
formulation, as explained in~\cite{Sikora:2004}.

It is possible to derive an expression for a general skein relation resolving the crossing of lines labeled by 
arbitrary $i,j\in\{0,1,\ldots, N-1\}$
\cite{MOY} (see~\cite{Tachikawa:2015iba} for the normalization):
\bea\label{generalCrossingMOY}
\includegraphics[width=0.93\textwidth]{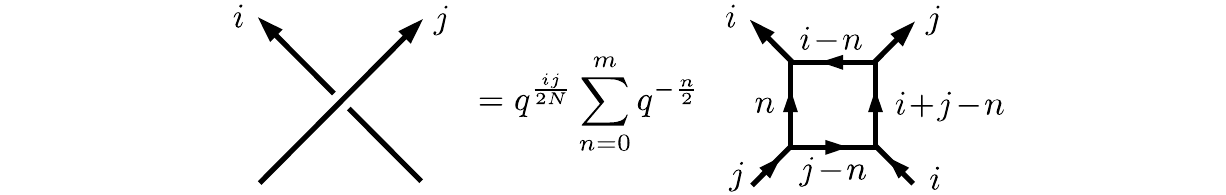}  
\eea
with $m = \text{min}\{i,j,N-i,N-j\}$. 
For the other ordering at the crossing, one should replace $q$ by $q^{-1}$.
When $i=j=1$ this expression reproduces the fundamental skein relation~\eqref{qSkeinSLN}.
We will also need skein relations for $N=4$ with $i=2$ (thick line) and $j=1$ or $j=2$:
\bea\label{SL4Skein12}
\includegraphics[width=0.93\textwidth]{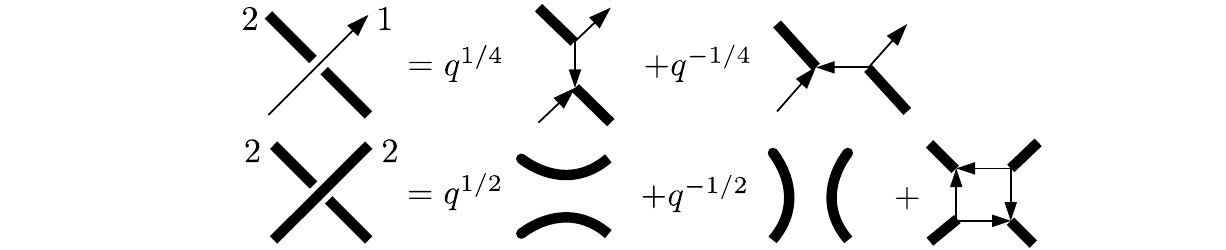} 
\eea

Let us finally note that the link invariants constructed in \cite{MOY} also correspond to 
simple special cases of the Reshetikhin-Turaev construction.
The label $i\in\{0,1,\dots,N-1\}$ assigned to an edge  of a colored $N$-web is 
identified with the label for one of the irreducible 
representation $M_i\equiv\wedge^i\square$ of the quantum group $\cU_q(sl_N)$
that is obtained as the $i$\ts{th} antisymmetric tensor power of the fundamental representation $\square$.
The linear map $R_{ij}:M_j\otimes M_i\rightarrow M_i\otimes M_j$ 
appearing on the left hand side of \eqref{generalCrossingMOY}
can be obtained from the universal R-matrix $\cR$ of $\cU_q(sl_N)$ via
\begin{equation}\label{Reval}
R_{ij}\,=\,P_{ij}(\pi_j\otimes \pi_i)(\cR)\,,
\end{equation} 
with $P_{ij} $ the permutation of tensor factors,
$P_{ij}:M_j\otimes M_i\rightarrow M_i\otimes M_j$.
The trivalent vertices in~\eqref{MOYvertices} are associated with the Clebsch-Gordan maps (with $k=i+j$):
\bea
C^{k}_{ij}: \ M_i \otimes M_j \rightarrow M_k ~,\qqq  C_{k}^{ij}: \ M_k \rightarrow M_i \otimes M_j~.
\eea
In the case where $i+j=N$, one edge carries the trivial representation and can thus be removed. This gives cap and cup maps:
\bea
& \includegraphics[width=0.93\textwidth]{CapCup}& \nn
& C^{0}_{i,N-i}: \ M_i \otimes M_{N-i} \rightarrow \bC  ~,\qqq  
C_{0}^{N-j,j}: \ \bC \rightarrow M_{N-j} \otimes M_j~. &
\eea
Quantum invariants of a network are then obtained by composing the intertwining maps
associated with the pieces of the network.

\section{Quantization of tinkertoys} \label{secExamples}

In this section we will describe the algebras  obtained by using skein quantization in 
some simple examples associated with surface $\cgn$ with 
$(g,n)$ being $(0,3)$, $(1,1)$ and $(0,4)$. As explained previously, it seems reasonable to 
regard the results as building blocks for the description of the algebras associated with more general
surfaces $\cgn$.

\subsection{Pants networks}\label{secPantsAlgebra}

\begin{figure}[tb]
\centering
\includegraphics[width=\textwidth]{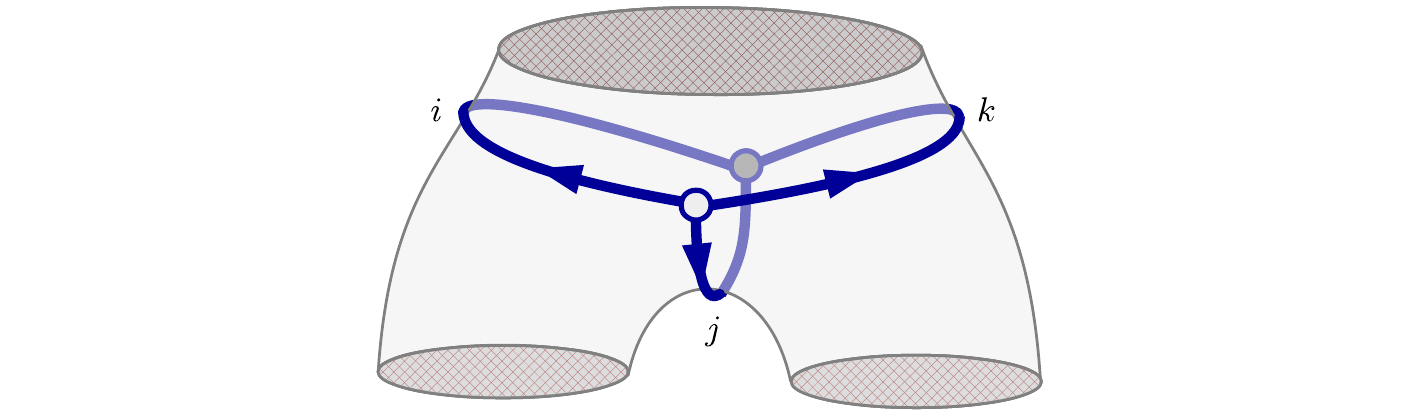}
\caption{Pants network on a pair of pants $\cC_{0,3}$.}
\label{C03panties}
\end{figure}

As the prototypical illustration of the role of network operators, we consider flat $SL(N,\bC)$-connections on a three-punctured sphere $\cC_{0,3}$, also known as the pair of pants, or pants for short. 
As we mentioned in subsection~\ref{comHam}, any Riemann surface $\cgn$ can be decomposed into pants by choosing a maximal set of simple loops that do not intersect. The pair of pants $\pants$ is hence not merely the simplest example, but also the most essential one, from which any other surface can in principle be understood.
The main novelties for the case $N>2$ will be apparent in this example.
Indeed, any simple loop on $\cC_{0,3}$ can be deformed into a loop surrounding a puncture, so networks are the only relevant objects in this case. A particularly important network has two trivalent junctions and three edges, running between every pair of punctures; we call it the \emph{pants network} (see figure~\ref{C03panties}).
The number of possible pants network operators is given by the partition of $N$ into three strictly positive integers $i+j+k=N$, which gives ${N-1 \choose 2}= (N-1)(N-2)/2$.

In terms of 4d $\cN=2$ gauge theories, $\pants$ corresponds to the theory $T_N$ studied by Gaiotto~\cite{Gaiotto:2009we}, which can be used as a fundamental building block for more general theories.
$T_N$ is a strongly coupled $\cN=2$ superconformal field theory with no known weakly-coupled Lagrangian description (except for $T_2$, which is free).
It has $SU(N)^3$ flavor symmetry and $SU(2)\times U(1)$ R-symmetry.
It contains operators $Q$ and $\tilde Q$ with scaling dimension $(N-1)$ that transform in the trifundamental representations $(\square,\square,\square)$ and $(\bar \square,\bar \square,\bar\square)$ of $SU(N)^3$.
There are also Higgs branch operators $\mu_1,\mu_2,\mu_3$, which have scaling dimension 2 and transform in the adjoint of one $SU(N)$.
Finally, there are Coulomb branch operators $u_k^{(i)}$ with dimension $k$;
the labels take the values $k=3,\ldots,N$ and $i=1,\ldots, k-2$, so their number is $(N-1)(N-2)/2$.
We see that this matches nicely the number of pants networks.

The fundamental group of the sphere $\pants$ with three punctures $A,B,C$ is represented by the loops $\gamma_{A}, \gamma_{B}, \gamma_{C}$ around each puncture satisfying one relation: 
\bea
\pi_1 (\mathcal{C}_{0,3}) = \{ (\gamma_{A}, \gamma_{B}, \gamma_{C} ) | \gamma_{A}\gamma_{B} \gamma_{C} = \gamma_\circ  \}~,
\eea
where $\gamma_\circ$ denotes a contractible loop.
The corresponding holonomy matrices $\bfA$, $ \bfB$, $\bfC$ satisfy $\bfA \bfB\bfC =(-1)^{N-1} \id$
(the sign is chosen for consistency with~\eqref{MOYvertices} and~\eqref{FGsnakeRectHex}).
The moduli space of flat $SL(N,\bC)$-connections has the dimension
\bea
\dim [\cM_{0,3}^N] = N^2-1~.
\eea
The functions coming from the loops around the punctures, $A_i$, $B_i$, $C_i$, with $i=1, \ldots, N-1$, and from the pants network with both orientations, $N_a$, $\bar N_a$, with $a=1, \ldots, (N-1)(N-2)/2$, provide the correct number of coordinates on $\cM_{0,3}^N$:
\bea
3(N-1) + 2 \frac{(N-1)(N-2)}{2} &=& N^2-1 ~.
\eea
Fixing the eigenvalues of the holonomies around the punctures then gives $3(N-1)$ constraints and leaves us with only the pants networks:
\bea
\dim [\bar\cM_{0,3}^N] =  (N-2)(N-1)~.
\eea

 \subsubsection*{SL(3)} 

The first non-trivial case is $N=3$. 
We will show how to obtain a closed Poisson algebra involving the loops and pants network, together with an extra six-junction network.
This gives 10 generators satisfying 2 polynomial relations, which can be quantized using quantum skein relations.

\paragraph{Loop and network functions:}

\begin{figure}[tb]
\centering
\includegraphics[width=\textwidth]{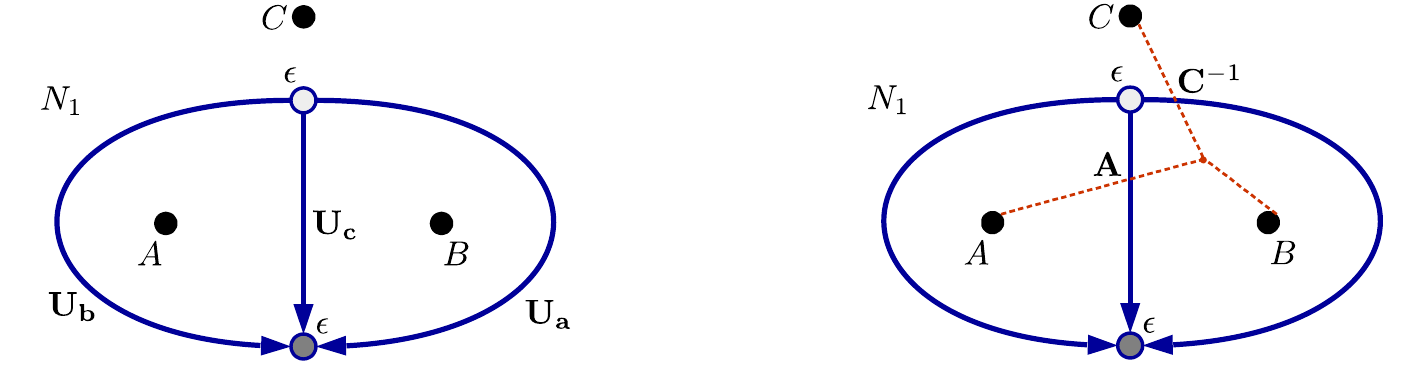}
\caption{\emph{Left}: Construction of the network operator $N_1$ on $\pants$ from the holonomy matrices $\mathbf{U_a}$, $\mathbf{U_b}$, $\mathbf{U_c}$ along its edges contracted with $\epsilon$-tensors at the junctions. \emph{Right}: The holonomy matrices $\mathbf{A}$, $\mathbf{B}$, $\mathbf{C}$ are associated with the branch-cuts (dashed) starting at the punctures $A$, $B$, $C$. Two edges of the network $N_1$ intersect the branch-cuts and are thus associated with $\mathbf{A}$ and $\mathbf{C}^{-1}$.}
\label{cutpants}
\end{figure}

There are two loop functions for each holonomy matrix, namely the coefficients of the characteristic polynomial, which can be expressed in terms of traces as in~\eqref{coeffCharPolAi}: $A_1=\tr \mathbf{A}$ and $A_2=\tr \mathbf{A}^{-1}$ (and similarly for $\bfB$ and $\bfC$).
The network function $N_1$ and its reverse $\bar N_1$ can be constructed as in~\eqref{networkdefined} by fusing the three edges at the two trivalent junctions with $\epsilon$-tensors (see figure~\ref{cutpants} left):\footnote{The overall sign is chosen for later convenience, so that $N_1$ will be expressed as a \emph{positive} Laurent polynomial in the Fock-Goncharov coordinates, as in~\eqref{N1fullExpr}.}
\bea\label{pantsN1VabcC03}
N_1 &=& - \epsilon_{mnp} \bfU_{{\bf a}\, r}^{m}\bfU_{{\bf b}\, s}^{n} \bfU_{{\bf c}\, t}^{p} \epsilon^{rst}~, \nn
\bar N_1 &=& - \epsilon_{mnp} 
(\bfU_{\bf a}^{-1})^m_r(\bfU_{\bf b}^{-1})^n_s(\bfU_{\bf c}^{-1})^p_t \epsilon^{rst}~.
\eea

\begin{figure}[h]
\centering
\includegraphics[width=\textwidth]{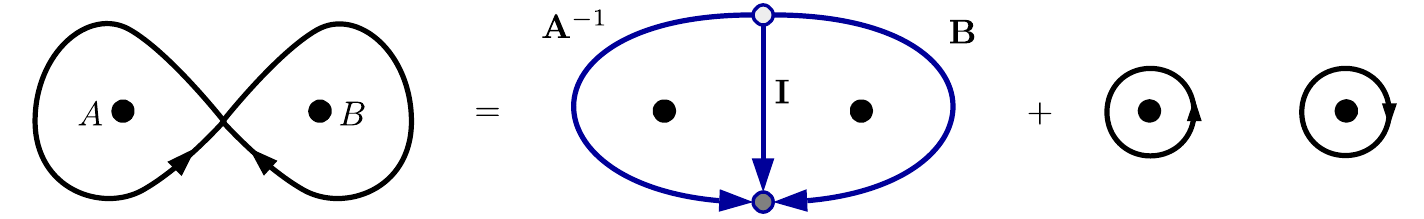}
\caption{Applying the skein relation to a self-intersecting loop going around two punctures produces a network and pair of loops around the punctures.}
\label{netFig8Skein2}
\end{figure}

Alternatively, we can construct the network functions as in \eqref{networkTraceRelation} by associating holonomy matrices with the edges of the network according to which branch-cuts they cross (see figure~\ref{cutpants} right). This gives the following expressions:
\bea \label{N1ACbC03}
N_1 &=& - \epsilon_{mnp} \mathbf{A}^{m}_{r}\delta^{n}_{s}(\mathbf{C}^{-1})^{p}_{t} \epsilon^{rst} = \tr \mathbf{A C^{-1}} -  A_1 C_2~, \nn
\bar N_1 &=& - \epsilon_{mnp} (\mathbf{A}^{-1})^{m}_{r}\delta^{n}_{s}\mathbf{C}^{p}_{t} \epsilon^{rst} =  \tr \mathbf{A^{-1} C} -  A_2 C_1~.
\eea
The several possible choices for the position of the branch-cuts all lead to the same network function:
\bea\label{network8loopC03}
N_1 &=& \tr \mathbf{A C^{-1}} -  A_1 C_2 = \tr \mathbf{B A^{-1}} -  B_1 A_2 = \tr \mathbf{C B^{-1}} -  C_1 B_2 ~.
\eea
A term such as $\tr \mathbf{B A^{-1}}$ corresponds to a self-intersecting figure-8 loop going around the punctures $B$ clockwise and $A$ anticlockwise.
Resolving the intersection with the skein relation in~\eqref{SkeinSL3} produces the relation~\eqref{network8loopC03} (see figure~\ref{netFig8Skein2}).

%

\paragraph{Closed Poisson algebra:}

\begin{figure}[tb]
\centering
\includegraphics[width=\textwidth]{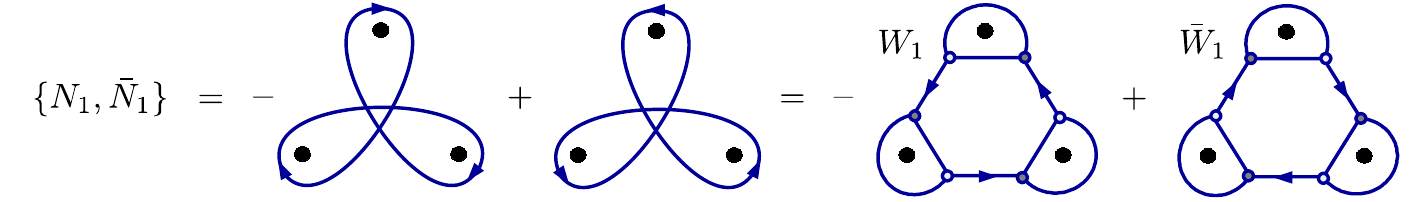}
\caption{The Poisson bracket of the pants networks $N_1$ and $\bar N_1$ can be expressed in terms of Pochhammer curves or in terms of the 6-junction network $W_1$ and its reverse $\bar W_1$.}
\label{C03PBN1N2}
\end{figure}

We would like to find a set of generators including $A_i$, $B_i$, $C_i$ and $N_1$, $\bar N_1$ that forms a closed Poisson algebra.
The Poisson brackets can be obtained from Goldman's formula~\eqref{GoldmanFormula}.
Since the Poisson bracket is proportional to the intersection number, the loop functions $A_i$, $B_i$, $C_i$ around the punctures obviously Poisson-commute with everything:
\bea
\{ A_i , \bullet \} = \{ B_i , \bullet \} =\{ C_i , \bullet \} =0~.
\eea
To apply Goldman's formula to the pants networks we use their expressions~\eqref{N1ACbC03} in terms of trace functions.
We get
\bea\label{PBN1N1b}
\{N_1,  \bar N_1\} &=& \{ \tr \bfA \bfC^{-1} , \tr \bfA^{-1}\bfC \} = -\tr\mathbf{CBA} + \tr(\mathbf{CBA})^{-1} =  -W_1 + \bar W_1 ~,
\eea
where the functions $W_1$ and $\bar W_1$ correspond to the six-junction networks shown in figure~\ref{C03PBN1N2} and are related to the so-called Pochhammer curves $\tr\mathbf{CBA}$ and $\tr(\mathbf{CBA})^{-1} $ via the skein relation and reductions in~\eqref{qSkeinSLN} and figure~\ref{ReductionSLN} (with $q=1$):
\bea
W_1 &=& \tr\mathbf{CBA}- A_1  A_2 -B_1  B_2 -C_1  C_2~, \nn
\bar W_1 &=&  \tr(\mathbf{CBA})^{-1} - A_1  A_2 -B_1  B_2 -C_1  C_2~.
\eea

If we want to obtain a closed Poisson algebra we thus need to add $W_1$ and $\bar W_1$ to the set of generators and compute their Poisson brackets. We find that they indeed close:\footnote
{Because of the large number of self-intersections to resolve, it is tedious to get the expressions in terms of networks (second lines) from applying skein relations. However, they can be derived easily with Mathematica in the explicit representation of loop and network functions as Fock-Goncharov polynomials that we will present in section~\ref{secFG}.}
\bea \label{PBC03SL3}
\{N_1,  W_1\}  &=&  \tr \mathbf{B^{-1}C^2BA} - \tr\mathbf{CB^{-1}CBA} + \tr\mathbf{A^{-1}B^{-1}AC} - \tr\mathbf{CB^{-1}} \nn
&=&  - N_1  W_1 + 3 \bar N_1^2 + 2 \bar N_1 ( A_1 B_2 + B_1 C_2 + C_1 A_2) - 6 N_1 + \Lambda ~,\nn
\{N_1,  \bar W_1\}   &=& - \tr\mathbf{CB^{-2}C^{-1}A^{-1}}  + \tr\mathbf{B^{-1}CB^{-1}C^{-1}A^{-1}}  - \tr\mathbf{A^{-1}B^{-1}AC}+\tr\mathbf{CB^{-1}}\nn
&=&  N_1  \bar W_1 - 3 \bar N_1^2 - 2 \bar N_1 (A_1 B_2 + B_1 C_2 + C_1  A_2 ) + 6 N_1 - \Lambda~, \nn
\{W_1,  \bar W_1\}   &=& \tr(\mathbf{BAC})^{-1}\mathbf{ACB} -\tr(\mathbf{CBA})^{-1}\mathbf{ACB} +  \tr(\mathbf{ACB})^{-1}\mathbf{CBA} \nn
&& -\tr(\mathbf{BAC})^{-1}\mathbf{CBA} +  \tr(\mathbf{CBA})^{-1}\mathbf{BAC} -\tr(\mathbf{ACB})^{-1}\mathbf{BAC} \nn
&=&  3 (N_1^3 -  \bar N_1^3) + 2 N_1^2 (A_2 B_1 + B_2 C_1 + C_2 A_1)\nn
 &&  -  2 \bar N_1^2 (A_1 B_2 + B_1 C_2 + C_1 A_2 ) +  
 N_1 \bar\Lambda -  \bar N_1  \Lambda~,
\eea
and the remaining Poisson brackets can be deduced by replacing every object by its reverse: $\bfA \to \bfA^{-1}$,  $A_1\to A_2$, $N_1\to \bar N_1$, $W_1\to \bar W_1$, and so on.
Here we have defined
\bea
\Lambda &\equiv& A_1 A_2 B_2 C_1 + B_1 B_2 C_2 A_1 + C_1 C_2 A_2 B_1 + A_1^2 B_1 +  B_1^2  C_1  +  C_1^2 A_1 \nn
 &&+  A_2 B_2^2 + B_2 C_2^2 + C_2  A_2^2 - 
 3 ( A_2 B_1 + B_2 C_1 + C_2 A_1)~.
\eea
In conclusion, we have obtained a closed Poisson algebra with the generators 
\bea\label{C03gener}
A_i,B_i,C_i, N_1,\bar N_1, W_1, \bar W_1 ~.
\eea
Since the dimension of the moduli space $\cM_{0,3}^3$ is 8, there must be 2 relations between these 10 generators.

\paragraph{Relations:}

\begin{figure}[tb]
\centering
\includegraphics[width=\textwidth]{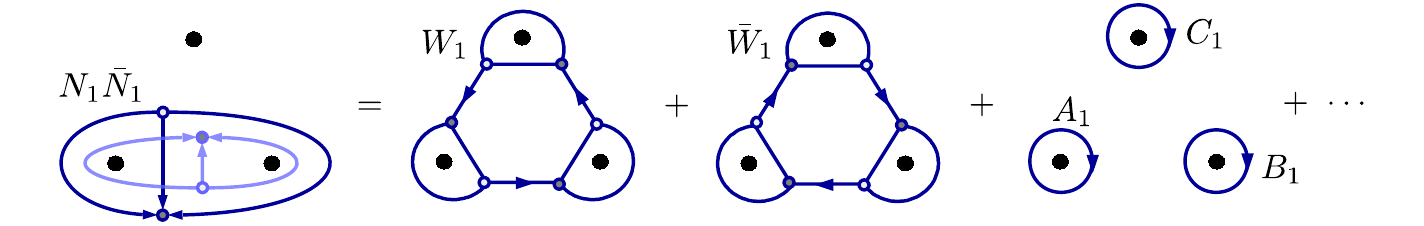}
\caption{The product $N_1 \bar N_1$ can be expressed in terms of the networks $W_1$ and $\bar W_1$ via the skein relation.}
\label{C03N1N1bSkein}
\end{figure}

A simple way to obtain a relation is to consider the product of $N_1$ and $\bar N_1$.
We can draw these networks on $\cC_{0,3}$ such that they have two intersections, which we resolve by applying the skein relation (see figure~\ref{C03N1N1bSkein}).
The resulting networks can be simplified via the square reduction of figure~\ref{ReductionSLN} (with $q=1$), and the polynomial relation is then $\cP_1=0$ with 
\bea\label{N1N1bskeinRelC03classical}
\cP_1 &=& N_1 \bar N_1 -\left( W_1 + \bar W_1 + A_1 B_1 C_1+  A_2  B_2 C_2 + A_1  A_2+B_1  B_2 +C_1  C_2 + 3\right) ~.
\eea
The second relation $\cP_2=0$ comes from the product of $W_1$ and $\bar W_1$:\footnote
{This complicated expression is also more readily obtained in the explicit representation presented in section~\ref{secFG}.}
\bea\label{sexticC03SL3}
\cP_2&=& (W_1+6)( \bar W_1+6) -\Big[ N_1^3   + N_1^2 (A_2 B_1 + B_2 C_1 + A_1 C_2)  + N_1 \bar \Lambda \nn
 &&       + A_1^3 + B_1^3 + C_1^3 + 
   A_1^2 A_2 B_1 C_1 + A_1 B_1^2 B_2 C_1 + A_1 B_1 C_1^2 C_2 \nn
   && + A_1^2 B_1^2 C_2 + 
   A_2 B_1^2 C_1^2 + A_1^2 B_2 C_1^2 - 
   2 ( A_1^2 B_2 C_2 + A_2 B_1^2 C_2 + A_2 B_2 C_1^2)  + \text{reverse}\nn
 &&+ A_1 A_2 B_1 B_2 C_1 C_2  + A_1 A_2 B_1 B_2 + B_1 B_2 C_1 C_2 + 
 C_1 C_2 A_1 A_2  \nn
   &&  - 
 3 ( A_1 B_1 C_1 + A_2 B_2 C_2)  - 9 (A_1 A_2 + B_1 B_2 + C_1 C_2) +27 \Big] ~,
\eea
Here ``$+\text{reverse}$'' means that all the previous terms should be added with reverse orientation.

We have therefore arrived at a description of the algebra $\cA_{0,3}^3$ in terms of the 10 generators~\eqref{C03gener} satisfying the 2 relations $\cP_1$ and $\cP_2$:
\bea
\cA_{0,3}^3  = \bC\left[ A_i,B_i,C_i, N_1,\bar N_1, W_1, \bar W_1 \right] / \{ \cP_1, \cP_2\}~.
\eea
It is straightforward to relate our description involving networks to the description in terms of trace functions as in~\eqref{SL3LawtonGenerators} and~\eqref{LawtonRelSumProd}.

Remarkably, we can write the Poisson brackets in terms of derivatives of the polynomial relations (this can be compared with~\cite{Lawton:2007}):
\bea
\{N_{1},  \bar N_{1}\}   &=& \frac{\del \cP_1}{\del  W_1}\frac{\del \cP_2 }{\del \bar W_1} -  \frac{\del \cP_1}{\del\bar W_1}  \frac{\del \cP_2}{\del W_1}   ~, \qquad \{ N_{1},   W_1\}   =   \frac{\del \cP_1}{\del \bar W_1} \frac{\del \cP_2}{\del\bar N_1}  -   \frac{\del \cP_1}{\del \bar N_1} \frac{\del \cP_2}{\del\bar W_1}  ~, \nn
\{ N_{1},  \bar W_1\}   &=&   \frac{\del \cP_1}{\del \bar N_{1}} \frac{\del \cP_2}{\del W_1}  -   \frac{\del \cP_1}{\del  W_1} \frac{\del \cP_2}{\del\bar N_{1}}  ~, \qquad  \{W_1,  \bar W_1 \}   = \frac{\del \cP_1}{\del  N_1}\frac{\del \cP_2}{\del\bar N_1}  -   \frac{\del \cP_1}{\del\bar N_1} \frac{\del \cP_2}{\del N_1}  ~.
\eea
This indicates that that the Poisson structure of $\cM_{0,3}^3$ is compatible with its structure as an algebraic variety.

\paragraph{Quantization:}

Quantum versions of the polynomial relations $\cP_1$ and $\cP_2$, in which the network functions are replaced by noncommuting operators, can be obtained by applying the quantum skein relations in~\eqref{qSkeinSLN} and figure~\ref{ReductionSLN}.

\begin{figure}[tb]
\centering
\includegraphics[width=\textwidth]{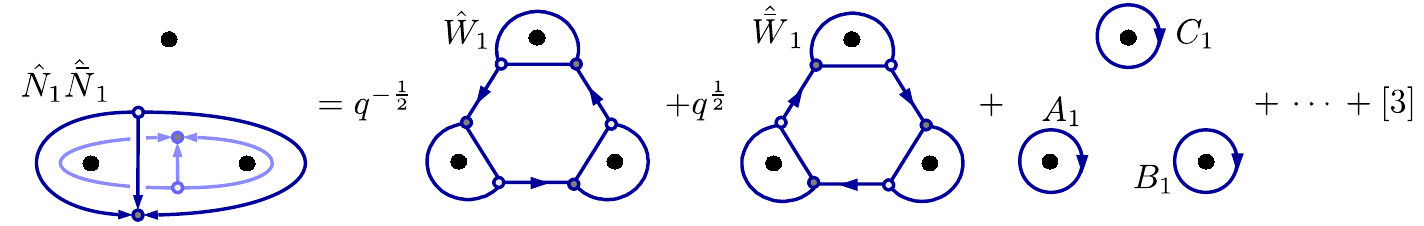}
\caption{The quantum product of $\hat N_1\hat{\bar N}_1$ resolved via the quantum skein relation.}
\label{C03quantumProductN1N1b}
\end{figure}

The quantum relation $\hat \cP_1$ is obtained by superposing the operators $\hat N_1$ and $\hat{\bar N}_1$ and resolving their two intersections via the quantum skein relation (see figure~\ref{C03quantumProductN1N1b}):
\bea \label{N1N1bc03}
\hat N_1 \hat {\bar N}_1 &=&  q^{-\frac12} \hat W_1 +  q^{\frac12}  \hat{ \bar W}_{1} +\hat A_1 \hat B_1 \hat C_1+  \hat A_2  \hat B_2 \hat C_2 + \hat A_1  \hat A_2 +\hat B_1  \hat B_2 +\hat C_1  \hat C_2 + [3]~. 
\eea
The operator $\hat N_1$, which appears first in the product, is drawn on top of the second operator $\hat {\bar N}_1$.
Note that the product $\hat {\bar N}_1 \hat N_1$ with inverted order would give the same expression but with the replacement $q\to q^{-1}$.

The quantization of the second relation $\cP_2$ gives
\bea\label{W1bW1quantum}
\left(\hat W_1+[6]\right)\left( \hat{\bar W}_{1}+[6]\right) &=&   q^{\frac32} \hat N_1^3   +q \hat  \hat N_1^2 (\hat A_2 \hat B_1 + \hat B_2 \hat C_1 + \hat A_1 \hat C_2)  +q^{\frac12}\hat  N_1 \hat{\bar \Lambda}' \nn
 &&       + \hat A_1^3 + \hat B_1^3 + \hat C_1^3 + 
   \hat A_1^2 \hat A_2 \hat B_1\hat  C_1 + \hat A_1 \hat B_1^2 \hat B_2 \hat C_1 + \hat A_1\hat  B_1 \hat C_1^2\hat  C_2 \nn
   && + \hat A_1^2\hat  B_1^2 \hat C_2 + 
   \hat A_2 \hat B_1^2 \hat C_1^2 + \hat A_1^2 \hat B_2 \hat C_1^2\nn
   &&  -   (q+q^{-1}) ( \hat A_1^2 \hat B_2 \hat C_2 + \hat A_2\hat  B_1^2 \hat C_2 + \hat A_2 \hat B_2\hat  C_1^2)  + \text{reverse}   \nn
    &&+  \hat A_1\hat  A_2 \hat B_1 \hat B_2 \hat C_1 \hat C_2 - (q-3+q^{-1})(\hat A_1\hat  A_2 \hat B_1 \hat B_2 + \hat B_1 \hat B_2 \hat C_1 \hat C_2 + 
 \hat C_1\hat  C_2 \hat A_1 \hat A_2)  \nn
 &&  - (2q^2+q-3+q^{-1}+2q^{-2}) ( \hat A_1 \hat B_1 \hat C_1 + \hat A_2 \hat B_2 \hat C_2) \nn
   &&  - (2q^2+q+3+q^{-1}+2q^{-2}) (\hat A_1 \hat A_2 +\hat  B_1 \hat B_2 + \hat C_1 \hat C_2)  \nn
 && +q^5+2 q^4+q^3+3 q^2+3 q+7 + 3q^{-1}+3q^{-2}+q^{-3}+2 q^{-4}+q^{-5}~,
\eea
where now ``+ reverse'' also implies the replacement $q\to q^{-1}$,
and in $\hat{\bar \Lambda}' $ we have replaced the coefficient of $3$ by $2q + q^{-2}$.

We also find the following quantum commutators:
\bea
\hat N_1 \hat {\bar N}_1 - \hat {\bar N}_1 \hat N_1 &=& (q^{\frac12} - q^{-\frac12}) (\hat{ \bar W}_{1} - \hat W_1)~, \nn
q^{\frac12}\hat N_1 \hat W_1  - q^{-\frac12}\hat W_1  \hat N_1   &=&  (q^{\frac32}-q^{-\frac32}) \hat{\bar N}_1^2 + (q-q^{-1}) \hat{\bar N}_1 ( \hat A_1 \hat B_2 + \hat B_1 \hat C_2 + \hat C_1\hat  A_2)\nn
&&  - (q^2+q-q^{-1}-q^{-2})\hat N_1 +  (q^{\frac12}-q^{-\frac12})\hat \Lambda~, \nn
q^{-\frac12}\hat N_1 \hat {\bar W}_{1}  - q^{\frac12}\hat{\bar W}_{1}  \hat N_1   &=&  (q^{-\frac32}-q^{\frac32}) \hat{\bar N}_1^2 + (q^{-1}-q) \hat{\bar N}_1 ( \hat A_1\hat  B_2 + \hat B_1 \hat C_2 + \hat C_1 \hat A_2)\nn
&&  - (q^{-2}+q^{-1}-q-q^{2})\hat N_1 +  (q^{-\frac12}-q^{\frac12})\hat \Lambda~, \nn
\hat W_1 \hat{\bar W}_{1}  - \hat{\bar W}_{1}\hat W_1   &=&   (q^{\frac32}-q^{-\frac32}) (\hat N_1^3 - \hat{ \bar N}_1^3) + (q-q^{-1})\hat N_1^2 (\hat A_2 \hat B_1 + \hat B_2 \hat C_1 + \hat C_2 \hat A_1)  \nn
 &&  -  (q-q^{-1}) \hat{\bar N}_1^2 (\hat A_1\hat  B_2 +\hat  B_1 \hat C_2 + \hat C_1 \hat A_2 ) \nn
 &&  + (q^{\frac12}-q^{-\frac12}) ( \hat N_1 \hat{\bar\Lambda} -   \hat{\bar N}_1 \hat \Lambda )~,
\eea
where in $\hat \Lambda$ we have made the replacement $3\to [3] $.
These relations reduce at first order in $\hbar$ to the Poisson brackets~\eqref{PBN1N1b} and~\eqref{PBC03SL3}.
For example, with a little bit of rewriting we obtain
\bea
[\hat N_1, \hat W_1]    &=& (1-q)\hat W_1  \hat N_1 +  (q-q^{-2}) \hat{\bar N}_1^2 + (q^{\frac12}-q^{-\frac32}) \hat{\bar N}_1 ( \hat A_1 \hat B_2 + \hat B_1 \hat C_2 + \hat C_1 \hat A_2)\nn
&& - (q^{\frac32}+q^{\frac12}-q^{-\frac32}-q^{-\frac52})\hat N_1 +  (1-q^{-1})\hat \Lambda  \nn
 &=&\hbar \Big[ -\hat W_1  \hat N_1 +  3\hat{\bar N}_1^2 + 2\hat{\bar N}_1 ( \hat A_1 \hat B_2 +\hat  B_1 \hat C_2 + \hat C_1 \hat A_2) - 6 \hat N_1 +  \hat \Lambda \Big] + \cO(\hbar^2) \nn
 &= & \hbar \{ N_1, W_1\}  + \cO(\hbar^2) ~.
\eea

\subsubsection*{SL(4)}

\begin{figure}[tb]
\centering
\includegraphics[width=\textwidth]{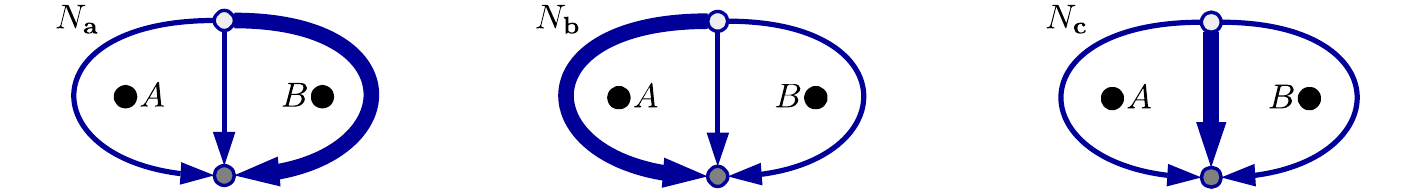}
\caption{The pants networks $N_{\bfa}$, $N_{\bfb}$, $N_{\bfc}$ differ from one another by the choice of the edge (thick) that carries the second antisymmetric representation of $SL(4)$.}
\label{C03SL4networks}
\end{figure}

We find a similar structure for $SL(4)$ loop and network operators.
The loop functions $A_i$ around the puncture $A$ are
\bea
A_1 &=& \tr\bfA ~, \qqq 
A_2 =  \frac12 \left[(\tr\bfA)^2 - \tr(\bfA^2)\right]  ~, \qqq 
A_3 = \tr\bfA^{-1} ~.
\eea
We can construct three pants networks $N_{\bfa}$, $N_{\bfb}$, $N_{\bfc}$, differing by the choice of the edge that carries the second antisymmetric representation of $SL(4)$ (see figure~\ref{C03SL4networks}):
\bea
N_{\bfa} &=& -\frac12  \epsilon_{mnpq} \bfU_{{\bf a}\, r}^{m}\bfU_{{\bf a}\, s}^{n}\bfU_{{\bf b}\, t}^{p} \bfU_{{\bf c}\, u}^{q} \epsilon^{rstu}= \tr \bfC \bfB^{-1} - C_1B_3~, \nn
N_{\bfb} &=& -\frac12  \epsilon_{mnpq} \bfU_{{\bf a}\, r}^{m}\bfU_{{\bf b}\, s}^{n}\bfU_{{\bf b}\, t}^{p} \bfU_{{\bf c}\, u}^{q} \epsilon^{rstu}= \tr \bfA \bfC^{-1} - A_1C_3~, \nn
N_{\bfc} &=& -\frac12  \epsilon_{mnpq} \bfU_{{\bf a}\, r}^{m}\bfU_{{\bf b}\, s}^{n}\bfU_{{\bf c}\, t}^{p} \bfU_{{\bf c}\, u}^{q} \epsilon^{rstu}= \tr \bfB \bfA^{-1} - B_1A_3~.
\eea
Pants networks with the same orientation Poisson-commute with each other:
\bea
\{N_{\bfa},N_{\bfb}\} = \{N_{\bfb},N_{\bfc}\} = \{N_{\bfc},N_{\bfa}\} &=& 0~, 
\eea
but they do not commute with their reverses:
\bea
\{ N_{\bfa},\bar N_{\bfa}\}  &=& \{ N_{\bfb},\bar N_{\bfb}\}  = \{ N_{\bfc},\bar N_{\bfc}\}  =  \tr  \bfC \bfB \bfA  - \tr\left( \bfC \bfB \bfA \right)^{-1} ~, \nn
\{ N_{\bfa},\bar N_{\bfc}\}  &=&   \tr \bfC \bfB^{-1} \bfA \bfB^{-1} - \tr \bfC \bfB^{-2} \bfA  ~,\nn
\{ N_{\bfc},\bar N_{\bfb}\}  &=&  \tr \bfB \bfA^{-1} \bfC \bfA^{-1} - \tr \bfB \bfA^{-2} \bfC  ~,\nn
\{ N_{\bfb},\bar N_{\bfa}\}  &=&  \tr \bfA \bfC^{-1} \bfB \bfC^{-1} - \tr \bfA \bfC^{-2} \bfB  ~.
\eea
As in the $SL(3)$ case, to obtain a closed Poisson algebra we would need to add the functions appearing in the Poisson brackets of the pants networks to the set of generators, compute their Poisson brackets, and so on.
Repeating this procedure until the Poisson algebra closes would lead to a large number of generators, satisfying many polynomial relations.
Ultimately, it should be possible to choose the set of 15 \emph{independent} generators of $\cA_{0,3}^4$ to be given by the loop functions around the punctures and by the pants networks: 
\bea
A_i,B_i,C_i, N_{\bfa}, \bar N_{\bfa}, N_{\bfb}, \bar N_{\bfb}, N_{\bfc}, \bar N_{\bfc} ~.
\eea
This can be compared with the 15 generators of the ring of invariants of two matrices given in~\cite{Teranishi}.

\begin{figure}[tb]
\centering
\includegraphics[width=\textwidth]{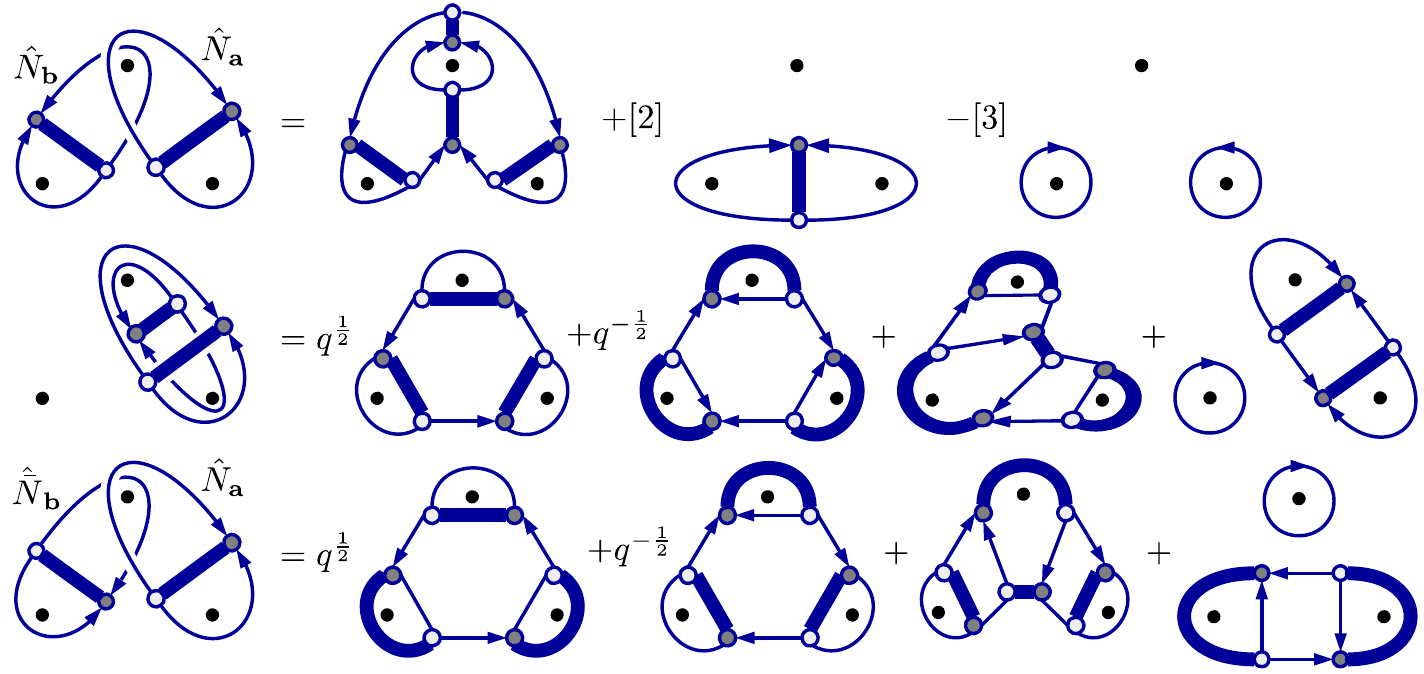}
\caption{Product of pants networks: $\hat N_{\bfa}\hat N_{\bfb}$ (top), $\hat N_{\bfa} \hat {\bar N}_{\bfa}$ (middle), $\hat N_{\bfa} \hat {\bar N}_{\bfb}$ (bottom).}
\label{C03SL4NaNb}
\end{figure}

Many quantum relations can be easily obtained by applying quantum skein relations to products of pants networks.
We show a few examples in figure~\ref{C03SL4NaNb}.

\subsection{One-punctured torus}

\begin{figure}[tb]
\centering
\includegraphics[width=\textwidth]{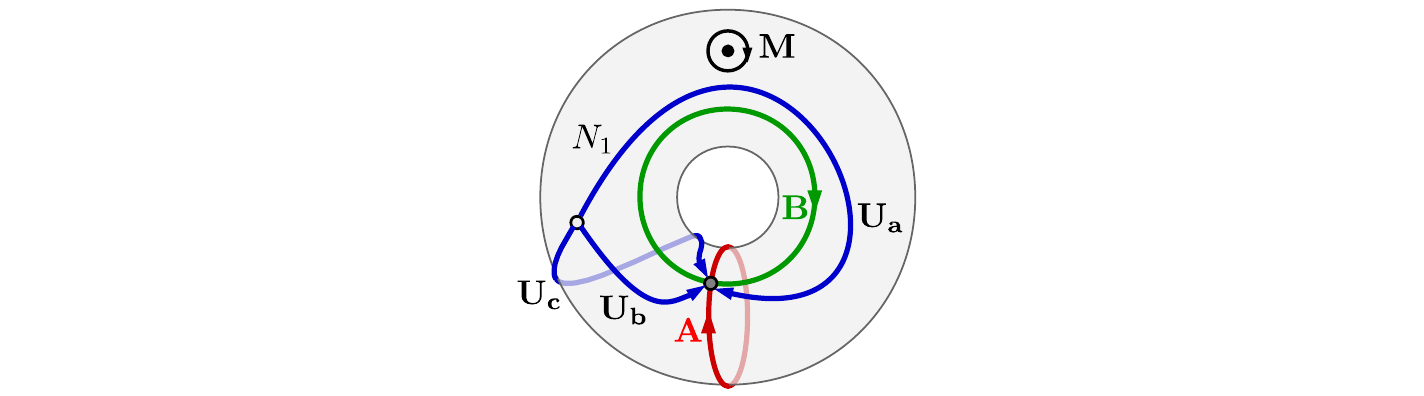}
\caption{One-punctured torus $\cC_{1,1}$. The A- and B-cycles are shown, together with a network operator $N_1$, consisting of three edges and two junctions.}
\label{C11triang}
\end{figure}

Our next simple example is the torus with one full puncture, denoted by $\cC_{1,1}$ (see figure~\ref{C11triang}). 
The corresponding 4d gauge theory is the so-called $\cN=2^*$ $SU(N)$ gauge theory,
which can be obtained from the $\cN=4$ theory by giving a mass to an adjoint hypermultiplet.

The fundamental group of $\cC_{1,1}$ consists of three loops, the A-cycle $\gamma_A$ (meridian), the B-cycle $\gamma_B$ (longitude), and the loop $\gamma_M$ around the puncture, subject to one relation: 
\bea
\pi_1 (\mathcal{C}_{1,1}) = \{ (\gamma_A, \gamma_B, \gamma_M) | \gamma_A \gamma_B \gamma_A^{-1}\gamma_B^{-1} =\gamma_M  \}~.
\eea
The holonomy matrices by $\bfA$, $\bfB$, and $\bfM $ satisfy $\bfA\bfB\bfA^{-1}\bfB^{-1}=\bfM$. 
We can also combine the matrices $\bfA$ and $\bfB$ into matrices $\bfC \equiv (\bfA\bfB )^{-1}$ and $\bfCp \equiv \bfA \bfB^{-1}$ associated with the curves $\gamma_C = (\gamma_A\gamma_B)^{-1} $ and $\gamma_{C'} =  \gamma_A\gamma_B^{-1}$, respectively.
Each of these holonomy matrices gives $(N-1)$ loop functions.

We also consider a particular two-junction network $N_1$ with two junctions whose edges go around the A- and B-cycles (in contrast to the case of $\pants$, there are many two-junction networks that we can consider on $\torus$).
The network $N_1$ and its reverse $\bar N_1$ each contributes $(N-1)(N-2)/2$ operators.
Together, the operators coming from $\bfA$, $\bfB$, $\bfC$ and from the networks $N_1$, $\bar N_1$ add up to the dimension~\eqref{dimMgnN} of the moduli space of flat $SL(N)$-connections on $\torus$:
\bea
3(N-1) + 2 \frac{(N-1)(N-2)}{2} &=& N^2-1  = \dim[\cM_{1,1}^N] ~.
\eea

\subsubsection*{SL(2)}

We start by briefly reviewing the well-studied case of flat $SL(2,\bC)$-connections on the one-punctured torus $\cC_{1,1}$~\cite{Gaiotto:2010be} (see also~\cite{Dimofte:2011jd}\cite{Nekrasov:2011bc}\cite{TV13}).
The Poisson bracket of the A-cycle function $A_1=\tr \bfA$ and the B-cycle function $B_1=\tr \bfB$ 
is (by Goldman's formula~\eqref{GoldmanFormula}):
\bea \label{PBS1T1sl2}
\{A_1, B_1 \} = \tr \bfA \bfB - \frac12 A_1 B_1 = -\tr \bfA \bfB^{-1} + \frac12 A_1 B_1 ~.
\eea
The extra traces $C_1 = \tr\bfA \bfB$ and $C_1'  =\tr \bfA \bfB^{-1} $ correspond to curves that go once around the A-cycle and once around the B-cycle (in different directions).
Applying the skein relation to products of loop functions gives
\bea
A_1 B_1 &=& C_1 + C_1'~, \qqq C_1 C_1' = A_1^2 + B_1^2+M_1 -2~,
\eea
which can be combined to obtain the relation $\cP_1 =0$ with 
\bea\label{C11SL2cubicrel}
\cP_1 = A_1 B_1 C_1 -( A_1^2 + B_1^2 + C_1^2 + M_1 -2 )~.
\eea
The Poisson bracket between the generators $A_1$, $B_1$, $C_1$ can be written as
\bea
\{A_1,B_1\} = -\frac12 \frac{\del \cP_1}{C_1}  ~,
\eea
and cyclic permutations of $A_1,B_1,C_1$.

To obtain quantum relations, we apply the quantum skein relation:
\bea\label{hatAhatBskeinSL2}
\hat A_1 \hat B_1 = q^{\frac14} \hat C_1 + q^{-\frac14} \hat C'_1~.
\eea
Inverting the order in the product amounts to exchanging $q \leftrightarrow q^{-1}$, so we can obtain the $q$-deformed commutation relation
\bea
q^{\frac14} \hat A_1 \hat B_1 - q^{-\frac14} \hat B_1 \hat A_1 = \left( q^{\frac12} - q^{-\frac12} \right) \hat C_1~. 
\eea
Note that this relation can be written as a quantum commutator
\bea
[\hat A_1, \hat B_1] = (q^{-\frac12} -1) \hat B_1 \hat A_1 + (q^{\frac14} - q^{-\frac34})\hat C_1~,
\eea
which reproduces the Poisson bracket~\eqref{PBS1T1sl2} at first order in $\hbar$.
Similarly, the quantum skein relation leads to the quantization of the cubic relation~\eqref{C11SL2cubicrel}:
\bea\label{C11SL2hatP1}
\hat \cP_1 = q^{\frac14} \hat A_1 \hat B_1 \hat C_1 -\left(  q^{\frac12} \hat A_1^2 +q^{-\frac12} \hat B_1^2 +q^{\frac12} \hat C_1^2 +  M_1 - [2]  \right) ~.
\eea

\subsubsection*{SL(3)}

\begin{figure}[tb]
\centering
\includegraphics[width=\textwidth]{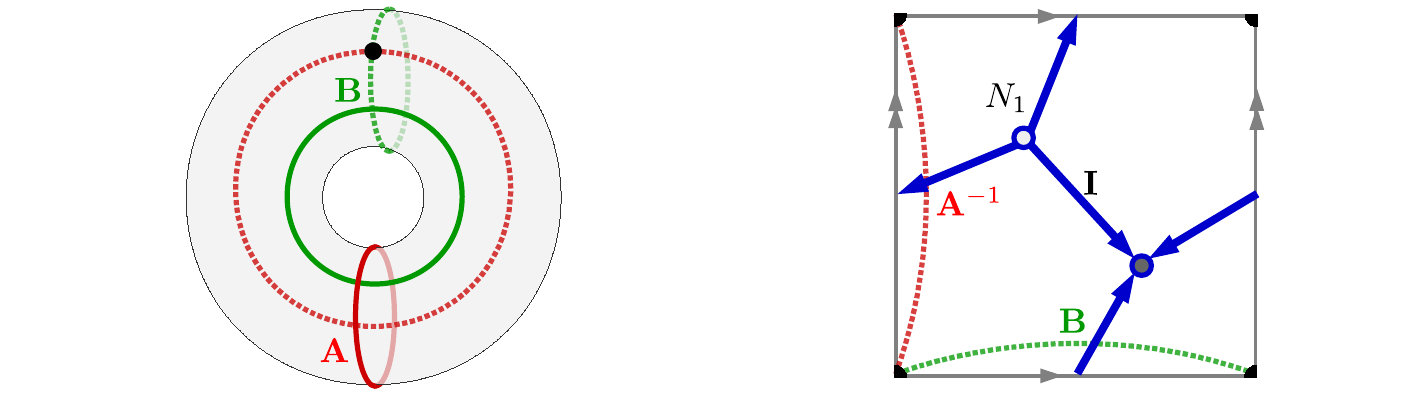}
\caption{\emph{Left}: Covering of $\cC_{1,1}$ with two branch-cuts (dashed lines). Each branch-cut is associated with a holonomy matrix $\bfA$ or $\bfB$. \emph{Right}: $\torus$ described as a square with opposite sides identified, and with the puncture at the corners. The holonomy matrices associated with the three edges of the network $N_1$ are determined by the branch-cuts that they cross.}
\label{C11cuts}
\end{figure}

\begin{figure}[tb]
\centering
\includegraphics[width=\textwidth]{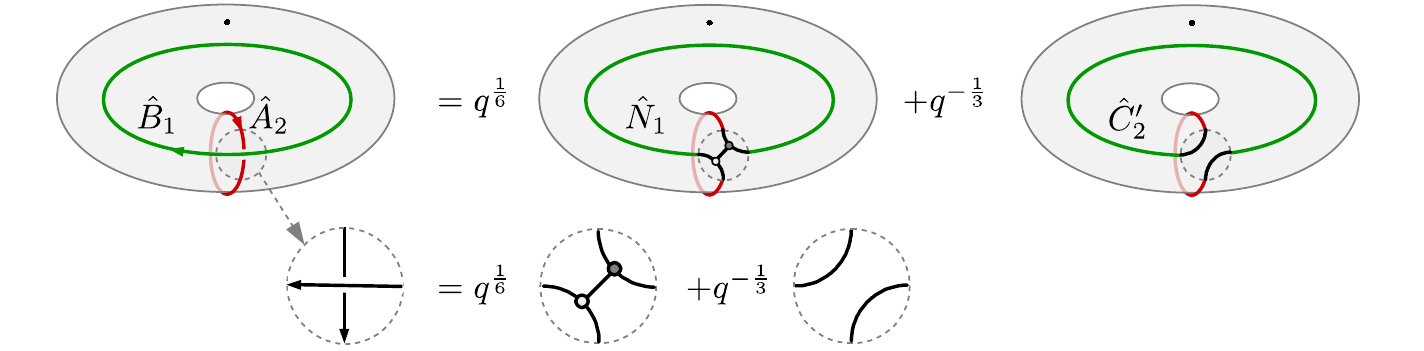}
\caption{Relations between loop and network operators arise from the skein relation.}
\label{C11SkeinAB}
\end{figure}

We now have the loop functions $A_1=\tr \mathbf{A}$ and $A_2=\tr \mathbf{A}^{-1}$ and similarly for $B_i$, $M_i$, $C_i$, and $C_i'$.
There are many different two-junction networks that can be put on $\torus$. 
A natural choice is the network that goes once around the A-cycle and once around the B-cycle.
The associated network function $N_1$ and its reverse $\bar N_1$ can be expressed as
\bea
N_1 &=&  \epsilon_{mnp} \bfU_{{\bf a}\, r}^{m}\bfU_{{\bf b}\, s}^{n} \bfU_{{\bf c}\, t}^{p} \epsilon^{rst}~, \nn
\bar N_1 &=&  \epsilon_{mnp} 
(\bfU_{\bf a}^{-1})^m_r(\bfU_{\bf b}^{-1})^n_s(\bfU_{\bf c}^{-1})^p_t \epsilon^{rst}~.
\eea
As explained in section~\eqref{secLoopNetwork}, the network functions $N_1$ and $\bar N_1$ can also be obtained from the transition functions in a covering of $\torus$.
We can for example cover $\cC_{1,1}$ with one patch that overlaps itself along two branch-cuts that go around the A- and B-cycles and intersect at the puncture (see figure~\ref{C11cuts}).
The network $N_1$ has one edge that crosses the branch-cut associated with $\bfA$ (with reverse orientation), one that crosses the branch-cut associated with $\bfB$, and one that does not cross any branch-cut.
This gives the following expressions
\bea\label{networkscutsC11SL3}
N_1 &=&  \epsilon_{mnp} (\bfA^{-1})^{m}_{\ r}\bfB^{n}_{\ s}\delta^{p}_{\ t} \epsilon^{rst} = A_2 B_1 - C'_2~, \nn
\bar N_1 &=&  \epsilon_{mnp} \bfA^{m}_{\ r}(\bfB^{-1})^{n}_{\ s}\delta^{p}_{\ t} \epsilon^{rst} = A_1 B_2 - C'_1 ~,
\eea
with $C'_1 = \tr\bfA\bfB^{-1} $ and $C'_2 = \tr\bfA^{-1}\bfB$.
Such relations between networks and products of intersecting loops can of course also be understood as arising from the skein relation~\eqref{SkeinSL3} (see figure~\ref{C11SkeinAB}):
\bea\label{torusSkeinA2B1}
A_2 B_1 &=& N_1 + C_2'~, \qqq  A_1 B_2 = \bar N_1 + C_1'~.
\eea

\begin{figure}[b]
\centering
\includegraphics[width=\textwidth]{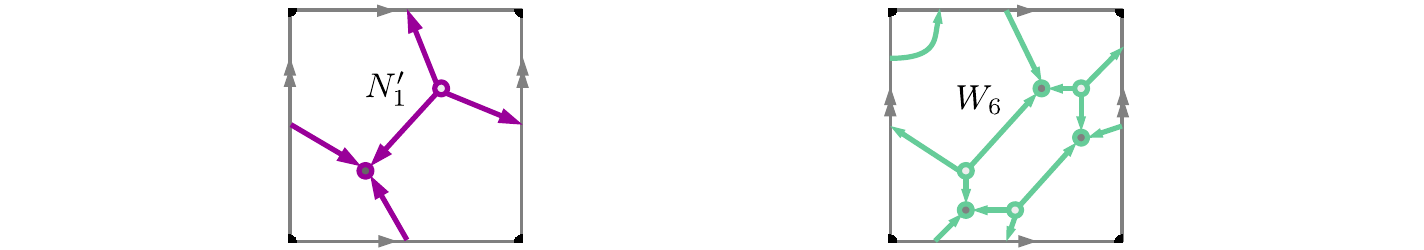}
\caption{\emph{Left}: Flipped network $N_1'$. \emph{Right}: Network $W_6$ appearing in the product $N_1 \bar N_1$.}
\label{C11flippedNetwork}
\end{figure}

The quantization of relations such as~\eqref{torusSkeinA2B1} can be obtained by applying quantum skein relations to resolve the intersection of the A- and B-cycles:
\bea\label{qskeinC11SL3AB}
\hat  A_2  \hat B_1 &=& q^{\frac16}  \hat  N_1+q^{-\frac13} \hat C'_2   ~, \qqq \hat   A_1 \hat  B_1= q^{-\frac16}   \hat N'_1+ q^{\frac13} \hat C_2  ~,\nn
\hat A_1 \hat  B_2 &=& q^{\frac16} \hat{\bar{N}}_1+ q^{-\frac13}\hat  C'_1 ~, \qqq  \hat A_2  \hat B_2 = q^{-\frac16} \hat{\bar{N}}'_1+q^{\frac13} \hat C_1  ~.
\label{skeinA2C11}
\eea
Here the operator $\hat N_1'$ corresponds to the flipped network shown on the left of figure~\ref{C11flippedNetwork}.

Changing the ordering in the product of two operators simply inverts $q$, so we can deduce expressions for commutators, whose leading terms in $\hbar$ correspond to the Poisson brackets. For example we obtain
\bea
[\hat  A_2,  \hat B_1 ] &=& (q^{\frac16}-q^{-\frac16} )  \hat  N_1+(q^{-\frac13}-q^{\frac13} ) \hat C'_2  \quad \to \quad \{A_2, B_1\} = \frac13 (N_1 - 2 C'_2)~.
\eea

We can also apply the quantum skein relation to the product of $\hat N_1$ and $\hat {\bar N}_1$:
\bea \label{N1N1bqtmRelation}
\hat N_1  \hat{\bar{N}}_1 = q^{-\frac12} \hat W_6 + q^{\frac12} \hat{\bar{W}}_6 + \hat A_1\hat A_2+\hat B_1\hat B_2+\hat C_1\hat C_2+\hat M_1+\hat M_2+[3]~,
\eea
where $W_6$ is a network with six junctions shown on the right of figure~\ref{C11flippedNetwork}.
The same network $W_6$ also appears in the product
\bea
\hat{\bar N}_1'\hat  C_2 = \hat W_6 + q^{\frac12} \hat A_1\hat A_2 + q^{-\frac12} \hat B_1\hat B_2~.
\eea
More complicated relations are impractical to derive in this way, but can be computed in the explicit representation of the loop and network operators in terms of Fock-Goncharov polynomials, as we will discuss in section~\ref{secFG}.
We have for example the classical relation (also given in~\cite{Lawton:2006a}\cite{Xie:2013lca})
\bea\label{sexticRelC11}
A_1A_2 B_1 B_2 C_1 C_2  &=&  \Big[ N_1^3   + N_1^2 (A_2 B_1 + B_2C_1+C_2A_1)  \\
&& + N_1 ( A_1 A_2 B_1 C_2 + A_1^2 C_1 + A_2^2B_2-3A_1B_2 + \cyc)   \nn
 &&  - ( A_1 B_2^2C_2^2 - 2A_1^2 B_2C_2 + A_1^3 + \cyc)  + \text{reverse}\Big]  \nn
   && - N_1 \bar N_1(A_1A_2+\cyc) - (A_1A_2B_1B_2+\cyc) \nn
 && +3(A_1B_1C_1+A_2B_2C_2) + M_1M_2+6(M_1+M_2)+9 ~.  \nonumber 
\eea
Here ``$+\cyc$'' means adding the terms obtained by cyclic permutation of $A,B,C$, and ``$+\rev$'' the terms obtained by reversing the orientation, $A_1 \leftrightarrow A_2$, $N_1 \leftrightarrow \bar N_1$, and so on.
Recall that at the classical level the algebraic structures of $\cM_{0,3}^N$ and $\cM_{1,1}^N$ are the same, so that the relation~\eqref{sexticRelC11} is the counterpart of the relation~\eqref{sexticC03SL3} for the 3-punctured sphere. The Poisson structures on $\cM_{0,3}^N$ and $\cM_{1,1}^N$ can also be related via a symplectic quotient, as explained in~\cite{Lawton:2008}.

In conclusion, we can describe the algebra $ \cA_{1,1}^3$ in terms of the 10 generators $ A_i$, $ B_i$, $ M_i$, $ C_i$, $ N_1$, $ {\bar N}_1$ satisfying the relations~\eqref{N1N1bqtmRelation} and~\eqref{sexticRelC11}.
Classically, these generators form a closed Poisson algebra (as noted in~\cite{Xie:2013lca}):
\bea
\{A_1, B_1 \} &=& C_2 - \frac13 A_1 B_1~, \qqq\qqq  \{A_2, B_1 \} = N_1 - \frac23 A_2 B_1 ~,  \nn
\{B_1, C_1 \} &=& A_2 - \frac13 B_1 C_1~, \qqq\qqq   \{B_2, C_1 \} =   N_1 - \frac23 B_2 C_1~,  \nn
\{C_1, A_1 \} &=& B_2 - \frac13 C_1 A_1~, \qqq\qqq   \{C_2, A_1 \} = N_1 - \frac23 C_2 A_1~,  \nn
\{A_1, N_1 \} &=& - \frac13 A_1 N_1 + A_2C_2 -B_1~, \qqq   \{A_1, \bar N_1 \} =   \frac13 A_1 \bar N_1  - A_2B_2+C_1~,  \nn
\{B_1, N_1 \} &=&  -\frac13 B_1 N_1 + A_2B_2 - C_1~, \qqq   \{B_1, \bar N_1 \} =  \frac13 B_1 \bar N_1 - B_2 C_2 +A_1~,  \nn
\{C_1, N_1 \} &=&  -\frac13 C_1 N_1 + B_2 C_2 - A_1~, \qqq   \{C_1, \bar N_1 \} =  \frac13 C_1 \bar N_1 - A_2 C_2 + B_1~, \nn
&&\qqq \{ N_1, \bar N_1 \} = A_1 B_1 C_1 - A_2 B_2 C_2~.
\eea
The remaining Poisson brackets can be obtained by reversing the orientation. In addition, the functions $M_i$ associated with the curve around the puncture are central elements of the Poisson algebra, $\{M_i, \bullet\}=0$.

\begin{figure}[tb]
\centering
\includegraphics[width=\textwidth]{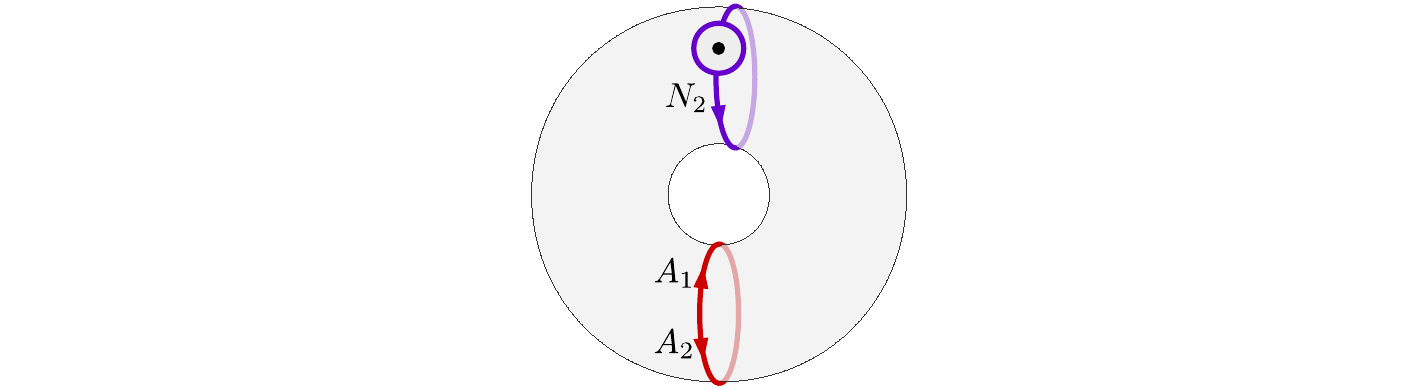}
\caption{A maximal set of commutting Hamiltonians on $\bar \cM_{1,1}^3$ is provided by the A-cycle functions $A_i$ and the pants network $N_2$.}
\label{C11meridianpanties}
\end{figure}

Fixing the values of the central elements $M_i$ leaves us with a 6-dimensional moduli space $\bar \cM_{1,1}^3$ which is symplectic.
A natural maximal set of commuting Hamiltonians consists of the A-cycle functions $A_i$ together with a network $N_2$ that surrounds the puncture and has one edge along the A-cycle (see figure~\ref{C11meridianpanties}):
\bea
N_2 = -\epsilon_{mnp}\bfA^{m}_{\ r}\bfM^{n}_{\ s}\delta^{p}_{\ t}\epsilon^{rst} = \tr \bfA  \bfM - A_1 M_1~.
\eea
This network is the pants network in the pants decomposition obtained by cutting $\cC_{1,1}$ along the A-cycle.
It is related to the network $N_1$ defined above via
\bea\label{panties2linesnet}
N_2&=& - N_1 C_1   - \bar N_1 B_1  + A_2 B_1 C_1  + B_2 C_2 - A_2^2+ A_1  ~.
\eea
The A-cycle functions $A_i$ and the pants network $N_2$ Poisson-commute, as is obvious from the fact that they do not intersect:
\bea
\{ A_1, A_2 \} = \{ A_1, P_1 \} = \{ A_2, P_1 \} = 0~.
\eea

\subsubsection*{SL(4)}

The A-cycle functions are $A_1 = \tr\bfA$, $A_2 =  \frac12 \left[(\tr\bfA)^2 - \tr(\bfA^2)\right] $, $A_3 =  \tr\bfA^{-1}$.
We define three networks which are going once around the A-cycle and once around the B-cycle, and differ by the choice of the branch that is doubled:
\bea
N_a &=& \tfrac12  \epsilon_{mnpq} \bfU_{{\bf a}\, r}^{m}\bfU_{{\bf a}\, s}^{n}\bfU_{{\bf b}\, t}^{p} \bfU_{{\bf c}\, u}^{q} \epsilon^{rstu} ~, \nn
N_{\bfb} &=& \tfrac12  \epsilon_{mnpq} \bfU_{{\bf a}\, r}^{m}\bfU_{{\bf b}\, s}^{n}\bfU_{{\bf b}\, t}^{p} \bfU_{{\bf c}\, u}^{q} \epsilon^{rstu} ~, \nn
N_{\bfc} &=& \tfrac12  \epsilon_{mnpq} \bfU_{{\bf a}\, r}^{m}\bfU_{{\bf b}\, s}^{n}\bfU_{{\bf c}\, t}^{p} \bfU_{{\bf c}\, u}^{q} \epsilon^{rstu} ~.
\eea

\begin{figure}[b]
\centering
\includegraphics[width=\textwidth]{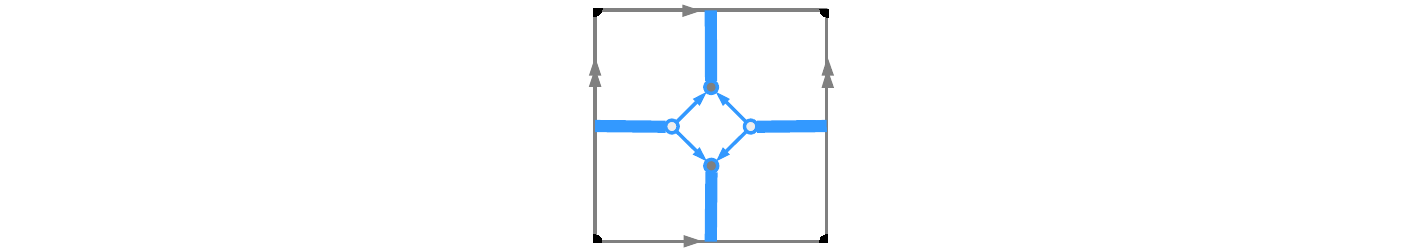}
\caption{Network $N_4$ that appears in the product $A_2 B_2$ (thick lines carry $\wedge^2 \square$)).}
\label{NetworkSquare}
\end{figure}

Quantum relations between the loop and network operators can be obtained by applying quantum skein relations:
\bea\label{qtRelsSL4C11}
\hat A_1\hat B_3&=&  q^{-\frac18}\hat{\bar N}_{\bfb} + q^{\frac38} \hat C'_1 ~, \qqq \hat A_1\hat B_1=  q^{\frac18}\hat{ N}'_{\bfb} + q^{-\frac38} \hat C_3 ~, \nn
\hat A_3\hat B_1&=&  q^{-\frac18}\hat{  N}_{\bfb} + q^{\frac38} \hat C'_3  ~, \qqq \hat A_3\hat B_3=  q^{\frac18}\hat{\bar  N}'_{\bfb} + q^{-\frac38} \hat C_1  ~, \nn
\hat A_1\hat B_2&=&  q^{-\frac14}\hat{\bar N}_{\bfa} + q^{\frac14} \hat N'_{\bfa} ~,  \qqq \hat A_2\hat B_1= q^{-1/4}\hat{  N}_{\bfc} + q^{\frac14} \hat N'_{\bfc} ~, \nn
\hat A_3\hat B_2&=& q^{-\frac14}\hat{ N}_{\bfa} + q^{\frac14} \hat{\bar  N}'_{\bfa} ~, \qqq  \hat A_2\hat B_3= q^{-1/4}\hat{\bar N}_{\bfc} + q^{\frac14} \hat{\bar  N}'_{\bfc} ~,\nn 
 \hat A_2\hat B_2&=& q^{\frac12} \hat C'_2 + q^{-\frac12} \hat C_2 + \hat N_4 ~.
\eea
Here the networks with a prime are flipped, and $N_4$ is the four-junction network shown in figure~\ref{NetworkSquare}.

\subsection{Four-punctured sphere}

\begin{figure}[tb]
\centering
\includegraphics[width=\textwidth]{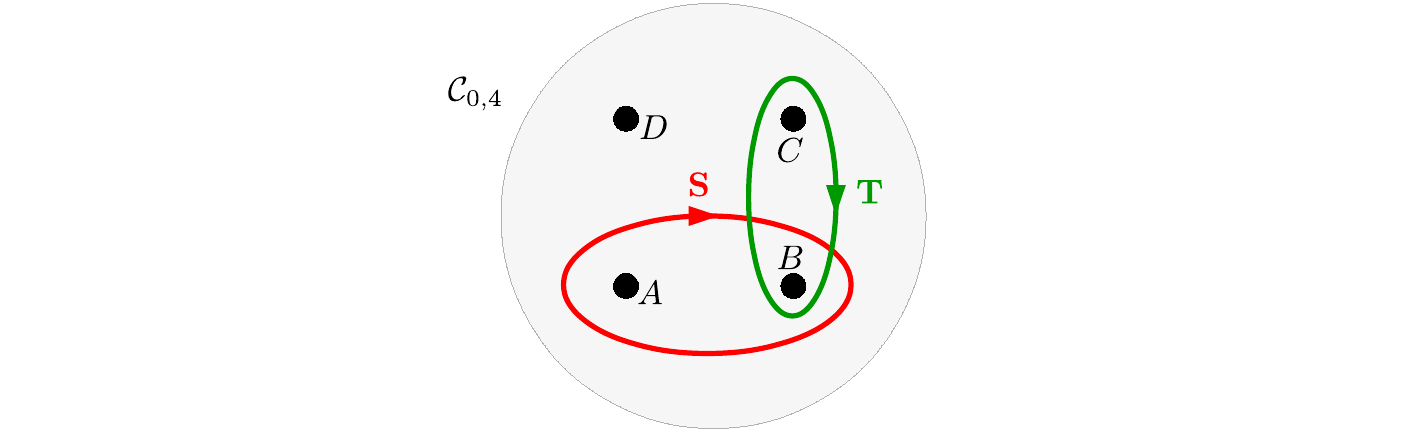}
\caption{Sphere $\mathcal{C}_{0,4}$ with 4 punctures $A,B,C,D$. The holonomy matrices $\bfS$ and $\bfT$ are associated with loops surrounding pairs of punctures.}
\label{C04ST}
\end{figure}

The next example (with $\chi=-2$) is the sphere $\mathcal{C}_{0,4}$ with four full punctures, denoted by $A$, $B$, $C$, $D$.
Its fundamental group can be expressed in terms of the loops $\gamma_A,\gamma_B,\gamma_C,\gamma_D$ surrounding the punctures clockwise, subject to one relation:
\bea
\pi_1 (\mathcal{C}_{0,4}) = \{ (\gamma_{A}, \gamma_{B}, \gamma_{C}, \gamma_{D}) | \gamma_{A}\gamma_{B} \gamma_{C} \gamma_{D}= \gamma_\circ \}~.
\eea
We associate the holonomy matrices $\bfA,\bfB,\bfC,\bfD$ to these loops, satisfying $\bfA\bfB\bfC\bfD=(-1)^{N-1} \id $.
In addition, we consider the loops $\gamma_S$ and $\gamma_T$ surrounding pairs of punctures: $\bfS \equiv \bfA\bfB$ and $\bfT \equiv \bfB\bfC$ (see figure~\ref{C04ST}).
We also define the two networks $N_{AB},N_{CD}$ (and their inverses $\bar N_{AB},\bar N_{CD}$) around punctures $A,B$,  and $C,D$, respectively, which are adapted to the pants decomposition determined by the curve $\gamma_S$.
Each of the holonomy matrices $\bfA,\bfB,\bfC,\bfD,\bfS,\bfT$ gives $(N-1)$ functions, while each of the four networks $N_{AB},N_{CD},\bar N_{AB}, \bar N_{CD}$ gives $(N-1)(N-2)/2$ functions.
This gives the following number of functions:
\bea
6(N-1) + 4  \frac{(N-1)(N-2)}{2} &=& 2(N^2-1 ) = \dim[\cM_{0,4}^N] ~.
\eea
Fixing the conjugacy classes of $\bfA,\bfB,\bfC,\bfD$ gives $4(N-1)$ constraints.

\subsubsection*{SL(2)}

The generators $A_1, B_1,C_1,D_1, S_1, T_1$ do not form a closed Poisson algebra on their own.
Indeed, applying the quantum skein relation gives
\bea\label{C04SL2ST}
\hat S_1 \hat T_1 &=& q^{-\frac12}\hat U_1  + q^{\frac12}  \hat U_1'+ \hat A_1\hat C_1 + \hat B_1 \hat D_1~,
\eea
with $U_1 \equiv \tr \bfB \bfD$ and $U'_1 \equiv \tr \bfA \bfC$ (up to an overall sign for later convenience).
The leading order in $\hbar$ gives the Poisson bracket
\bea
\{ S_1 , T_1 \} = - U_1 + U_1'~.
\eea
We can eliminate $U_1'$ via the relation~\eqref{C04SL2ST}, but then we must include $U_1$ in the set of generators in order to obtain a closed Poisson algebra.
The closure of the Poison algebra with the 7 generators $A_1, B_1,C_1,D_1, S_1, T_1,U_1$ is implied by the following $q$-deformed commutators:
\bea
q^{-\frac12}\hat S_1 \hat T_1 -  q^{\frac12} \hat T_1 \hat S_1 &=& (q^{-1}-q) \hat U_1 - (q^{\frac12} - q^{-\frac12})(\hat A_1 \hat C_1 + \hat B_1\hat D_1)~, \nn
q^{-\frac12}\hat T_1 \hat U_1 -  q^{\frac12} \hat U_1 \hat T_1 &=&(q^{-1}-q) \hat S_1 - (q^{\frac12} - q^{-\frac12})(\hat A_1 \hat B_1 + \hat C_1\hat D_1)~, \nn
q^{-\frac12}\hat U_1 \hat S_1 -  q^{\frac12} \hat S_1 \hat U_1 &=& (q^{-1}-q)\hat T_1 - (q^{\frac12} - q^{-\frac12})(\hat B_1 \hat C_1 + \hat A_1\hat D_1)~.
\eea
Since $\dim [\cM_{0,4}^2]=6$, there must be one relation between the 7 generators.
It is provided by the product of $U_1$ and $U_1'$:
\bea
\hat U_1 \hat U'_1 &=& q\hat S_1^2  + q^{-1} \hat T_1^2 + q^{\frac12} \hat S_1(  \hat A_1 \hat B_1 + \hat C_1 \hat D_1) + q^{-\frac12} \hat T_1 ( \hat B_1 \hat C_1 + \hat A_1 \hat D_1) \nn
&&  +\hat A_1\hat  B_1 \hat C_1 \hat D_1 +\hat A_1^2 +\hat B_1^2+ \hat C_1^2 +\hat D_1^2 -[2]^2~.
\eea
Eliminating $U_1'$ with the relation~\eqref{C04SL2ST} we obtain the familiar cubic relation (see for example~\cite{TV13})
\bea\label{relP1C04}
\hat \cP_1 &=& q^{-\frac12}\hat S_1\hat T_1\hat U_1 - q^{-1}\hat S_1^2 -q \hat T_1^2 -q^{-1}\hat U_1^2 \nn
&& - q^{-\frac12}\hat S_1 ( \hat A_1 \hat B_1 + \hat C_1 \hat D_1) -q^{\frac12} \hat T_1(\hat B_1 \hat C_1 + \hat A_1 \hat D_1)  -q^{-\frac12} \hat U_1 ( \hat A_1 \hat C_1 + \hat B_1 \hat D_1) \nn
&&  -\hat A_1 \hat B_1 \hat C_1 \hat D_1 -\hat A_1^2 -\hat B_1^2- \hat C_1^2 -\hat D_1^2 +[2]^2 ~.
\eea
Note that the same relation holds with $U_1$ replaced by $U_1'$ and $q$ by $ q^{-1}$.
We thus have a presentation of the algebra $\cA_{0,4}^2$ in terms of 7 generators satisfying the cubic relation $\cP_1$:
\bea
\cA_{0,4}^2 = \bC \left[ A_1, B_1,C_1,D_1, S_1, T_1,U_1\right] /  \cP_1~.
\eea
The Poisson brackets can be expressed as derivatives of $\cP_1$, for example
\bea
\{ S_1,T_1 \} &=& \frac{\del \cP_1}{\del U_1} = S_1 T _1-2 U_1 -( A_1 C_1 + B_1 D_1) ~.
\eea

\subsubsection*{SL(3)}

\begin{figure}[tb]
\centering
\includegraphics[width=\textwidth]{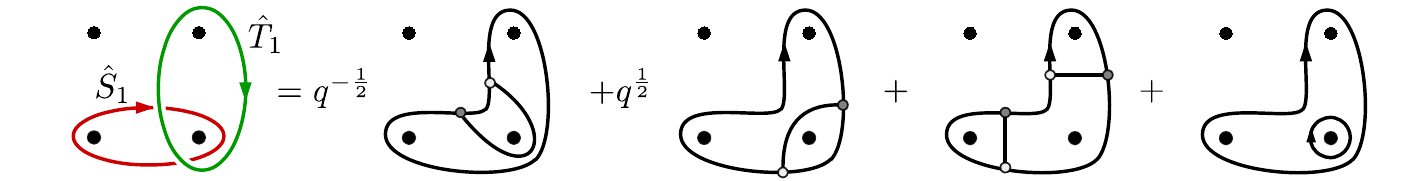}
\caption{The product $\hat S_1 \hat T_1$ generates networks via skein relations.}
\label{C04SL3ST}
\end{figure}

\begin{figure}[tb]
\centering
\includegraphics[width=\textwidth]{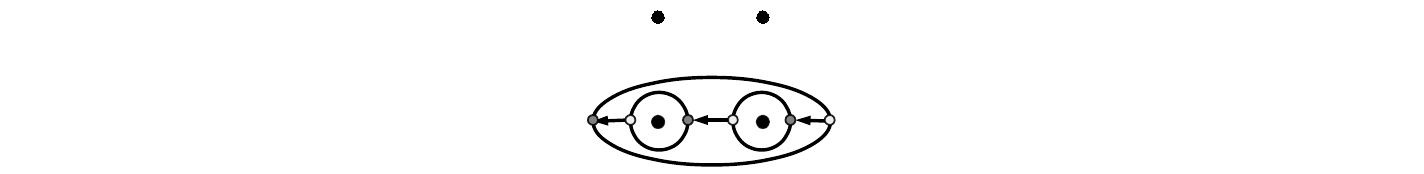}
\caption{The six-junction network $N_{6}$ that appears in the product $N_{AB} \bar N_{AB}$.}
\label{NABAbBb}
\end{figure}

For Riemann surfaces with $\dim[\pi_1(\cgn)] = 3$ such as $\cC_{0,4}$, 
the $SL(3)$ character variety has dimension 16, and is generated by a minimal number of 45 trace functions,  see~\cite{Teranishi}\cite{Lawton:2007ma}.
This implies the existence of 29 relations between the generators. 
Choices for the 16 \emph{independent} generators were given in~\cite{Lawton:2008ai}.
In terms our description, we can take the following loop and pants network functions:
\bea
 A_i, B_i, C_i, S_i, T_i, U'_i, N_{AB}, \bar N_{AB}, N_{AC}, \bar N_{AC} ~,
\eea
with
\bea 
N_{AB} &=& \tr \bfA \bfB^{-1}  - A_1B_2~, \qqq
N_{AC} = \tr \bfA \bfC^{-1}  - A_1C_2~.
\eea
We can then apply quantum skein relations to products of generators to obtain relations.
For example we get (see figure~\ref{C04SL3ST})
\bea\label{S1T1C04q}
\hat S_1 \hat T_1 &=& q^{-\frac12}\hat N_{BD} +q^{\frac12}\hat N'_{BD} + \hat N_{ABC} +  \hat B_1  \hat D_2~,
\eea
with 
\bea 
N_{BD} &=& \tr \bfB \bfD^{-1}  - B_1D_2~, \qqq
N'_{BD} = \tr \bfA \bfB^{-1} \bfC - B_2 U'_1~, \nn
N_{ABC} &=& \tr \bfC \bfB^{-1} \bfA - A_1 \bar N_{BC} - C_1 N_{AB} - A_1 B_2 C_1~.
\eea
We also find
\bea\label{NABNABbqC04}
\hat N_{AB} \hat {\bar N}_{AB} &=& q^{-\frac12}\hat N_{6} +q^{\frac12} \hat {\bar N}_{6} \nn
&&  + \hat S_1 \hat A_2\hat  B_2 + \hat S_2 \hat A_1 \hat B_1 + \hat S_1\hat  S_2 + \hat A_1\hat A_2 + \hat B_1 \hat B_2+[3]~,
\eea
with (see figure~\ref{NABAbBb})
\bea 
N_{6} &=& \tr \bfA \bfB \bfA^{-1} \bfB^{-1}  -  S_1 S_2 - A_1 A_2 - B_1 B_2~.
\eea

\newpage

\section{Fock-Goncharov holonomies}\label{secFG}

We now give an explicit representation of the algebra $\cA_{g,n}^N$ in terms of polynomials in the coordinates defined by Fock and Goncharov in their seminal work on higher Teichm\"uller theory~\cite{FG}.
The holonomy matrices constructed with their methods have some nice positivity properties, which imply that all loop and network functions are given by Laurent polynomials with positive integral coefficients.
Relations between generators can then be easily obtained (with the help of Mathematica for higher rank).
Thanks to the natural quantization of the Fock-Goncharov coordinates, these relations can be quantized uniquely. 
In all examples we have studied
perfect agreement is found with the results of skein quantization presented in the previous section.
The most non-trivial part of the quantization concerns the positive integral coefficients in the loop and network polynomials.
We will see that they get quantized to positive integral Laurent polynomials in $q^{1/2}$ --- as expected from their interpretation as the framed protected spin characters of Gaiotto, Moore, and Neitzke~\cite{Gaiotto:2010be}.
We give many examples up to $N=4$ for the surfaces $\cC_{0,3}$, $\torus$, and $\cC_{0,4}$.

\subsection{Fock-Goncharov coordinates}

\begin{figure}[b]
\centering
\includegraphics[width=\textwidth]{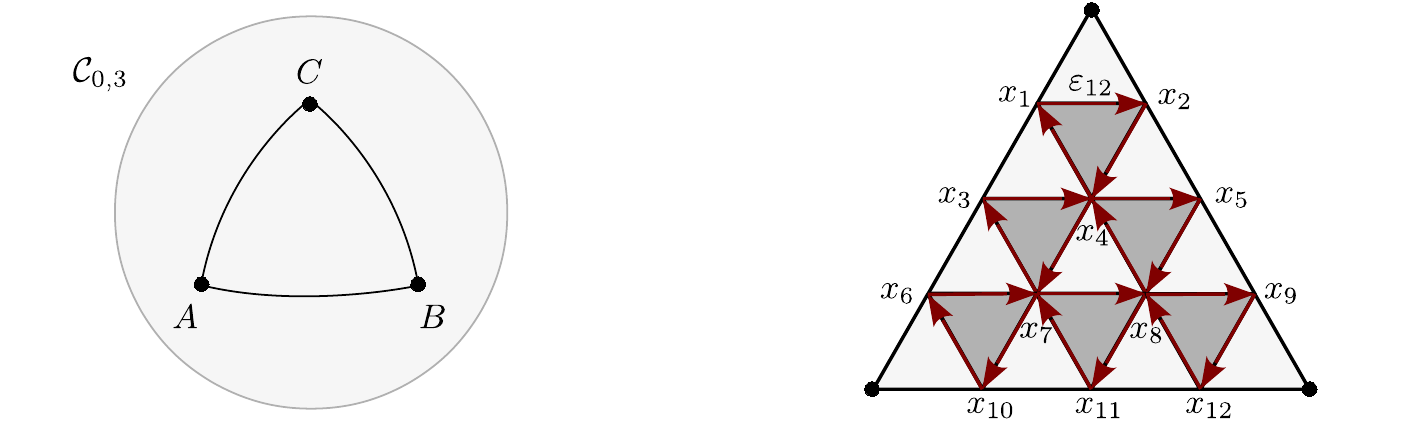}
\caption{\emph{Left}: ideal triangulation of the three-punctured sphere $\cC_{0,3}$ into two triangles.
\emph{Right}: $N$-triangulation of an ideal triangle into $N^2$ small black and white triangles (here for $N=4$). The Poisson structure $\varepsilon$ is encoded in the arrows circulating clockwise around the small black triangles.}
\label{Poisson}
\end{figure}

Fock and Goncharov defined useful systems of coordinates for $\cM_{g,n}^N$ associated with triangulations of $\cgn$.
Provided that $\cgn$ is a hyperbolic surface with at least one puncture, it can be decomposed into triangles with vertices at the punctures. There are $-2\chi$ triangles and $-3\chi$ edges in this ideal triangulation.
Each ideal triangle can then be further decomposed into $N^2$ small triangles, which produces a so-called $N$-triangulation (see figure~\ref{Poisson}).
The Fock-Goncharov coordinates $x_\alpha$, with $\alpha =1, \ldots, d$, are associated with the vertices of these small triangles (excluding the punctures of $\cgn$).
There are $(N-1)$ coordinates on each edge, and ${ N-1}\choose{2}$ coordinates inside each face, which add up to $d\equiv\dim[\cM_{g,n}^N]$
(see appendix~\ref{AppFG} for more details).

The Poisson structure on $\cM_{g,n}^N$ can be neatly encoded in a system of oriented arrows on the edges of the small triangles of the $N$-triangulation (see figure~\ref{Poisson} right).
The Poisson bracket between two Fock-Goncharov coordinates is given by
\bea\label{PoissonFG}
\{ x_\alpha, x_\beta \} = \varepsilon_{\alpha\beta} x_\alpha x_\beta ~,
\eea
with
\bea\label{PoissonTensorFG}
\varepsilon_{\alpha\beta}  = \#(\text{arrows from $x_\alpha$ to $x_\beta$}) - \#(\text{arrows from $x_\beta$ to $x_\alpha$}) \quad \in \{0,\pm 1,\pm 2\}~.
\eea
A monomial $x_1^{\ta_1} \cdots x_d^{\ta_d}$ can be encoded in a vector of exponents $\bfa = (\ta_1, \ldots, \ta_d)$, called tropical $\ta$-coordinates.\footnote
{The $\ta_i$ are coordinates for the tropicalization of the $\cA$-space defined by Fock and Goncharov, which is dual to the $\cX$-space parameterized by the $x_i$. 
The $\cA$-space is isomorphic to the space of laminations on $\cgn$.}
The Poisson bracket between two monomials $x_\bfa \equiv \prod_\alpha x_\alpha^{\ta_\alpha}$ and 
$x_\bfb \equiv \prod_\alpha x_\alpha^{\tb_\alpha}$ is given by
\bea\label{PoissonBxaxb}
\{ x_\bfa, x_\bfb \} = \sum_{\alpha,\beta} (\ta_\alpha \varepsilon_{\alpha\beta} \tb_\beta) x_\bfa x_\bfb = (\bfa^\text{t} \varepsilon \bfb) x_\bfa x_\bfb = -\sum_\alpha  (\tx_\alpha \tb_\alpha) x_\bfa x_\bfb  ~.
\eea
In the last expression, the combinations $\tx_\alpha \equiv \sum_\beta \varepsilon_{\alpha\beta} \ta_\beta$ are the tropical $\tx$-coordinates of the monomial $x_\bfa$.
Clearly, monomials with $\tx_\alpha=0$ for all $\alpha$ are central elements of the Poisson algebra (we will see below that they correspond to traces of holonomies around the punctures).
The moduli space $\cM_{g,n}^N$ is a symplectic fibration over the space of central monomials.

\subsection{Holonomies}

Fock and Goncharov constructed holonomies on the triangulated surface $\cgn$ using the snake matrices reviewed in appendix~\ref{AppFG}.
The general procedure to obtain the holonomy for a curve $\gamma$ is to choose a curve homotopic to $\gamma$ on the graph $\Gamma$ that is dual to the triangulation and multiply the matrices assigned to the corresponding edges and vertices of the dual graph.

\begin{figure}[tb]
\centering
\includegraphics[width=\textwidth]{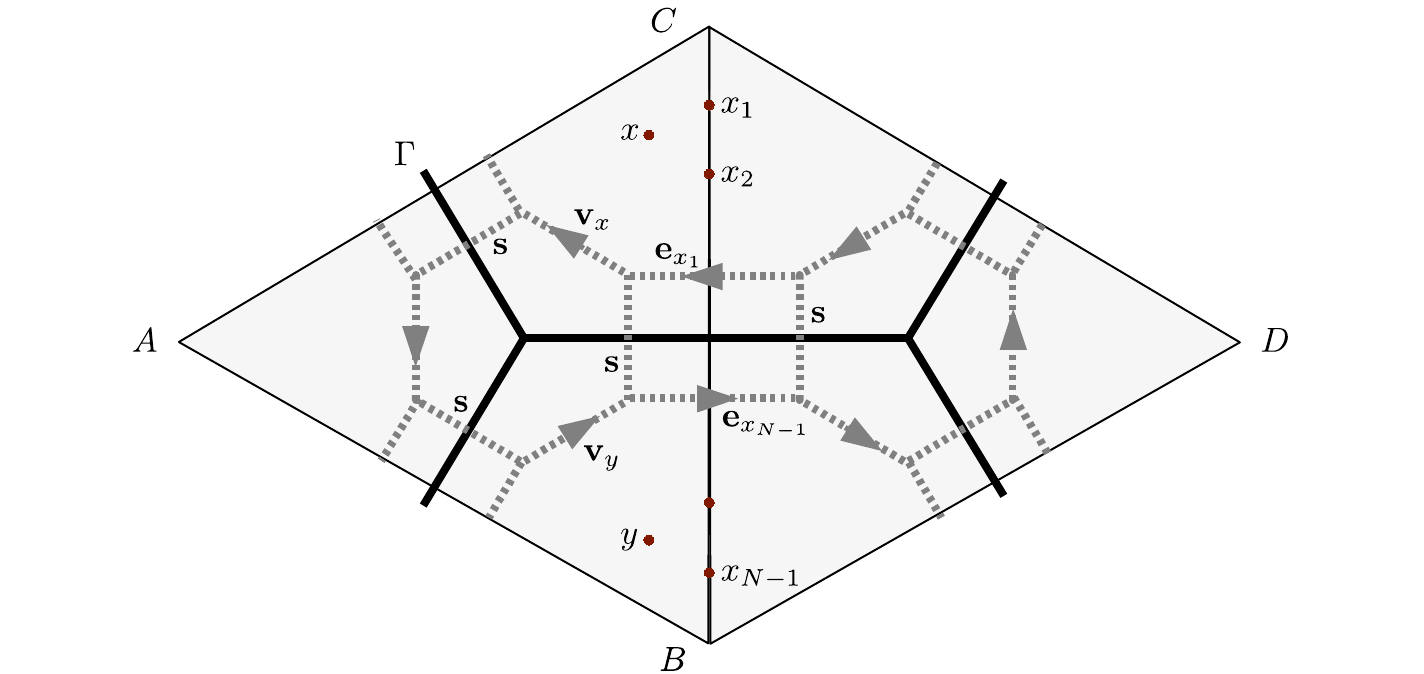}
\caption{Two triangles in a triangulation and the corresponding dual graph $\Gamma$ (black). The fat dual graph (dashed) is decomposed into rectangles and hexagons, and its segments are associated with matrices $\mathbf e$, $\mathbf s$, $\mathbf v$.}
\label{dualgraph}
\end{figure}

More precisely, the dual graph $\Gamma$ must be fattened and decomposed into rectangles along its edges and hexagons around its vertices (see figure~\ref{dualgraph}).
There are three types of segments in the decomposed fat graph: the segments $e$ crossing an edge of the triangulation, the segments $s$ intersecting the dual graph, and the segments $v$ around the vertices of the dual graph. The segments $e$ and $v$ are oriented clockwise around the punctures, while the segments $s$ are not oriented.
The matrices $ {\bf e,s,v}\in SL(N,\bC)$ assigned to the segments $e,s,v$ of the decomposed fat graph are the snake matrices defined in appendix~\ref{AppFG} (normalized to have unit determinant):
\bea
 {\bf e}_{x_1} &=& H_1(x_1) H_2(x_2) \cdots H_{N-1}(x_{N-1})~, \qqq
  {\bf s} =  S~, \qqq 
 {\bf v} _x=\cF~. 
\eea
Here the matrix ${\bf e}_{x_1} $ depends on the $(N-1)$ coordinates $x_1, \ldots, x_{N-1}$ along the relevant edge in the triangulation (for conciseness we only indicate the coordinate that is closest to the puncture around which it rotates).
Similarly, the matrix ${\bf v} _x$ depends on all the coordinates inside the relevant face in the triangulation.
For example, for $N=3$ we have
\bea
{\bf e} _{x_1} =  \begin{pmatrix} 1&0&0\\ 0&x_1&0\\  0&0& x_1 x_2 \end{pmatrix}~, \qqq
{\bf s} =  \begin{pmatrix} 0&0&1\\ 0&-1&0\\  1&0& 0 \end{pmatrix}~, \qqq 
{\bf v} _x =  \begin{pmatrix} 1&0&0\\ 1&1&0\\ 1&1+x & x \end{pmatrix}~.
\eea

The holonomy for a path on the fat graph with successive segments $(s_1, s_2, \cdots, s_m)$, where $s_i \in \{e,s,v\}$, is given by the product of the corresponding matrices: ${\bf s_m}\cdots {\bf s_2}{\bf s_1}$.
The holonomy around a rectangle or an hexagon is (projectively) equal to the identity:
\bea\label{FGsnakeRectHex}
{\bf eses} = {\bf vsvsvs} = (-1)^{N-1} \id~.
\eea
Note that this sign is consistent with our conventions for the contraction of a fundamental loop in~\eqref{qSkeinSLN} (in the classical case $q=1$).

The holonomy for any curve $\gamma$ on the surface $\cgn$ is obtained by choosing a curve homotopic to $\gamma$ on the fat graph and multiplying the corresponding matrices (the properties~\eqref{FGsnakeRectHex} ensure that the choice of curve on the graph is irrelevant). 
Fock and Goncharov showed that the resulting holonomy matrix is conjugate to a matrix whose minors are given by Laurent polynomials with positive integral coefficients (in any coordinate system, that is for any triangulation).\footnote
{In the case of a holonomy for a loop running around a puncture, the resulting matrix is moreover conjugate to a triangular matrix.}
It follows that the loop functions $A_i$ that we defined in~\eqref{AiSumsOfMinors} as sums of principal minors (invariant under conjugation) will also be given by positive integral Laurent polynomials in the variables $x_\alpha^{1/N}$.
We will moreover observe in explicit examples below that the network functions~\eqref{networkdefined} also turn out to be positive integral Laurent polynomials, but we did not find an easy derivation of this property from the positivity of the minors.

A loop or network function $L$ will always contain a highest term with unit coefficient, that is a monomial $x_\bfa=\prod_\alpha x_\alpha^{\ta_\alpha}$ such that any other monomial $x_\bfb=\prod_\alpha x_\alpha^{\tb_\alpha}$ has $\tb_\alpha \leq \ta_\alpha$ for all $\alpha$:
\bea
L = x_1^{\ta_1} x_2^{\ta_2}\cdots x_d^{\ta_d} +  \cdots
\eea
In contrast, other monomials in $L$ have integral coefficients $ \overline{\underline{\Omega}}$ that can be larger than 1:
\bea
L =  x_\bfa +   \overline{\underline{\Omega}}_\bfb x_\bfb +  \overline{\underline{\Omega}}_\bfc x_{\bfc} +  \cdots  ~.
\eea

The product of two loop or network functions $L$ and $L'$ can be expanded as
\bea\label{LLpProductFG}
L L' = \sum_{L''} c(L,L' ;L'')   L''  ~.
\eea
For $SL(2)$, Fock and Goncharov proved the positivity of the coefficients $c(L,L' ;L'')$ by applying the skein relation shown in figure~\ref{SkeinSL2} to the intersections between the loops associated with $L$ and $L'$ and by reducing contractible loops as in~\eqref{qUnknotReide1}.\footnote
{More precisely, their proof does not apply to the loop functions $A_i$ that we are using, but to their slightly different ``canonical maps'' $\bfI=\tr ( \bfA^i )$ from the space of integral $\cA$-laminations to the space of positive Laurent polynomials.}
Note that the positivity of the $SL(N)$ skein relation~\eqref{SkeinSL3} immediately implies that a product of loop or network functions can be written as a finite sum with positive coefficient.
However, reduction moves such as those shown in figure~\ref{ReductionSLN} can spoil this positivity since they involve negative signs.

\subsection{Quantization} \label{SECquantizationFG}

The Fock-Goncharov coordinates admit a natural quantization.
The algebra $\cA_{g,n}^N$ of functions on $\cM_{g,n}^N$ can be $q$-deformed into a noncommutative algebra $\cA^N_{g,n}(q)$ by promoting the coordinates $x_\alpha$ to operators $\hat x_\alpha$ satisfying the relations
\bea
\hat  x_\alpha \hat x_\beta = q^{\varepsilon_{\alpha\beta}} \hat x_\beta\hat x_\alpha~.
\eea
It is convenient to work with logarithmic coordinates $X_\alpha$, defined via
$x_\alpha = \exp X_\alpha$, and the corresponding operators $\hat X_\alpha$, which satisfy the commutation relation (recall $q = \exp\hbar$)
\bea
[ \hat X_\alpha,\hat  X_\beta ] = \hbar  \{ X_\alpha, X_\beta \}  = \hbar\varepsilon_{\alpha\beta} ~.
\eea
A monomial in the Fock-Goncharov coordinates can be quantized by first expressing it as an exponential of a sum of logarithmic coordinates, and then promoting them to operators (as in~\cite{Dimofte:2011jd} for example):
\bea \label{FGmonomQt}
x_{\bfa}=\prod_{\alpha=1}^n x_\alpha^{\ta_\alpha}  = \exp \sum_\alpha \ta_\alpha  X_\alpha   \qquad \stackrel{q}{\longrightarrow} \qquad \hat x_{\bfa}= \exp  \sum_\alpha \ta_\alpha \hat X_\alpha   = q^{-\frac12\sum_{\alpha<\beta} \ta_\alpha \varepsilon_{\alpha\beta} \ta_\beta } \prod_\alpha \hat x_\alpha^{\ta_\alpha} ~,  \nonumber \\ 
\eea
where in the last step we used the Baker-Campbell-Hausdorff formula.
Similarly, the quantum product of two monomials is given by
\bea \label{quantumProdFGmonomials}
\hat x_{\bfa}  * \hat x_{\bfb} &=& \exp \sum_\alpha \ta_\alpha \hat X_\alpha  *   \exp \sum_\beta \tb_\beta \hat X_\beta  =  \ex^{\frac12\sum_{\alpha,\beta} [\ta_\alpha \hat X_\alpha  ,\tb_\beta \hat X_\beta ]}  \exp  \sum_\alpha (\ta_\alpha   +\tb_\alpha) \hat X_\alpha \nn
& =& q^{\frac12 \bfa^\text{t} \varepsilon \bfb}\hat x_{\bfa+\bfb} ~.
\eea

The loop and network functions that we want to quantize are positive integral Laurent polynomials in the Fock-Goncharov coordinates. 
The monomials $x_\bfa$ that they involve can simply be quantized as in~\eqref{FGmonomQt}.
It is much less obvious to determine how to quantize the positive integral coefficients $\overline{\underline{\Omega}}$ of the monomials. 
Fock and Goncharov conjectured that these quantum coefficients are positive Laurent polynomials in $q^{1/2}$.
They also conjectured that the highest term, which has a unit coefficient classically, has a unit coefficient in the quantum operator too.
We will make the assumption that all the unit coefficients in a loop or network polynomial remain unit coefficients in the quantized operator (as expected from the interpretation of these coefficients as protected spin characters in~\cite{Gaiotto:2010be}).
What remains to find is how the non-unit coefficients $\overline{\underline{\Omega}}$ get quantized:
\bea
L=  x_\bfa +   \overline{\underline{\Omega}}_\bfb x_\bfb +     \cdots    \qquad \stackrel{q}{\longrightarrow} \qquad  \hat L= \hat x_\bfa +   \overline{\underline{\Omega}}^q_\bfb \hat  x_\bfb +  \cdots      ~.
\eea
Our strategy is to demand that the classical loop and network functions, which satisfy some relations of the form~\eqref{LLpProductFG} (such as~\eqref{N1N1bskeinRelC03classical} and~\eqref{sexticC03SL3} for $\pants$), get quantized into operators satisfying quantized versions of these relations. This requirement turns 
out to be powerful enough to determine uniquely the  coefficients in expansions of 
the loop and network generators
into monomials of quantised Fock-Goncharov-coordinates.
We will illustrate this quantization procedure in many examples in the following subsections.

We first need to determine how the quantized relations look like.
The classical relations are typically of the form~\eqref{LLpProductFG} and get quantized to 
\bea\label{LLpProductFGq}
\hat L *  \hat L' = \sum_{\hat L''} c^q(\hat L,\hat L' ;\hat L'') \hat  L''  ~,
\eea
where the $c^q(\hat L,\hat L' ;\hat L'')$ are some functions of $q$.
On the left-hand side, the quantum product $\hat L \hat L' $ generates some powers of $q$ as in~\eqref{quantumProdFGmonomials}.
Focusing on the monomials in $\hat L$, $\hat L'$, and $\hat L''$ with unit coefficients, we can then read off the quantum coefficients $c^q(\hat L,\hat L' ;\hat L'')$.
We will see that they are integral Laurent polynomials in $q^{1/2N}$.
The resulting quantum relations agree with the ones that we could obtain from skein quantization, such as~\eqref{N1N1bc03}.

We can then determine the quantum coefficients $\overline{\underline{\Omega}}^q$ by comparing
coefficients of monomials with the same exponents on both sides of the quantum relations.
We will find that the $\overline{\underline{\Omega}}^q$ are always finite positive integral Laurent polynomials in $q^{1/2}$ which are invariant under $q \leftrightarrow q^{-1}$.
This is in agreement with the positivity conjectures made in~\cite{Gaiotto:2010be} for the protected spin characters.

\subsection{Pants networks}

We come back to the basic example of flat $SL(N,\bC)$-connections on the pair of pants $\cC_{0,3}$.
The abstract structure of the algebra of loop and network operators was discussed in subsection~\ref{secPantsAlgebra}.

 \subsubsection*{SL(3)} 

\begin{figure}[tb]
\centering
\includegraphics[width=\textwidth]{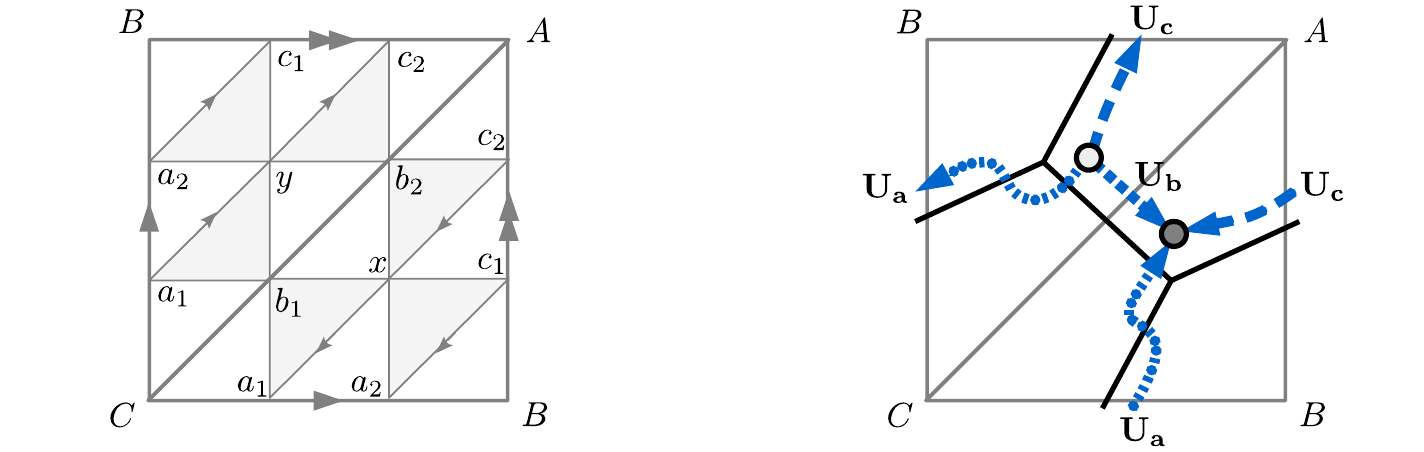}
\caption{\emph{Left}: 3-triangulation of the sphere 
$\mathcal{C}_{0,3}$ with three punctures labeled by $A,B,C$. The edges carry two 
coordinates each, $a_i, b_i,c_i$ and the faces one each, $x$, $y$. \emph{Right}: The dual graph, and the branches ${\bf U_a,\bf U_b,\bf U_c}$ projected on the fat graph. The white junction indicates the base point for the loops in $\pi_1(\cC_{0,3})$.}
\label{C03triangdual}
\end{figure}

A 3-triangulation of the pair of pants $\pants$ is shown in figure~\ref{C03triangdual} left.
We denote the Fock-Goncharov coordinates on the edges of the triangles by $a_i, b_i,c_i$, with $i=1,2$, and on the faces by $x$, $y$.

\paragraph{Loop functions:}

\begin{figure}[tb]
\centering
\includegraphics[width=\textwidth]{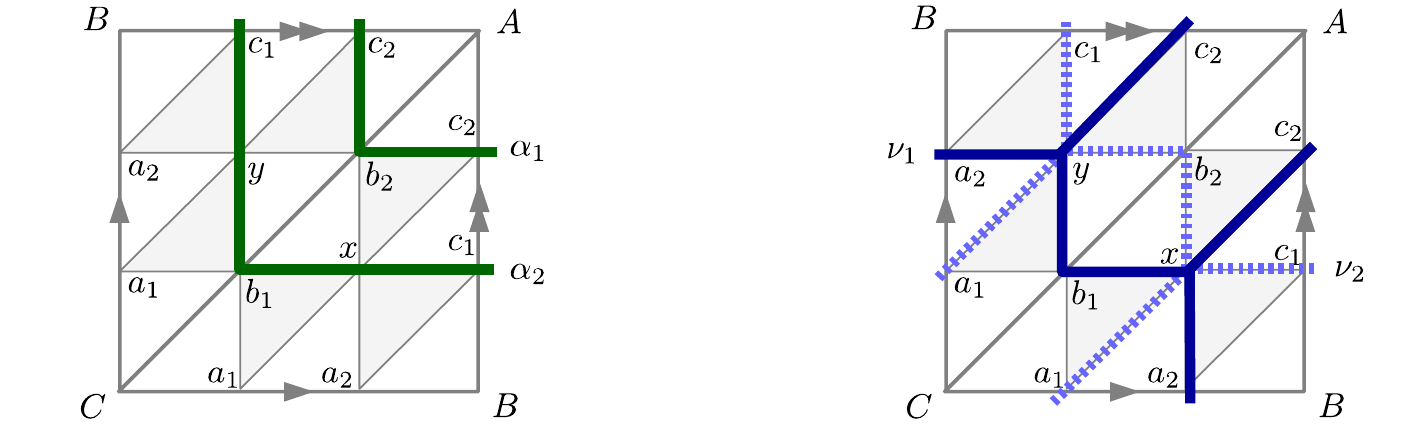}
\caption{\emph{Left}: Paths homotopic to the loop around the puncture $A$ on the 3-triangulation of $\pants$, which correspond to the monomials $\alpha_1 = b_2 c_2$ and $\alpha_2=b_1c_1xy$. \emph{Right}:~Paths homotopic to the network $N_1$, corresponding to $\nu_1= a_2b_1c_2 xy$ (dark) and $\nu_2 = a_1 b_2 c_1 xy$ (dashed).}
\label{C03networkstriang}
\end{figure}

The dual graph consists of three edges ${\bf U_a,\bf U_b,\bf U_c}$, which we can project on 
segments of the fat graph to obtain the following holonomy matrices:
\bea
{\bf U_a} &=& {\bf s v}^{-1}_x{\bf s e}_{a_2}{\bf sv}^{-1}_y{\bf s}~, \qqq
{\bf U_b} = {\bf e}^{-1}_{b_2}~, \qqq
{\bf U_c}  = {\bf v}_x{\bf e}_{c_2}{\bf v}_y~. 
\eea
The holonomy matrices for the three clockwise loops around the punctures can then be expressed as
\bea
\mathbf{A} = \bfU_{\bf b}^{-1} {\bf U_c}~, \qqq
\mathbf{B}= \bfU_{\bf c}^{-1}{\bf U_a} ~, \qqq
\mathbf{C} = \bfU_{\bf a}^{-1}{\bf U_b}~, 
\eea
so that they satisfy the relation
$\mathbf{A B C} = \id$ from $\pi_1(\cC_{0,3})$.
The eigenvalues of these matrices correspond (up to normalizations) to products of coordinates along parallel loops around the punctures in the 3-triangulation (see figure~\ref{C03networkstriang} left):
\bea\label{evalABC}
\mathbf{A } &:& \qqq  ( 1, \alpha_1 , \alpha_1 \alpha_2) ~,\qqq  \alpha_1 = b_2c_2~, \quad  \alpha_2 = b_1 c_1 xy~,  \nn
\mathbf{B}  &:& \qqq  ( 1, \beta_1 , \beta_1 \beta_2) ~, \qqq \beta_1 = c_1a_2~, \quad  \beta_2=  c_2a_1 xy~,\nn
\mathbf{C} &:& \qqq  ( 1, \gamma_1 , \gamma_1 \gamma_2)  ~,\qqq  \gamma_1  = a_1 b_1~, \quad \gamma_2 = a_2 b_2 xy~.
\eea
Defining the loop functions $A_1=\tr \mathbf{A}$ and $A_2=\tr \mathbf{A}^{-1}$, we can write the compact expression 
\bea
A_i =  \prod_j  \alpha_j^{-\kappa^{-1}_{ij}} (1 + \alpha_i + \alpha_1\alpha_2 )~,
\eea
where the normalization factor is determined by the Cartan matrix $\kappa$ of $SL(3)$:
\bea \label{CartanSL3}
\kappa = \begin{pmatrix} 2 & -1 \\ -1 &2 \end{pmatrix}~, \qqq \kappa^{-1} = \frac13 \begin{pmatrix} 2  & 1  \\ 1  &2  \end{pmatrix}~.
\eea
This simple interpretation of the loop functions comes from the fact that for a path around a puncture the Fock-Goncharov holonomy matrix can be written as a triangular matrix:
\bea
\bfA =\prod_j  \alpha_j^{-\kappa^{-1}_{1j}}  \left(
\begin{array}{ccc}
 1 & 0 & 0 \\
b_2+\alpha_1  & \alpha_1  & 0 \\
 b_1  b_2(1 +  c_2   +  c_2  x  +  c_1  c_2  x )  & b_1\alpha_1( 1  +x   + c_1  x   +c_1xy ) & \alpha_1\alpha_2  \\
\end{array}
\right)~.
\eea
The normalization factor ensures that $\det \bfA=1$.
Note that the tropical $\tx$-coordinates all vanish for $A_i$, $B_i$, $C_i$, which implies that they are central elements of the Poisson algebra (recall~\eqref{PoissonBxaxb}).
%

\paragraph{Network functions:}

We construct the network function $N_1$ and its reverse $\bar N_1$ by fusing the three edges at the two trivalent junctions with $\epsilon$-tensors (see~\eqref{pantsN1VabcC03}, or alternatively~\eqref{N1ACbC03}): 
\bea
N_1 &=& - \epsilon_{mnp} \bfU_{{\bf a}\, r}^{m}\bfU_{{\bf b}\, s}^{n} \bfU_{{\bf c}\, t}^{p} \epsilon^{rst}~, \nn
\bar N_1 &=& - \epsilon_{mnp} 
(\bfU_{\bf a}^{-1})^m_r(\bfU_{\bf b}^{-1})^n_s(\bfU_{\bf c}^{-1})^p_t \epsilon^{rst}~.
\eea
This gives polynomials with 25 terms each.
Three of these terms stand out: the highest term, the lowest term, and the middle term (which also happens to be the only term with a non-unit coefficient).
They can be written in terms of monomials $\nu_1$ and $\nu_2$ that have a geometric interpretation as paths homotopic to the network on the 3-triangulation (see figure~\ref{C03networkstriang} right):
\bea\label{N1nu1nu2}
N_1 &\ni  &\prod_i  \nu_i^{-\kappa^{-1}_{1i}} \left(  1 + 2 \nu_1 + \nu_1\nu_2 \right)~, \qqq 
\nu_1= a_2b_1c_2 xy~, \nn
\bar N_1 &\ni & \prod_i \nu_i^{-\kappa^{-1}_{2i}} \left(  1 + 2 \nu_2 + \nu_1\nu_2 \right)~, \qqq  \nu_2 = a_1 b_2 c_1 xy ~.
\eea
The full expression for $N_1$ is 
\bea\label{N1fullExpr}
N_1 &=& \prod_i  \nu_i^{-\kappa^{-1}_{1i}}  \big( 
1 + y + a_2 y + b_1 y + c_2 y + a_2 b_1 y + a_2 c_2 y + b_1 c_2 y + 
 a_2 b_1 c_2 y + a_2 b_1 x y \nn
 &&\qqq + a_2 c_2 x y + b_1 c_2 x y + 2 \nu_1 + 
 a_1 a_2 b_1 c_2 x y + a_2 b_1 b_2 c_2 x y + a_2 b_1 c_1 c_2 x y   \nn
 && \qqq+ 
 a_2 b_1 c_2 x^2 y + a_1 a_2 b_1 c_2 x^2 y + a_2 b_1 b_2 c_2 x^2 y  + 
 a_1 a_2 b_1 b_2 c_2 x^2 y + a_2 b_1 c_1 c_2 x^2 y  \nn
 && \qqq+ a_1 a_2 b_1 c_1 c_2 x^2 y + 
 a_2 b_1 b_2 c_1 c_2 x^2 y + a_1 a_2 b_1 b_2 c_1 c_2 x^2 y + 
\nu_1\nu_2
\big)~.
\eea
This is consistent with the expression obtained in~\cite{Xie:2013lca} from products of trace functions (see also~\cite{Saulina:2014dia}).
The tropical $\ta$-coordinates of $N_1$
(the exponents of its highest term) and its tropical $\tx$-coordinates are given by 
\bea
\bfa(N_1) &=& \frac13 (2,1,1,2,2,1,3,3)~, \qqq
\bfx (N_1) = (0,0,0,0,0,0,1,-1)~.
\eea

\paragraph{Poisson brackets:}

The Poisson bracket for the pants networks is computed by using~\eqref{PoissonBxaxb}. 
The Poisson tensor~\eqref{PoissonTensorFG} can be read off from the 3-triangulation in figure~\ref{C03triangdual}, and is given in the basis $\{a_1,a_2, b_1,b_2,c_1,c_2,x,y\}$ by
\bea
\varepsilon  = \left(
\begin{array}{cccccccc}
 0 & 0 & 0 & 0 & 0 & 0 & -1 & 1 \\
 0 & 0 & 0 & 0 & 0 & 0 & 1 & -1 \\
 0 & 0 & 0 & 0 & 0 & 0 & 1 & -1 \\
 0 & 0 & 0 & 0 & 0 & 0 & -1 & 1 \\
 0 & 0 & 0 & 0 & 0 & 0 & -1 & 1 \\
 0 & 0 & 0 & 0 & 0 & 0 & 1 & -1 \\
 1 & -1 & -1 & 1 & 1 & -1 & 0 & 0 \\
 -1 & 1 & 1 & -1 & -1 & 1 & 0 & 0 \\
\end{array}
\right)
\eea
This gives the Poisson bracket~\eqref{PBN1N1b}
\bea\label{PBN1N1b2}
\{N_1,  \bar N_1\} &=& -W_1 + \bar W_1 ~.
\eea
The network functions $W_1$ and $\bar W_1$ have 187 terms each, with the following highest, lowest, and middle terms:
\bea 
W_1 &\ni  &\prod_j  \sigma_j^{-\kappa^{-1}_{1j}} \left(  1 + 8 \sigma_1 + \sigma_1\sigma_2 \right)~, \qqq 
\sigma_1= a_1a_2b_1b_2c_1c_2y^3~, \nn
\bar W_1&\ni & \prod_j  \sigma_j^{-\kappa^{-1}_{2j}} \left(  1 + 8 \sigma_2 + \sigma_1\sigma_2 \right)~, \qqq  \sigma_2 = a_1a_2b_1b_2c_1c_2x^3~.
\eea
All the Poisson brackets~\eqref{PBC03SL3} between the generators $N_1,\bar N_1, W_1, \bar W_1$ can be easily computed in this way.

\paragraph{Classical relations:}

It is easy to obtain the polynomial relations between the generators $A_i,B_i,C_i, N_1,\bar N_1, W_1, \bar W_1$.
One useful method is to start with a product, say $N_1 \bar N_1$, and to look for a combination of generators with the same highest term, in this case $A_1 B_1 C_1$, in order to cancel it. 
Then we find that the highest terms in  $N_1 \bar N_1 - A_1 B_1 C_1$ can be cancelled by $W_1$ and $
\bar W_1$. Repeating this procedure a few more times leads to the relation $\cP_1$ given in~\eqref{N1N1bskeinRelC03classical}.
After implementing this algorithm in Mathematica we can obtain relatively complicated relations such as $\cP_2$ given in~\eqref{sexticC03SL3}.
In contrast, it would be very laborious to derive this relation purely from applying skein relations, because it would require the resolution of many intersections in the products $W_1 \bar W_1$, $N_1^3$, and $N_1^2$.
Of course, the fact that $\cP_2$ is a combination of several product expansions of the form~\eqref{LLpProductFG} implies that its coefficients can appear somewhat unnatural (this comment applies even more for the quantized relation~\eqref{W1bW1quantum}).

\paragraph{Quantization of the relations:}

We now want to obtain quantum versions of the polynomial relations $\cP_1$ and $\cP_2$, in which the network functions are replaced by noncommuting operators.
Each term in the classical relations can acquire at the quantum level a coefficient that is an arbitrary function of the quantization parameter $q=\ex^\hbar$ and that reduces to the classical integral coefficient in the limit $q\to1$.

The quantum product of polynomials in the Fock-Goncharov coordinates can be obtained by applying the product~\eqref{quantumProdFGmonomials} to each pair of monomials.
For example, the quantum product $\hat N_1\hat{\bar N}_1$ will produce a certain power of $q$ for each pair of monomials coming from $\hat N_1$ and $\hat{\bar N}_1$.
Let us consider first  the highest terms $x_{\bfa}$ and $x_{\bar \bfa}$ in $N_1$ and $\bar N_1$ with tropical $\ta$-coordinates  
\bea
\bfa &=&\frac13 ( 2, 1, 1, 2, 2, 1, 3, 3 )~, \qqq   \bar\bfa = \frac13 (1, 2, 2, 1, 1, 2, 3, 3)~.
\eea
These two monomials Poisson-commute, $\{ x_{\bfa},x_{\bar \bfa}\}=0$, which implies that their quantum product does not produce any power of $q$:
\bea
\hat x_{\bfa} \hat x_{\bar \bfa}&=&  \hat x_{\bfa+\bar\bfa} ~.
\eea
This implies that the corresponding term in $\hat\cP_1$ with tropical $\ta$-coordinate 
\bea
\bfa+\bar\bfa=(1, 1, 1, 1, 1, 1, 2, 2)
\eea 
must have the same coefficient as $\hat N_1\hat{\bar N}_1$.
This term turns out to be the highest term in $\hat A_1\hat B_1\hat C_1$, and so, by an overall rescaling, we can set the coefficients of $\hat N_1\hat{\bar N}_1$ and $\hat A_1\hat B_1\hat C_1$ to one in the quantum relation $\hat \cP_1$.
The next highest terms in the relation have tropical $\ta$-coordinates
$(1, 1, 1, 1, 1, 1, 2, 1)$ and $(1, 1, 1, 1, 1, 1, 1, 2)$.
They correspond respectively to the highest terms of the operators $\hat W_1$ and $\hat{\bar W}_1$, as well as to two products of monomials in $\hat N_1\hat{\bar N}_1$ with coefficients $q^{-1/2}$ and $q^{1/2}$.
This fixes the quantum coefficients of $\hat W_1$ and $\hat{\bar W}_1$ in $\hat \cP_1$.

Repeating this procedure for the next highest terms in the relation allows us to determine all the quantum coefficients, except for the constant term 3 in $\cP_1$.
We find that it should quantize as $Q_2 \bar Q_2 -1$, where $Q_2$ and $\bar Q_2$ are the quantizations of the coefficients of 2 that appear in the expansion of the network operators $\hat N_1$ and $\hat{\bar N}_1$ (recall~\eqref{N1nu1nu2}).
We will show momentarily that $Q_2=\bar Q_2 = [2]$ and thus the 3 should quantize as $[2]^2-1=[3]$.
The quantum relation $\hat \cP_1$ finally takes the form
\bea \label{N1N1bc032}
\hat N_1 \hat {\bar N}_1 &=&  q^{-\frac12} \hat W_1 +  q^{\frac12}  \hat{ \bar W}_{1} +\hat A_1 \hat B_1 \hat C_1+ \hat  A_2  \hat B_2 \hat C_2 + \hat A_1  \hat A_2 +\hat B_1 \hat  B_2 +\hat C_1 \hat C_2 + [3]~. 
\eea
Pleasingly, this quantum relation agrees exactly with the result~\eqref{N1N1bc03} that we obtained in section~\ref{secPantsAlgebra} by applying the quantum skein relation.

Applying the same procedure (with the help of Mathematica) to the second relation $\cP_2$ leads to the quantum relation given in~\eqref{W1bW1quantum}.

\paragraph{Quantization of the generators:}

Having quantized the relations, we would now like to quantize the generators themselves, as described in subsection~\ref{SECquantizationFG}.
The generators are polynomials in the Fock-Goncharov coordinates with integer coefficients.
The unit coefficients should not be affected by the quantization, but we need to find how to quantize the non-integral coefficients that appear in front of some monomials.
We will see how the quantum relation~\eqref{N1N1bc03} can be used to determine these quantum coefficients uniquely.

There is only one term in the expansion for $N_1$ with a non-unit coefficient, namely the middle term with the factor of 2 in~\eqref{N1nu1nu2}.
In the quantum operator $\hat N_1$, this coefficient of 2 will be replaced by a function $Q_2$ of $q$.
We can determine $Q_2$ by finding a term in the quantum relation that is linear in it.
For example, the monomial $1/a_1$ appears in $\hat N_1 \hat{\bar N}_1$ with the coefficient $Q_2$, and also in $\hat W_{1}$ and $\hat {\bar W}_{1}$ with unit coefficient. Since $W_1$ and $\bar W_1$ appear in the quantum relation~\eqref{N1N1bc03} with factors of $q^{1/2}$ and $q^{-1/2}$, we deduce that the coefficient of 2 in $N_1$ simply becomes a quantum [2] in $\hat N_1$:
\bea
2 \in N_1  \quad \stackrel{q}{\to} \quad  Q_2 = q^{\frac12}+q^{-\frac12} \equiv [2]  \in \hat N_1~.
\eea 
The coefficient of 2 in $\bar N_1$ similarly quantizes to $\bar Q_2 = [2]$.
So we have (up to normalization)
\bea
N_1 &\ni  & 1 + 2 \nu_1 + \nu_1\nu_2  \qqq \stackrel{q}{\to} \qqq \hat N_1 \ni   1 + [2] \hat \nu_1 + \hat \nu_1\hat \nu_2~, \nn
\bar N_1 &\ni  & 1 + 2 \nu_2 + \nu_1\nu_2  \qqq \stackrel{q}{\to} \qqq \hat N_1 \ni   1 + [2] \hat \nu_2 + \hat \nu_1\hat \nu_2~.
\eea

Among the 187 monomials in the network $W_{1}$, 24 have a coefficient 2, 4 a 3, 12 a 4, 6 a 5, 2 a 6, and 1 an 8.
Let us focus on the monomial $8y/x$ in $W_{1}$.
In the classical relation $\cP_1$ given in~\eqref{N1N1bskeinRelC03classical}, the coefficient of 8 is cancelled by 8 pairs of monomials in $N_1 \bar N_1$, $ x_{\bfa_i}$ from $N_1$ and $ x_{\bar\bfa_i}$ from $ \hat{\bar N}_1$, whose products give $y /x$.
For example we have
\bea
\bfa_1&=&-\frac{1}{3} (1,2,2,1,1,2,3,0)~, \qqq \bar \bfa_1= \frac{1}{3}( 1,2,2,1,1,2,0,3 )
\eea 
satisfying $\bfa_1+ \bar \bfa_1 = (0,0,0,0,0,0,-1,1)$.
In the quantum relation, however, the quantum product of the corresponding monomials in $\hat N_1 \hat{\bar N}_1$ produces some power of $q$ as in~\eqref{quantumProdFGmonomials}:
\bea
\hat x_{\bfa_1} \hat x_{\bar\bfa_1} &=& q^{\frac12 \bfa_1^\text{t} \varepsilon \bar\bfa_1}\hat x_{\bfa_1+ \bar\bfa_1} 
= q \hat x_{ \bfa_1+ \bar\bfa_1} ~.
\eea
Summing over the contributions of all such pairs $(\bfa_i, \bar\bfa_i)$, we obtain the expression for the quantization of the term $8 y/x$ in $\hat W_{1}$ (recall there is a factor of $q^{-1/2}$ in front of $\hat W_{1}$ in the quantum relation~\eqref{N1N1bc03}):
\bea
8\in W_1  \quad \stackrel{q}{\to} \quad q^{\frac12} \sum_{i=1}^8 q^{\frac12 \bfa_i^\text{t} \varepsilon \bar\bfa_i}= q^{\frac32}+3 q^{\frac12}+3 q^{-\frac12}+ q^{-\frac32} =   [2]^3   \in \hat W_1~.
\eea
The quantization of all the non-unit coefficients in the network operators $\hat W_1$ and $\hat{\bar W}_1$ can be determined in the same way.
We find
\bea
&& 2 \to q^{\frac12}+q^{-\frac12} = [2]~, \qqq  \qquad\;\; 8 \to q^{\frac32}+3 q^{\frac12}+3 q^{-\frac12}+ q^{-\frac32}= [2]^3~, \nn 
&& 3\to q+1+q^{-1} = [3]~, \qqq  \;\; \;\;\; \;  4\to q+2+q^{-1} =[3]+1= [2]^2~,  \nn
&& 5 \to q+3+q^{-1} = [3] + 2~, \qqq 6\to q+4+ q^{-1} = [3] + 3~.
\eea
We see that all the quantized coefficients in the loop and network operators are positive integral Laurent polynomials in $q^{1/2}$, and are also invariant under $q \to q^{-1}$.

This provides further evidence for the positivity conjectures of Gaiotto, Moore, and Neitzke~\cite{Gaiotto:2010be} about the framed protected spin characters~\eqref{PSCbarOmegay}, which should take the form
\bea 
\overline{\underline{\Omega}}(L, \gamma;q) = \tr_{\cH_{L}^\text{BPS}} q^{J_3} ~,
\eea
with $J_3$ a generator of $so(3)$.

\subsubsection*{SL(4)}

\begin{figure}[tb]
\centering
\includegraphics[width=\textwidth]{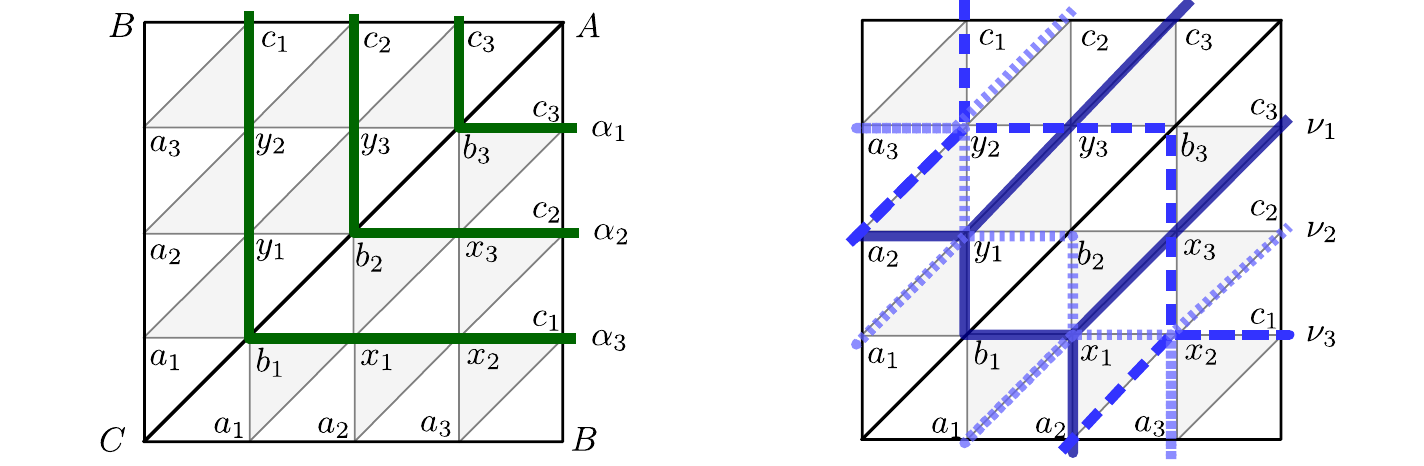}
\caption{4-triangulation of $\cC_{0,3}$. \emph{Left}: The green lines represent the eigenvalues $\alpha_i$ of the holonomy matrix $\mathbf{A}$ around the puncture $A$. \emph{Right}: Paths corresponding to $\nu_1$ (continuous), $\nu_2$ (fine dashed), and $\nu_3$ (dashed) appearing in $N_1$.}
\label{C03SL4}
\end{figure}

We find a similar structure for $SL(4)$ loop and network operators.
For example, the holonomy matrix $\mathbf{A}$ around the puncture $A$ has the following eigenvalues (see figure~\ref{C03SL4}):
\bea\label{evalCi}
\mathbf{A} &:& \qqq  \prod_i  \alpha_i^{-\kappa^{-1}_{1i}}  ( 1, \alpha_1 , \alpha_1 \alpha_2, \alpha_1 \alpha_2\alpha_3) ~,\nn
&&  \alpha_1= b_3 c_3 ~, \quad \alpha_2 = b_2 c_2x_3 y_3 ~,  \quad \alpha_3 = b_1 c_1 x_1x_2y_1y_2~,
\eea
with the $SL(4)$ Cartan matrix
\bea
\kappa = \begin{pmatrix} 2 & -1 & 0  \\ -1 &2 &-1 \\ 0&-1&2 \end{pmatrix}~, \qqq \kappa^{-1} =\frac14 \left(
\begin{array}{ccc}
 3 & 2 & 1 \\
 2 & 4 & 2 \\
 1 & 2 & 3 \\
\end{array}
\right)~.
\eea
The loop functions $A_i$ can be expressed in terms of the $\alpha_i$:
\bea
A_1 &=& \tr\bfA =  \prod_i  \alpha_i^{-\kappa^{-1}_{1i}}  ( 1+ \alpha_1+ \alpha_1 \alpha_2+ \alpha_1 \alpha_2\alpha_3)~,\nn
A_2 &=&  \frac12 \left[(\tr\bfA)^2 - \tr(\bfA^2)\right]  = \prod_i  \alpha_i^{-\kappa^{-1}_{2i}}  ( 1+ \alpha_2+ \alpha_1 \alpha_2+ \alpha_2 \alpha_3+ \alpha_1 \alpha_2\alpha_3+ \alpha_1 \alpha_2^2 \alpha_3)~,\nn
A_3 &=&  \tr\bfA^{-1} =  \prod_i  \alpha_i^{-\kappa^{-1}_{3i}}  ( 1+ \alpha_3+ \alpha_2 \alpha_3+ \alpha_1 \alpha_2\alpha_3)~.
\eea

We can construct three pants networks $N_{\bfa}$, $N_{\bfb}$, $N_{\bfc}$ (and their reverses), differing by the choice of the edge that carries the second antisymmetric representation of $SL(4)$ (recall figure~\ref{C03SL4networks}):
\bea
N_{\bfa} &=& -\frac12  \epsilon_{mnpq} \bfU_{{\bf a}\, r}^{m}\bfU_{{\bf a}\, s}^{n}\bfU_{{\bf b}\, t}^{p} \bfU_{{\bf c}\, u}^{q} \epsilon^{rstu}= \tr \bfC \bfB^{-1} - C_1B_3~, \nn
N_{\bfb} &=& -\frac12  \epsilon_{mnpq} \bfU_{{\bf a}\, r}^{m}\bfU_{{\bf b}\, s}^{n}\bfU_{{\bf b}\, t}^{p} \bfU_{{\bf c}\, u}^{q} \epsilon^{rstu}= \tr \bfA \bfC^{-1} - A_1C_3~, \nn
N_{\bfc} &=& -\frac12  \epsilon_{mnpq} \bfU_{{\bf a}\, r}^{m}\bfU_{{\bf b}\, s}^{n}\bfU_{{\bf c}\, t}^{p} \bfU_{{\bf c}\, u}^{q} \epsilon^{rstu}= \tr \bfB \bfA^{-1} - B_1A_3~.
\eea
These pants network functions contain 176 terms each, all with unit coefficient apart from 12 that have a coefficient of 2. 
Some of the terms have an interpretation as paths homotopic to the network (see figure~\ref{C03SL4}), and we can write for example
\bea\label{C03SL4N1el}
&& N_{\bfa} \ni \prod_i  \nu_i^{-\kappa^{-1}_{1i}}  ( 1+2\nu_1 +2 \nu_1 \nu_2+\nu_1 \nu_2\nu_3) ~,\\
&& \bar N_{\bfa} \ni \prod_i  \nu_i^{-\kappa^{-1}_{3i}}  ( 1+2\nu_3 + 2\nu_2 \nu_3+\nu_1 \nu_2\nu_3) ~,
\eea
with $ \nu_1= a_2b_1c_3x_1x_3y_1y_3 $, $ \nu_2 = a_1a_3b_2c_2x_1x_2y_1y_2 $, and $ \nu_3 =  a_2b_3c_1x_2x_3y_2y_3$.

Quantum relations can be obtained by using the quantum skein relations for $\cU_q(sl_4)$ (see figure~\ref{C03SL4NaNb}).
The quantization of the network operators can then be obtained from these relations.
In particular, the 12 coefficients of 2 appearing in each pants operator get quantized to $[2] = q^{1/2} + q^{-1/2}$, much in the same way as for $SL(3)$.
A more elaborate illustration is provided by the network that appears with a factor of $q^{1/2}$ in the relation for $\hat{N}_a\hat{\bar N}_b$ (see figure~\ref{C03SL4NaNb}).
This network operator contains 2344 terms with a coefficient of 2, 184 with a 3, 815 with a 4, 123 with a 5, 91 with a 6, 115 with an 8, 6 with a 9, 14 with a 10, 8 with a 12, and 4 with a 16.
Every coefficient gets quantized in a unique way. Note however that coefficients that are the same classically can quantize in different ways:
\bea
2  & \to & q^{\frac12}+ q^{-\frac12}\qquad  \text{ or } \qquad 2 ~, \nn
3  & \to & q+1+ q^{-1}  \qquad  \text{ or } \qquad 3  ~, \nn
4  & \to & q+2+ q^{-1}\qquad  \text{ or } \qquad 2q^{\frac12}+2 q^{-\frac12} ~, \nn
5  & \to & q+3+ q^{-1}   ~, \nn
6&\to& q^{\frac32}+2 q^{\frac12}+2 q^{-\frac12}+ q^{-\frac32}  \qquad  \text{ or } \qquad 3q^{\frac12}+3 q^{-\frac12}   \qquad  \text{ or } \qquad q+4+ q^{-1}  ~, \nn 
8  & \to & 2q+4 +2 q^{-1}  \qquad  \text{ or } \qquad  q^{\frac32} + 3q^{\frac12} + 3 q^{-\frac12} + q^{-\frac32}~, \nn
9  & \to & 2q+5+2 q^{-1}  ~, \nn
10 &\to& q^{\frac32}+4 q^{\frac12}+4 q^{-\frac12}+ q^{-\frac32}~, \nn 
12  & \to & q^2 +3q+4 +3 q^{-1} + q^{-2} \qquad  \text{ or } \qquad 2 q^{\frac32} + 4q^{\frac12} + 4 q^{-\frac12} + 2 q^{-\frac32}~, \nn
16  & \to & q^2 +4q+6 +4 q^{-1} + q^{-2} \qquad  \text{ or } \qquad 2 q^{\frac32} + 6q^{\frac12} + 6 q^{-\frac12} + 2 q^{-\frac32}~, 
\eea

\subsection{One-punctured torus}

\begin{figure}[tb]
\centering
\includegraphics[width=\textwidth]{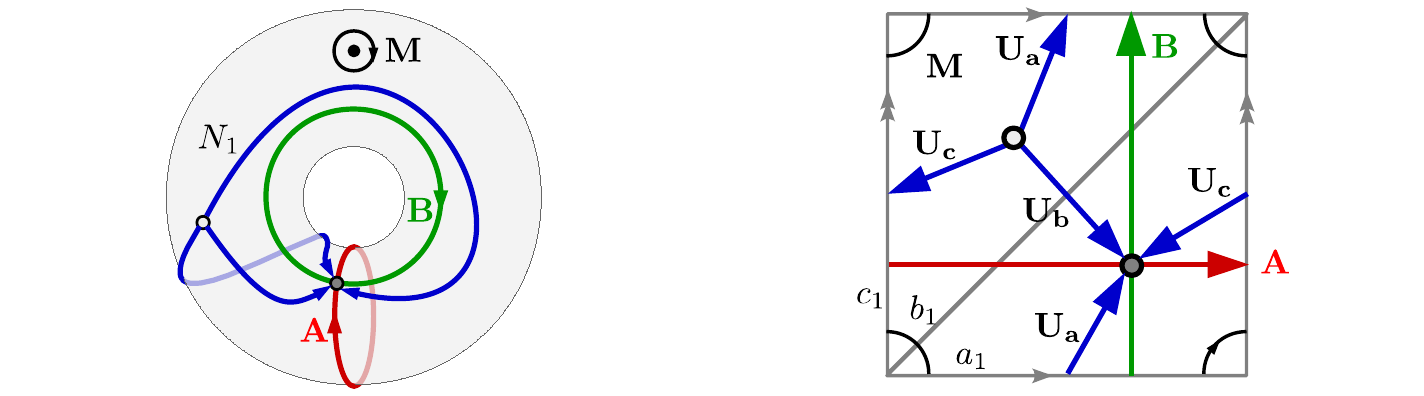}
\caption{\emph{Left}: One-punctured torus $\cC_{1,1}$. The A- and B-cycles are shown, together with a network operator $N_1$. \emph{Right}: A triangulation of $\torus$ into two triangles. The dual graph consists of the edges $\bfU_a$, $\bfU_b$, $\bfU_c$, and corresponds to the network $N_1$.}
\label{C11triang2}
\end{figure}
%

\subsubsection*{SL(2)}

We start by briefly reviewing the well-studied case of flat $SL(2,\bC)$-connections on the one-punctured torus $\cC_{1,1}$~\cite{Gaiotto:2010be} (see also~\cite{Dimofte:2011jd}).
A triangulation of $\torus$ and its dual graph are shown on the right of figure~\ref{C11triang2}.
The three edges of the dual graph can be used to express all the loop and network functions. 
After projecting the edges on the fat graph, we can express the corresponding holonomy matrices as products of  (normalized) snake matrices:
\bea
{\bf U_a} = {\bf s v}^{-1} {\bf s e}_{a }{\bf v}_y~,\qqq
{\bf U_b} = {\bf e}^{-1}_{b }~, \qqq
{\bf U_c} = {\bf v} {\bf e}_{c } {\bf s v}^{-1} {\bf s}~. 
\eea
The holonomy matrices for the A- and B-cycles, and for the clockwise closed loop around the puncture (all based at the black junction in figure~\ref{C11triang}) are expressed as
\bea  \label{STAbranches}
\bfA = {\bf U_b U}_{\bf c}^{-1}~, \qqq \bfB= {\bf U_a U}_{\bf b}^{-1}~, \qqq \bfM = {\bf U_b U}_{\bf c}^{-1} {\bf U_a U}_{\bf b}^{-1} {\bf U_c U}_{\bf a}^{-1}~.
\eea
In terms of the Fock-Goncharov coordinates $a,b,c$ on the edges of the triangulation, these loop functions are given by 
\bea\label{S1T1A1SL2C11}
A_1 &=& \sqrt{bc}+\sqrt{\frac bc}+ \frac1{\sqrt{bc}} ~,\quad 
B_1 = \sqrt{ab}+\sqrt{\frac ab}+\frac1{\sqrt{ab}} ~, \quad 
M_1 = abc+\frac1{abc}~.  
\eea
These results were already derived by Gaiotto, Moore, and Neitzke~\cite{Gaiotto:2010be}, who emphasized that they were unexpected from classical reasoning.
Indeed, the vev of a Wilson line operator in a representation $\cR$ of the gauge group would naively correspond in the IR theory (where the gauge group is broken to its Cartan subgroup) to a sum of vevs of Wilson lines labeled by the weights of $\cR$. 
In the case of $A_1$ in~\eqref{S1T1A1SL2C11}, these IR Wilson lines correspond to the terms $\sqrt{bc}$ and $1/\sqrt{bc}$.
However, extra contributions, such as $\sqrt{b/c}$, come as a surprise.
They were attributed to interesting bound states in~\cite{Gaiotto:2010be}.

The Poisson bracket of $A_1$ and $B_1$ can be obtained by direct calculation from~\eqref{PoissonFG} and agrees with~\eqref{PBS1T1sl2}:
\bea \label{PBS1T1sl22}
\{A_1, B_1 \} = \tr \bfA \bfB - \frac12 A_1 B_1 = -\tr \bfA \bfB^{-1} + \frac12 A_1 B_1 ~.
\eea
The traces $C_1 = \tr\bfA \bfB$ and $C_1'  =\tr \bfA \bfB^{-1} $ correspond to curves that go once around the A-cycle and once around the B-cycle (in different directions) and take the form
\bea \label{defC1C1p}
C_1 &=&  \sqrt{ca}+\sqrt{\frac ca}+ \frac1{\sqrt{ca}} ~, \qqq
C_1'  =\sqrt{\frac ac} \left( bc +b +2 +\frac 1b + \frac1{ab}   \right) ~.
\eea

As explained in subsection~\ref{SECquantizationFG}, in order to quantize the generators we first write them in terms of logarithmic coordinates $A,B,C$ defined via $a = \ex^A$, $b = \ex^B$, $c = \ex^C$, and then promote these coordinates to operators $\hat A,\hat B,\hat C$ satisfying the commutation relations
\bea
[\hat A, \hat B] = [\hat B, \hat C] = [\hat C, \hat A] = 2 \hbar ~.
\eea
For example, the quantized A-cycle operator is given by
\bea\label{AcycOpSL2C11}
\hat A_1 = \ex^{-\frac12(\hat B+\hat C)} + \ex^{\frac12(\hat B - \hat C)} + \ex^{\frac12(\hat B+ \hat C)}~. 
\eea
We can explicitly check that the quantum relation~\eqref{C11SL2hatP1} is satisfied by using the quantum product~\eqref{quantumProdFGmonomials}.

A key point is that the loop function $C_1'$ contains a coefficient of 2 in~\eqref{defC1C1p}, and it is not clear a priori how to quantize it.
However, the quantum relation~\eqref{hatAhatBskeinSL2} imposes that this 2 in $C_1'$ be replaced by the quantum integer $[2] = q^{1/2}+q^{-1/2}$ in $\hat C_1'$.

\subsubsection*{SL(3)}

\begin{figure}[tb]
\centering
\includegraphics[width=\textwidth]{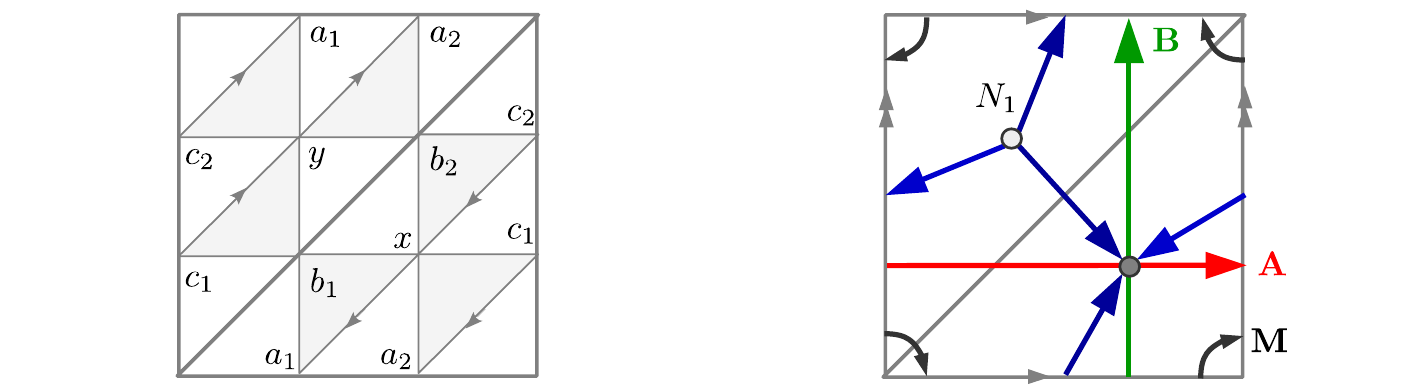}
\caption{\emph{Left}: 3-triangulation of $\mathcal{C}_{1,1}$. The edges carry two coordinates each, $a_i$, $b_i$, $c_i$, and the faces one each, $x$, $y$. 
\emph{Right}: the A-cycle (red), the B-cycle (green), the loop around the puncture (black), and the network $N_1$ corresponding to the dual graph (blue).}
\label{C11SL3triangul}
\end{figure}

A 3-triangulation of $\torus$ is shown in figure~\ref{C11SL3triangul}.
There are now two Fock-Goncharov coordinates for each edge, $a_i$, $b_i$, $c_i$, with $i=1,2$, and one for each triangle, $x$, $y$.
We express the holonomy matrices corresponding to the three branches of the dual graph as
\bea
{\bf U_a} = {\bf s v}^{-1}_x{\bf s e}_{a_2}{\bf v}_y~,\qqq
{\bf U_b} = {\bf e}^{-1}_{b_2}~, \qqq
{\bf U_c} = {\bf v}_x{\bf e}_{c_2} {\bf s v}^{-1}_y{\bf s}~. 
\eea
These three branches can be used to construct all the loop and network functions, exactly as in~\eqref{STAbranches} for the $SL(2)$ case.

\begin{figure}[tb]
\centering
\includegraphics[width=\textwidth]{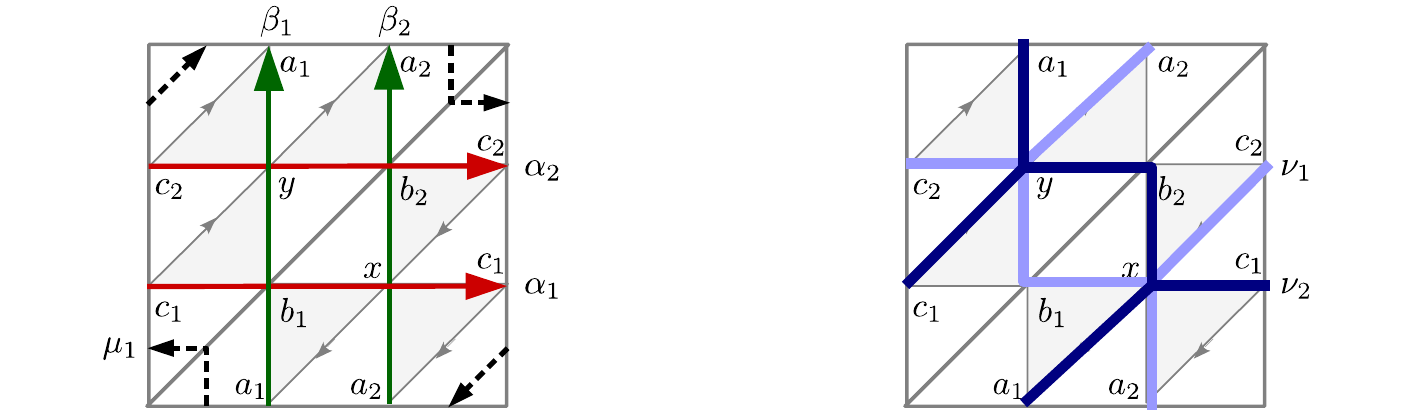}
\caption{\emph{Left}: Paths corresponding to the monomials $\alpha_1 = b_1c_1 x$, $ \alpha_2 =  b_2c_2 y$, $\beta_1 = a_1b_1 x $, $\beta_2 =  a_2b_2 y$, and $\mu_1 = a_1a_2b_1b_2c_1c_2 $. \emph{Right}: Paths corresponding to $\nu_1 =a_2 b_1 c_2 x y$ and $\nu_2 =  a_1 b_2 c_1 xy$.}
\label{C11networkselec}
\end{figure}

The loop functions $M_1=\tr \mathbf{M}$ and $M_2=\tr \mathbf{M}^{-1}$ can be expressed compactly as
\bea
M_i =  \prod_j  \mu_j^{-\kappa^{-1}_{ij}} (1 + \mu_i + \mu_1\mu_2 )~,
\eea
where $\mu_1 = a_1a_2b_1b_2c_1c_2 $ and $ \mu_2 = a_1a_2b_1b_2c_1c_2 x^3y^3 $
correspond to products of coordinates along paths surrounding the puncture.
The A-cycle functions $A_1=\tr \bfA$ and $A_2=\tr \bfA^{-1}$ similarly involve the monomials $\alpha_1 = b_1c_1 x$ and $ \alpha_2 =  b_2c_2 y$, which are products of coordinates along paths homotopic to the A-cycle (see figure~\ref{C11networkselec}):
\bea
A_1 &=& \prod_j  \alpha_j^{-\kappa^{-1}_{1j}}\left[1 +b_1 + b_1x + b_1b_2x +  \alpha_1 (1 +  b_2 + b_2y+ \alpha_2 ) \right] ~,  \nn
A_2 &=& \prod_j  \alpha_j^{-\kappa^{-1}_{2j}} \left[1 +b_2 + b_2 y + b_1b_2y +  \alpha_2 (1 +  b_1 + b_1 x+ \alpha_1 ) \right] ~.
\eea
Interestingly, the tropical $\tx$-coordinates of the terms in $A_i$ involving only the $\alpha_i$ (namely the highest, middle, and lowest terms) reproduce the weight systems of the fundamental and anti-fundamental representations.
More specifically, the tropical $\tx$-coordinates $(\tb_1,\tb_2)$ are
\be \label{C11SL3A1A2weightSys}
A_1 \ni  \prod_j  \alpha_j^{-\kappa^{-1}_{1j}}  \left\{ 
  \begin{array}{l l}
   \alpha_1 \alpha_2 & :\quad \; (0,1) \\
  \alpha_1 &  :\quad  (1,-1) \\
  1 & :\quad   (-1,0) ~,
  \end{array} \right. \qqq 
  A_2 \ni  \prod_j  \alpha_j^{-\kappa^{-1}_{2j}}  \left\{ 
  \begin{array}{l l}
   \alpha_1 \alpha_2 & :\quad \; (1,0) \\
  \alpha_2 &  :\quad  (-1,1) \\
  1 & :\quad   (0,-1) ~.
  \end{array} \right.
\ee
The B-cycle functions $B_i$ have similar expressions, involving the monomials
$\beta_1 = a_1b_1 x $ and $\beta_2 =  a_2b_2 y$.
The Poisson bracket between $\beta_i$ and $\alpha_j$ can be neatly expressed in terms of the Cartan matrix $\kappa$ (as in~\cite{Dimofte:2013iv}):
\bea\label{betaAlphaPBC11}
\{ \beta_i, \alpha_j \} = \kappa_{ij} \beta_i \alpha_j~, \qqq \{\beta_i, \beta_j\} = \{\alpha_i, \alpha_j\} = 0~.
\eea

The network function $N_1$ and its reverse $\bar N_1$ defined as
\bea
N_1 &=&  \epsilon_{mnp} \bfU_{{\bf a}\, r}^{m}\bfU_{{\bf b}\, s}^{n} \bfU_{{\bf c}\, t}^{p} \epsilon^{rst}~, \nn
\bar N_1 &=&  \epsilon_{mnp} 
(\bfU_{\bf a}^{-1})^m_r(\bfU_{\bf b}^{-1})^n_s(\bfU_{\bf c}^{-1})^p_t \epsilon^{rst}
\eea
contain 28 terms each, only one of which has a non-unit coefficient of 2.
The full expression for $N_1$ is precisely the same as the expression~\eqref{N1fullExpr} for the pants network function on $\cC_{0,3}$, apart from three extra terms:
\bea
N_1 (\cC_{1,1}) =  N_1 (\cC_{0,3})  + \prod_j  \nu_j^{-\kappa^{-1}_{1j}} \left(  a_2 b_1b_2xy+b_1c_1c_2xy  +c_2 a_1a_2xy\right)  ~.
\eea

The loop functions $C'_1 = \tr\bfA\bfB^{-1} $ and $C'_2 = \tr\bfA^{-1}\bfB$ are polynomials with 27 terms, 8 of which have a coefficient of 2.

The quantization of relations such as~\eqref{torusSkeinA2B1} can be obtained by using the quantum product:
\bea\label{qskeinC11SL3AB2}
\hat  A_2  \hat B_1 &=& q^{\frac16}  \hat  N_1+q^{-\frac13} \hat C'_2   ~, \qqq \hat   A_1 \hat  B_1= q^{-\frac16}   \hat N'_1+ q^{\frac13} \hat C_2  ~,\nn
\hat A_1 \hat  B_2 &=& q^{\frac16} \hat{\bar{N}}_1+ q^{-\frac13}\hat  C'_1 ~, \qqq  \hat A_2  \hat B_2 = q^{-\frac16} \hat{\bar{N}}'_1+q^{\frac13} \hat C_1  ~.
\eea
Here the network operator $\hat N_1'$ corresponds to the flipped dual graph (see figure~\ref{C11flippedNetwork}).
This agrees with the relations~\eqref{qskeinC11SL3AB} obtained by applying the quantum skein relation.

In order for these quantum relations to hold, the coefficients of 2 appearing in the classical functions $ N_1$, $ C'_i$, and $ N'_1$ must be replaced by the quantum integer $[2] = q^{1/2} + q^{-1/2} $ in the operators $\hat N_1$, $\hat C'_i$, and $\hat N'_1$.

We can also reproduce the quantum relations~\eqref{N1N1bqtmRelation}:
\bea 
\hat N_1  \hat{\bar{N}}_1 = q^{-\frac12} \hat W_6 + q^{\frac12} \hat{\bar{W}}_6 + \hat A_1\hat A_2+\hat B_1\hat B_2+\hat C_1\hat C_2+\hat M_1+\hat M_2+[3]~,
\eea
where $W_6$ is a network with six junctions shown in figure~\ref{C11flippedNetwork}.
The same network $W_6$ also appears in the product
\bea
\hat{\bar N}_1'\hat  C_2 = \hat W_6 + q^{\frac12} \hat A_1\hat A_2 + q^{-\frac12} \hat B_1\hat B_2~.
\eea
We can further compute the classical relation (also obtained in~\cite{Lawton:2006a} and~\cite{Xie:2013lca})
\bea\label{sexticRelC11bis}
A_1A_2 B_1 B_2 C_1 C_2  &=&  \Big[ N_1^3   + N_1^2 (A_2 B_1 + B_2C_1+C_2A_1)  \\
&& + N_1 ( A_1 A_2 B_1 C_2 + A_1^2 C_1 + A_2^2B_2-3A_1B_2 + \cyc)   \nn
 &&  - ( A_1 B_2^2C_2^2 - 2A_1^2 B_2C_2 + A_1^3 + \cyc)  + \text{reverse}\Big]  \nn
   && - N_1 \bar N_1(A_1A_2+\cyc) - (A_1A_2B_1B_2+\cyc) \nn
 && +3(A_1B_1C_1+A_2B_2C_2) + M_1M_2+6(M_1+M_2)+9 ~.  \nonumber 
\eea
Here ``$+\cyc$'' means adding the terms obtained by cyclic permutation of $A,B,C$, and ``$+\rev$'' the terms obtained by reversing the orientation, $A_1 \leftrightarrow A_2$, $N_1 \leftrightarrow \bar N_1$ and so on.
We also managed to quantize the relation~\eqref{sexticRelC11bis} but the result is not very enlightening so we omit it (in a basis where each monomial is ordered alphabetically, we need to add terms that vanish in the classical limit $q\to 1$).

\subsubsection*{SL(4)}

\begin{figure}[tb]
\centering
\includegraphics[width=\textwidth]{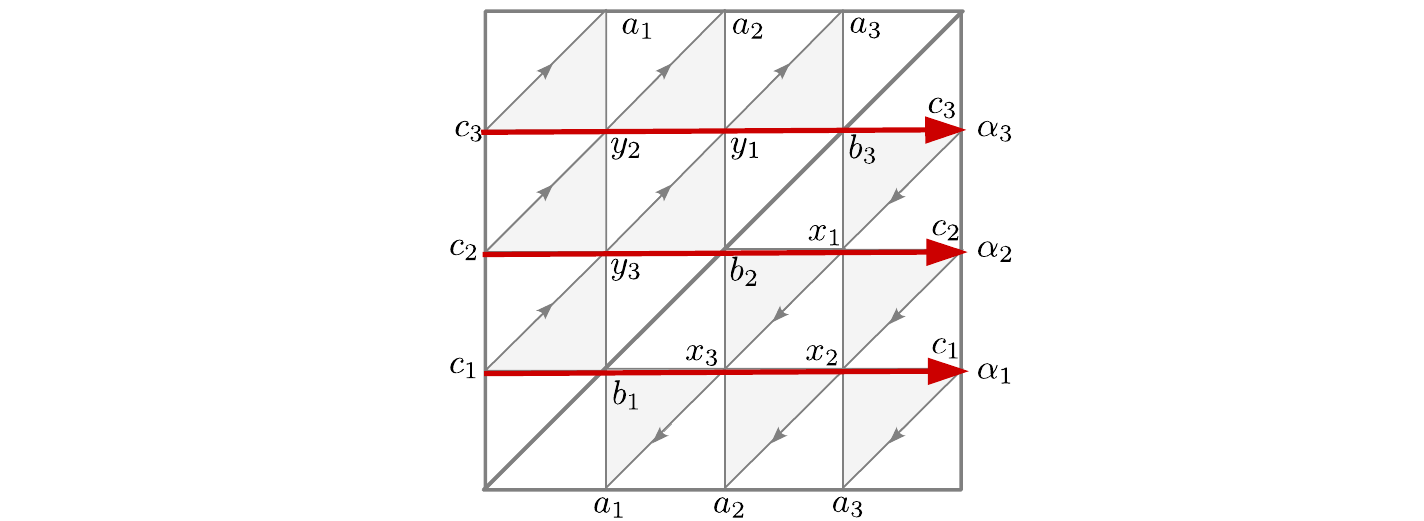}
\caption{4-triangulation of $\mathcal{C}_{1,1}$. The edges carry three coordinates each, $a_i$, $b_i$, $c_i$, and the faces three each, $x_i$ and $y_i$. The paths corresponding to the monomials $\alpha_1 = b_1 c_1 x_2x_3$, $\alpha_2 = b_2 c_2 x_1 y_3$, and $ \alpha_3 = b_3 c_3 y_1 y_2$ appearing in the loop functions $A_i$ are indicated.}
\label{SL4C11}
\end{figure}

The 4-triangulation of $\cC_{1,1}$ has 15 coordinates, 3 for each edge, $a_i,b_i,c_i$, and 3 for each face, $\{x_i,y_i\}$, with $i=1,2,3$ (see figure~\ref{SL4C11}). 
The A-cycle functions $A_1$ and $A_3$ have 21 terms each, while $A_2$ has 56 terms, some of which can be expressed as
\bea\label{AcyclesSL4}
A_1 &\ni&  \prod_i  \alpha_i^{-\kappa^{-1}_{1i}}  ( 1+ \alpha_1+ \alpha_1 \alpha_2+ \alpha_1 \alpha_2\alpha_3)~,\nn
A_2 &\ni&   \prod_i  \alpha_i^{-\kappa^{-1}_{2i}}  ( 1+ \alpha_2+ \alpha_1 \alpha_2+ \alpha_2 \alpha_3+ \alpha_1 \alpha_2\alpha_3+ \alpha_1 \alpha_2^2 \alpha_3)~,\\
A_3 &\ni&   \prod_i  \alpha_i^{-\kappa^{-1}_{3i}}  ( 1+ \alpha_3+ \alpha_2 \alpha_3+ \alpha_1 \alpha_2\alpha_3)~, \nonumber
\eea
with $\alpha_1 = b_1 c_1 x_2x_3$, $\alpha_2 = b_2 c_2 x_1 y_3$, and $ \alpha_3 = b_3 c_3 y_1 y_2$.
The tropical $\tx$-coordinates $\{ \tb_1, \tb_2, \tb_3 \}$ of these terms in $A_1$ and $A_3$ reproduce the weight systems of the fundamental and anti-fundamental representations (similarly to~\eqref{C11SL3A1A2weightSys}), while those in $A_2$ form the weight system of the second antisymmetric representation:
\be
  A_2 \ni  \prod_i  \alpha_i^{-\kappa^{-1}_{2i}}  \left\{ 
  \begin{array}{l l}
       \alpha_1\alpha_2^2 \alpha_3  & : \qqq \; (0,1,0)  \\
     \alpha_1\alpha_2 \alpha_3  & :   \qqq  (1,-1,1) \\
   \alpha_1 \alpha_2,  \alpha_2 \alpha_3  & :  \qquad  (-1,0,1) , ( 1,0,-1) \\
  \alpha_2 &  :  \qqq  (-1,1,-1)  \\
  1 & :    \qqq   (0,-1,0) ~.
  \end{array} \right.
\ee
We define three networks homotopic to the dual graph, differing by the choice of the branch that is doubled (their expansions contain 223 monomials each):
\bea
N_a &=& \frac12  \epsilon_{mnpq} \bfU_{{\bf a}\, r}^{m}\bfU_{{\bf a}\, s}^{n}\bfU_{{\bf b}\, t}^{p} \bfU_{{\bf c}\, u}^{q} \epsilon^{rstu} ~, \nn
N_b &=& \frac12  \epsilon_{mnpq} \bfU_{{\bf a}\, r}^{m}\bfU_{{\bf b}\, s}^{n}\bfU_{{\bf b}\, t}^{p} \bfU_{{\bf c}\, u}^{q} \epsilon^{rstu} ~, \nn
N_c &=& \frac12  \epsilon_{mnpq} \bfU_{{\bf a}\, r}^{m}\bfU_{{\bf b}\, s}^{n}\bfU_{{\bf c}\, t}^{p} \bfU_{{\bf c}\, u}^{q} \epsilon^{rstu} ~.
\eea

The quantum relations in~\eqref{qtRelsSL4C11} allow to uniquely determine how all the integral coefficients of 2, 4, and 8 appearing in the Fock-Goncharov expansions of the network functions get quantized:
\bea
&2\to [2]= q^{\frac12} + q^{-\frac12} ~, \qqq 4\to  [2]^2 = q+2+q^{-1} ~,  &\nn
& 8 \to   [2]^3 = q^{\frac32} + 3q^{\frac12}+3q^{-\frac12} + q^{-\frac32}~.&
\eea

\subsection{Four-punctured sphere}

Another example is a sphere $\mathcal{C}_{0,4}$ with four full punctures, $A$, $B$, $C$, $D$. 
It can be triangulated into four triangles, as shown in figure~\ref{C04triangul}.

\subsubsection*{SL(2)}

The loop around the punctures give the following trace functions:
\bea
A_1 &=& \sqrt{acf}+ \frac1{\sqrt{acf}} ~, \qqq B_1 = \sqrt{bce}+\frac1{\sqrt{bce}}  ~, \nn
C_1 &=&  \sqrt{def}+ \frac1{\sqrt{def}} ~, \qqq D_1 =\sqrt{abd}+ \frac1{\sqrt{abd}} ~.
\eea
We also consider loops surrounding pairs of punctures, $\bfS = \bfA\bfB$, $\bfT =  \bfB\bfC$,  and $\bfU =  \bfB\bfD$   (see~\cite{Gaiotto:2010be}):
\bea
S_1 &=& \frac1{\sqrt{a b e f}} ( 1 + a + e + a e + a b e + a e f + a b e f)~,\nn
T_1 &=& \frac1{ \sqrt{ b c d f}} ( 1 + b + f + b f + b c f + b d f + b c d f )  ~, \nn
U_1 &=& \frac1{\sqrt{a c d e}} ( 1 + c + d + c d + a c d + c d e + a c d e ) ~.
\eea
These polynomials do not contain any non-unit coefficients, so their quantization is straightforward.
In particular, we can reproduce the quantum relation $\hat\cP_1$~\eqref{relP1C04}.

The loop function associated with the holonomy $\bfU' =  \bfA\bfC$ does however contain coefficients of 2 and 4:
\bea
U_1' &=& \frac1{\sqrt{a} b \sqrt{c} \sqrt{d} \sqrt{e} f} (  a b^2 c d e f^2+a b^2 c e f^2+a b^2 c e f+a b^2 d e f^2+a b^2 d e f+a b^2 e f^2\nn
&&\qqq\qqq +2 a b^2 e f+a b^2 e+a b c e f^2+a b c e f+a b d e f^2+a b d e f+2 a b e f^2\nn
&&\qqq\qqq  +4 a b e f+2 a b e+a b f+a b+a e f^2\nn
&&\qqq\qqq  +2 a e f+a e+a f+a+b e f+b e+e f+e+1 ) ~.
\eea
The quantization of these non-unit coefficients can be determined by demanding that the quantum relation~\eqref{C04SL2ST} holds. We find that we must make the following replacements:
\bea
2 \to   q^{\frac12} + q^{-\frac12} ~, \qqq 4 \to   q + 2 + q^{-1}~.
\eea

\subsubsection*{SL(3)}

\begin{figure}[tb]
\centering
\includegraphics[width=\textwidth]{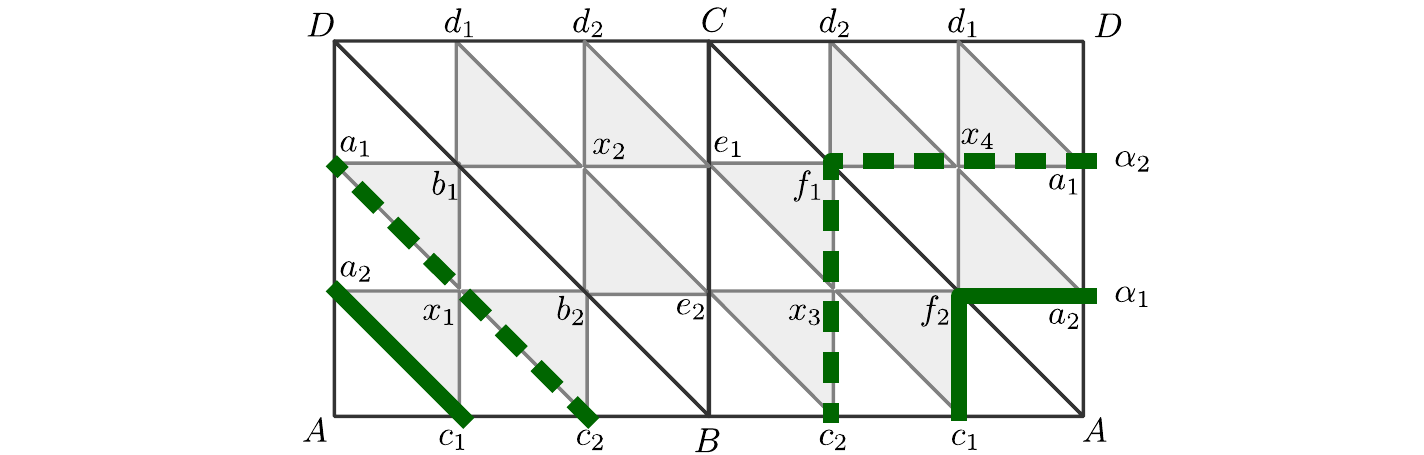}
\caption{$3$-triangulation of $\mathcal{C}_{0,4}$. The edges on the left and on the right are glued together. The paths corresponding to the monomials $\alpha_1 = a_2 c_1 f_2$ and $\alpha_2 = a_1 c_2 f_1 x_1 x_3x_4$ are indicated.}
\label{C04triangul}
\end{figure}

A 3-triangulation of $\mathcal{C}_{0,4}$ is shown in figure~\ref{C04triangul}.
Each holonomy matrix gives two loop functions, for example
\bea
A_i =  \prod_j  \alpha_j^{-\kappa^{-1}_{ij}} (1 + \alpha_i + \alpha_1\alpha_2 )~,\qqq \alpha_1 = a_2 c_1 f_2~, \quad \alpha_2 = a_1 c_2 f_1 x_1 x_3x_4~.
\eea
The monomials $\alpha_i$ are products of coordinates along paths surrounding the puncture on the 3-triangulation (see figure~\ref{C04triangul}).
The operators coming from ${\bf S,T,U }$ have 48 terms each. 
The operators $S_i$ contain the terms
\bea
S_i \ni   \prod_j  \sigma_j^{-\kappa^{-1}_{ij}} (1 + \sigma_i + \sigma_1\sigma_2 )~,\qqq \sigma_1 = a_2 b_2 e_2 f_2 x_1 x_3~, \quad \sigma_2 = a_1 b_1 e_1 f_1 x_2 x_4~.
\eea

We can reproduce the quantum relations~\eqref{S1T1C04q} and~\eqref{NABNABbqC04}, provided that we quantize the non-unit integer coefficients appearing in the network operators as follows:
\bea
&&   2 \to  q^{\frac12} + q^{-\frac12} ~, \qqq\quad\;\;  3 \to  q+1 +q^{-1}  , \qqq 4 \to  q + 2+q^{-1}~,  \nn
&& 5 \to q + 3+q^{-1}~,   \qqq 6 \to  q + 4+q^{-1}\quad \text{ or } \quad  q^{\frac32} + 2q^{\frac12}+ 2q^{-\frac12} + q^{-\frac32}~,  \\
&&  7 \to q + 5+q^{-1}~, \qquad   8 \to q^{\frac32} + 3q^{\frac12}+ 3q^{-\frac12} + q^{-\frac32}~, \qquad 10 \to  q^{\frac32} + 4q^{\frac12}+ 4q^{-\frac12} + q^{-\frac32}~.   \nonumber
\eea
Note that among the 21 coefficients of 6 appearing in $N_{6} $, 13 are quantized to $q + 4+q^{-1}$ and 8 to $q^{3/2} + 2q^{1/2}+ 2q^{-1/2} + q^{-3/2}$.

\newpage
\section{Relation to conformal field theory} \label{secCFT}

We will now explain how the quantized algebras $\cA_{g,n}^N(q)$ of functions on $\cM_{g,n}^N$ 
arise naturally in conformal field theory. 
We first review several aspects of $sl(N)$ Toda field theory.\footnote{A somewhat different approach leading to
important results is described in 
\cite{Fateev:2007ab} and references therein.}
We then define Verlinde network operators, which are natural generalizations of Verlinde loop operators.
We show that their algebra is equivalent to the quantum skein algebra described in section~\ref{SECquantization}.
Our arguments are based on the fact that the braid matrix is twist-equivalent to the R-matrix of the quantum group $\cU_q(sl_N)$ defining the skein algebra.

\subsection{Toda field theory and $W_N$-algebra}

The Lagrangian for $sl(N)$ Toda field theory has the form\bea
\cL = \frac1{8\pi} (\del_a \phi,\del^a \phi) + \mu \sum_{i=1}^{N-1} \ex^{b(e_i, \phi)}~,
\eea
where $\phi=(\phi_1,\ldots,\phi_{N-1})$ is a two-dimensional scalar field and 
$(\cdot,\cdot)$ denotes the scalar product in $\mathbb{R}^{N-1}$.
The vectors $e_i$ are the simple roots of the Lie algebra $sl(N)$. Our conventions on the Lie algebra
$sl(N)$ are summarized in Appendix \ref{Lieconv}.
The parameters are the dimensionless coupling constant $b$ and the scale parameter $\mu$. 

The example of Liouville theory corresponding to the case $N=2$ \cite{T01,T03} suggests that we
should to be able to construct the Toda field $\phi(z,\bar{z})$ using 
chiral free fields $\varphi_j(z)$, $j=1,\dots,N-1$ with
mode expansion
\bea\label{modesphi}
\varphi_j(z) = q_j -\ii  p_j \ln z + \sum_{n\neq 0} \frac\ii n a_n^{j} z^{-n}~,
\eea
where the modes satisfy 
\bea\label{ccr}
[a_n^{i}, a_m^{j}] =  n \delta_{ij} \delta_{n+m}~, \qquad a_n^{i\dagger} = a^i_{-n}~,\qquad
[p_j, q_k ] = -  \ii \delta_{jk}\,.
\eea
In order to build the non-chiral fields $\phi_i(z,\bar{z})$
out of the chiral free fields $\varphi_j(z)$ one must of course supplement them with a similar collection of 
anti-chiral fields $\bar\varphi_j(\bar z)$.

The chiral algebra of Toda conformal field theory is a $W_N$-algebra. It is an 
associative algebra generated by holomorphic currents $W^j(z)$ of spin 
$j=2,\ldots , N$ (see \cite{Bouwknegt:1992wg} for a review). These can be constructed
in the free-field representation via the Miura transformations 
\bea
\prod_{i=0}^{N-1}:\! \left(Q\del +(h_{N-i},\del\varphi)\right)\! : \ =\sum_{k=0}^N W^{N-k}(z)(Q\del)^k~,
\eea
where $: \ :$ denotes Wick ordering, $Q\equiv b+b^{-1}$, and $W^0(z)=1$, $W^1(z)=0$.  
The currents $W^k(z)$, $k=2,\dots,N$, have mode expansions 
\bea 
W^k(z)=\sum_{n=-\infty}^{\infty} W^k_n z^{-n-k}~.
\eea 
The modes $L_n\equiv W^2_n$ of the stress energy tensor 
$T(z)\equiv W^2(z)=(Q \rho,\del^2\varphi)-\frac{1}{2}(\del\varphi,\del\varphi)$ generate a Virasoro 
subalgebra 
\bea
\left[L_n,L_m\right]=(n-m)L_{n+m}+\frac{c}{12}(n^3-n)\delta_{n+m} ~.
\eea
The modes of the currents $W^k(z)$ satisfy nonlinear commutation relations which 
contain a quadratic term. The $W_N$-algebra 
is therefore not a Lie algebra.

Irreducible representations $\cV_\alpha$ of the $W_N$-algebra are labeled by an 
$(N-1)$-component vector $\alpha$ in the Cartan subalgebra of $A_{N-1}$. 
The representations $\cV_\alpha$ are generated from highest-weight vectors $v_\alpha$
which are annihilated by the
positive modes of $W^k(z)$, and eigenvectors of $W^k_0$ with eigenvalues determined by $\alpha$.
The representation space of $\cV_\alpha$ is generated by acting on $v_\alpha$ with the modes
$W_{-n}^k$ for $n>0$, as usual. 

It will be useful to distinguish three types of representations. 
A distinguished role will be played by the representations 
which have  $\alpha=-b\omega_i-b^{-1}\omega_j$, where
$\omega_i$ and $\omega_j$ are weights of finite-dimensional representations of $sl(N)$. Such representations are called 
\emph{fully degenerate} to reflect the fact that the vectors in these representations
satisfy the maximal possible number of inequivalent 
relations of the form $\mathcal{P}_l(W^k_{-n})v_\alpha=0$ for certain
polynomials $\mathcal{P}_l$. 
For generic $\alpha$ one has no such relations in the representations $\cV_{\alpha}$, which are then called \emph{fully non-degenerate}.
There are various intermediate cases, called \emph{semi-degenerate}, in which there do exist relations of the form
$\mathcal{P}_l(W^k_{-n})v_\alpha=0$, but the number of inequivalent relations
of this type is smaller than in fully degenerate representations. This happens for example if
$\alpha=\kappa\omega_1$, with $\kappa\in\mathbb{C}$.

\subsection{Conformal blocks}

Conformal blocks can be introduced elegantly as certain invariants in the tensor product of 
representations of $W_N$-algebras associated with Riemann surfaces $\cgn$. The
definition will be spelled out explicitly only for the case $g=0$, more general cases are discussed in~
\cite{FBZ:04}. We shall associate a highest-weight 
representation $\cV_r\equiv \cV_{\alpha_r}$ of the $W_N$-algebra with the $r$\ts{th} puncture 
$P_r$ of $\cgn$ for $r=1,\ldots , n$. Let $t_r$ be local coordinates around $P_r$ such that
$t_r(P_r)=0$. Conformal blocks can be defined as linear functionals 
\bea
\cF_{0,n} : \otimes_{r=1}^n \cV_{r} \rightarrow \bC 
\eea
satisfying an invariance condition of the form 
\bea
\label{defininginvarianceproperty}
\cF_{0,n}\big(W^k[\eta_k] \cdot v\big)=0\,, \qqq v\in \otimes_{r=1}^n \cV_r \, , 
\eea
where $W^k[\eta_k]$ is defined for meromorphic $(1-k)$-differentials $\eta_k(t)$ on $\cC_{0,n}$ by
expanding 
\begin{equation} 
\eta_k(t)=\sum_{n\in\mathbb{Z}}\eta_{k,n}^{(r)}\,t_r^{n-(1-k)}\,,
\end{equation}
and setting 
\begin{equation}
W^k[\eta_k]=\sum_{r=1}^n\sum_{n\in\mathbb{Z}}\eta_{k,n}^{(r)}\,
\Big[\,{\rm id}\otimes\dots\otimes
\underset{{\text{$r$\ts{th} }}}{W_n^k}\otimes\dots\otimes{\rm id}\,\Big]\,.
\end{equation}
The conditions (\ref{defininginvarianceproperty}) generalize
the conformal Ward identities \cite{BPZ}. The value of a 
conformal block on vectors $\otimes_{r=1}^n w_r$ can be interpreted as a generalized 
correlation function of $n$ chiral vertex operators $V_{\alpha_r}(w_r|P_r)$. 
If $w_r=v_{\alpha_r}$ is the highest-weight state in the representation $\cV_{\alpha_r}$, one calls
$V_{\alpha_r}(P_r)\equiv V_{\alpha_r}(v_{\alpha_r}|P_r)$ a primary field, otherwise a descendant.

\paragraph{Space of conformal blocks:} 

The invariance conditions \eqref{defininginvarianceproperty} represent an infinite system of linear equations
defining a subspace of the dual vector space to $\otimes_{i=1}^n \cV_i$.
The vector space defined in this way is called the space of conformal blocks 
${\rm CB}(\cV_{[n]},\cC_{0,n})$ associated with the Riemann surface $\cC_{0,n}$ with 
representations $\cV_i$ at the punctures. This space is infinite-dimensional,  in general. 
We now want to get a first idea about the ``size'' of this space.

In the case $N=2$, the conformal block for $\cC_{0,3}$ is known to be defined uniquely up to 
normalization by the invariance property \eqref{defininginvarianceproperty}. Using 
this same equation one may express all values 
$\cF_{0,4}(w_4\otimes w_3\otimes w_2\otimes w_1)$
associated with $\cC_{0,4}$ in terms of  
\bea \label{CB4ptRK1}
\cF_{0,4}(v_{\alpha_4}\otimes v_{\alpha_3}\otimes L_{-1}^k v_{\alpha_2}\otimes v_{\alpha_1})\equiv F_k~, 
\qquad F_k\in\bC~, \qquad k\in\bZ_{>0}~.
\eea
One therefore finds that the space of conformal blocks associated with $\cC_{0,4}$ is 
infinite-dimensional and isomorphic as a vector space to the space of formal 
power series in one variable. This space is far too big to be interesting for physical applications,
as stressed in  \cite{TV13,T14a}. Only if the growth of the complex numbers $F_k$
ensures convergence of series like $F(z)=\sum_k\frac{1}{k!}F_kz^k$, can one integrate the canonical connection
on spaces of conformal blocks defined by the energy-momentum tensor at least locally.
Further conditions like existence of an analytic continuation of $F(z)$ to the Teichm\"uller space $\mathcal{T}_{0,4}$,
and reasonable growth at the boundaries of $\mathcal{T}_{0,4}$ characterize the subspaces of 
${\rm CB}(\cV_{[4]},\cC_{0,4})$ of potential physical interest.

In the case $N=3$, one generically finds an infinite-dimensional space of conformal blocks 
already for the three-punctured sphere. 
Using the invariance property \eqref{defininginvarianceproperty}, it is possible to 
express the values of the conformal blocks associated with $\cC_{0,3}$ in terms of \cite{Bowcock:1993wq}
\bea
\label{3ptCBW3}
\cF_{0,3}(v_{\alpha_3}\otimes (W_{-1}^3)^{k} v_{\alpha_2}\otimes v_{\alpha_1})\in\bC ~.
\eea 
Therefore, similarly to \eqref{CB4ptRK1}, the space of conformal blocks associated with 
$\cC_{0,3}$ is infinite-dimensional. It can also be identified with the space of formal Taylor 
series in one variable. In analogy to the case $N=2$, $n=4$ mentioned above
we may expect that the physically relevant
subspaces of ${\rm CB}(\cV_{[3]},\cC_{0,3})$ may have a representation as spaces of 
analytic functions in one variable.

At general rank $(N-1)$, the number of extra variables required to get similar representations
for the spaces of conformal blocks associated with $\cC_{0,3}$ is equal to half of the 
dimension\footnote{This number is equal to the dimension of the Coulomb branch of 
the $T_N$ gauge theory. Together with the $3(N-1)$ parameters of the 
momenta $\alpha$, it gives the $(N+4)(N-1)/2$ parameters of the $T_N$ theory~
\cite{Kozcaz:2010af}.} of the moduli space of flat connections $(N-1)(N-2)/2$. One 
way to understand this is by considering, as in \cite{Kozcaz:2010af}
\cite{Bao:2013pwa}, the difference between the number of basic 3-point functions 
and the number of constraints from the Ward identities 
\eqref{defininginvarianceproperty} and corresponding to generators 
of the $W_N$-algebra. There are $N(N-1)/2$ descendant operators constructed by 
acting with the modes $(W_{-l}^{k})^n$ on a highest-weight state, for 
$l=1,2,\ldots,k-1$ and $k=2,3,\ldots,N$, which lead to $3N(N-1)/2$ basic 3-point 
functions after taking into account the three primaries. Subtracting $N^2-1$ 
constraints corresponding to generators $W_l^{k}$, for $l=-k+1,-k+2,\ldots,k-1$ 
and $k=2,\ldots,N,$ gives precisely the number $(N-1)(N-2)/2$ of unconstrained 
positive integers parameterizing the conformal block associated with $\cC_{0,3}$.

\paragraph{Gluing construction:}  Given a possibly disconnected Riemann
surface with two marked points $P_0^i$, $i=1,2$, surrounded by parameterized discs 
one can construct a new Riemann surface by
pairwise identifying the points in suitable annular regions around the two marked points, respectively. 
Having conformal blocks
associated with two surfaces $\cC_i$ with $n_i+1$ punctures $P_0^i,P_1^i,\dots,P_{n_i}^i$ one may
construct a conformal block associated with the surface $\cC_{12}$ obtained by gluing annular 
neighborhoods of $P_0^i$, $i=1,2$ as follows
\begin{equation}\label{gluing}\begin{aligned}
\cF_{\cC_{12}}(v_{1}& \otimes\dots\otimes  v_{n_1}\otimes w_1\otimes\dots\otimes w_{n_2})
=\\
=&\sum_{\nu\in\mathcal{I}_\beta}\cF_{\cC_1}(v_{1}\otimes\dots\otimes v_{n_1}\otimes v_\nu)
\,\cF_{\cC_2}(\ex^{2\pi i\tau L_0}v_\nu^{\vee}\otimes w_{1}\otimes\dots\otimes w_{n_2})\,.
\end{aligned}\end{equation}
The vectors $v_\nu$ and $v_\nu^{\vee}$ are elements of bases $\{v_\nu;\nu\in\mathcal{I}_\beta\}$
and $\{v_\nu^\vee;\nu\in\mathcal{I}_\beta\}$
for the representation $\cV_\beta$
which are dual with respect to the invariant bilinear form $(.,.)_\beta$ on $\cV_{\beta}$. The parameter $\tau$ in (\ref{gluing})
is the modulus of the annular region used in the gluing construction of $\cC_{12}$.
The rest of the notations in (\ref{gluing}) are hopefully
self-explanatory. 
The case where $P_0^i$, $i=1,2$, are on a connected surface can be treated in a similar 
way. 

A general Riemann surface $\cgn$ can be obtained by 
gluing $2g-2+n$ pairs of pants $\cC_{0,3}^v$, $v=1,\dots,2g-2+n$.  It is possible to 
construct conformal blocks for the resulting Riemann surface from the conformal blocks 
associated with the pairs of pants $\cC_{0,3}^v$ by recursive use of the gluing construction outlined
above. This yields families of conformal blocks parameterized by (i) the choices of representations
$\cV_{\beta}$ used in the gluing construction, and (ii) the choices of elements of the spaces
${\rm CB}(\cC_{0,3}^v)$, $v=1,\dots,2g-2+n$.

\begin{figure}[t]
\centering
\includegraphics[width= \textwidth]{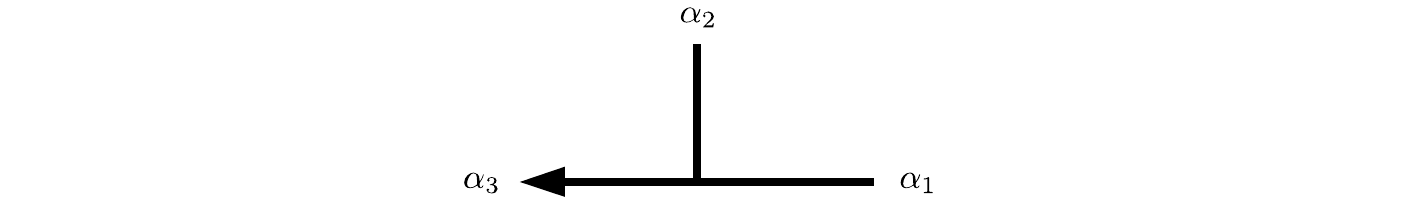}
\caption{Vertex operator as an intertwiner of highest-weight representations $\cV_{\alpha_i}$.}
\label{VertexAlpha123}
\end{figure}

Closely related to the gluing construction of conformal blocks are constructions 
of  conformal blocks using chiral vertex operators. 
To any 3-point conformal block $\cF\in{\rm CB}(\cC_{0,3})$ we may associate a chiral vertex operator 
$V_{\cF}^{}\big({}^{\;\alpha_2}_{\alpha_3\alpha_1}\big)(v_2 | z):\cV_{\alpha_1}\rightarrow\cV_{\alpha_3}$ such that
\bea\label{CVOvsblocks}
\big(\, v_3\,,\, V_{\cF}^{}\big({}^{\;\alpha_2}_{\alpha_3\alpha_1}\big) (v_2 | 1) \,v_1 \,\big)_{\alpha_3}
= \cF(v_3\otimes v_2\otimes v_1)
\eea
holds for all $v_i\in\cV_{\alpha_i}$, $i=1,2,3$. The field 
$V_{\cF}^{}\big({}^{\;\alpha_2}_{\alpha_3\alpha_1}\big)(v_{\alpha_2} | z)$ associated with the highest-weight 
vector $v_{\alpha_2}\in\cV_{\alpha_2}$ is called a primary field, and all other fields are called descendants.
A graphical
representation is given in figure~\ref{VertexAlpha123}. Conformal blocks associated with 
$n$-punctured spheres $\cC_{0,n}$ can be constructed using compositions of 
chiral vertex operators in the form
\begin{equation}
\big( v_{n}\;,\;
V_{\cF_{n-1}}^{}\big({}^{\;\;\alpha_{n-1}}_{\alpha_n,\beta_{n-3}}\big)(v_{n-1}|z_{n-1})\cdots\, 
V_{\cF_2}^{}\big({}^{\;\alpha_{2}}_{\beta_1,\alpha_{1}}\big)(v_2|z_{2})\,v_{1}
\big)_{\alpha_n}\,,
\end{equation}
assuming that $z_n=\infty$ and $z_1=0$.

\subsection{Free-field construction of chiral vertex operators}

There is a useful way to construct chiral vertex operators in terms of the free fields $\varphi_j(z)$. 
Basic building blocks are the normal-ordered exponential fields 
\bea \label{noe}
V_\alpha (\sigma) = \ex^{(\alpha, \varphi(\sigma))}~.
\eea
The normal-ordered exponential represents densely defined unbounded operators 
on $\cM\equiv\cF\otimes L^2(\mathbb{R}^{N-1})$ for ${\rm Im}(\sigma)>0$, $\cF$ being the 
Fock space generated by the oscillators $a_n^i$.
Representing the momenta $p_j$ as multiplication 
operators on $L^2(\mathbb{R}^{N-1})$ leads to a representation of the operators
$\ex^{(\alpha,q)}$ appearing in (\ref{noe}) as finite difference operators on functions $\psi(p)$.
For our goals it will often suffice to adopt the closely related definition of
the operators $\ex^{(\alpha,q)}$ as formal shift
operators mapping the highest-weight vector $v_{\beta}\in\cV_{\beta}$ to 
$v_{\beta+\alpha}\in\cV_{\alpha+\beta}$. 
The fields $V_\alpha(\sigma)$ may then be identified with the chiral
primary fields $V\big({}^{\;\;\;\,\alpha}_{\beta+\alpha,\beta}\big)(z)$ for the algebra $W_N$ via
\begin{equation}
V_\alpha(\sigma)v_{\beta}=z^{\Delta_{\alpha}}V\big({}^{\;\;\;\,\alpha}_{\beta+\alpha,\beta}\big)(z)v_{\beta}~, \qqq  v_{\beta}\in\cV_{\beta}\,,
\end{equation}
assuming that
$z=\ex^{i\sigma}$.
 For later use, let us note that the normal-ordered exponentials 
satisfy the exchange relations 
\bea\label{normalOrderFormula}
V_{\beta} (\sigma') V_{\alpha} (\sigma) = \ex^{-\pi \ii (\alpha ,\beta) 
\sgn(\sigma'-\sigma)} V_{\alpha}(\sigma) V_{\beta}(\sigma')~.
\eea
An important role is furthermore played by the 
screening charges which are defined  as  
\bea\label{scr-def}
Q_i (\sigma)
= \int_\sigma^{\sigma+2\pi} \dd \sigma' \;V_{b e_i}(\sigma') ~.
\eea
Powers of the screening charges like $Q_i^n(\sigma)\equiv (Q_i(\sigma))^n$ make sense as unbounded 
self-adjoint operators on $\cM$ if the power $n$ is sufficiently small compared to $b^{-2}$. Higher powers can be defined
by analytic continuation in $b^2$, and the result
can be represented explicitly using suitable modifications of the contours of integration in (\ref{scr-def}).
The key property of the 
screening charges $Q_i(\sigma)$ follows from the fact that the 
commutator of the fields $V_{b e_i}(\sigma)$ with the generators of the $W_N$-algebra
can be represented as a total derivative.  This implies 
that the screening charges commute with the generators $W_n^k$.

More general primary fields can therefore be constructed by multiplying 
normal-ordered exponentials with monomials formed out of the screening charges.
In the case $N=2$ one may consider composite fields of the form
$V_{\alpha}^s(\sigma)=V_{\alpha}(\sigma)Q^s(\sigma)$, $s\in\mathbb{Z}_{\geq 0}$.
For $N=3$ one could consider, more generally, fields of the form
$V_{\alpha}Q_1^{s_1}Q_2^{s_2}Q_1^{s_3}Q_2^{s_4}\cdots$.  One should note, however,
that there exist many linear relations among these fields, as follows from the 
fact that the screening charges $Q_i$ satisfy the Serre relations of the quantum 
group $\cU_q(sl_N)$ \cite{BMP},
\bea\label{Serre}
Q_i^2Q_j = (q+q^{-1})Q_iQ_jQ_i - Q_jQ_i^2~, \qqq \text{with} \quad |i-j|=1~.
\eea
Using these relations it is elementary to show  that in the case $N=3$
arbitrary screened vertex operators can be expressed as
linear combinations of  
\bea 
\label{N3screenedvertexoperator}
V_{\alpha}^\bfs(\sigma)=V_{\alpha}(\sigma)
Q_1^{s_1-s}(\sigma)Q_2^{s_2-s}(\sigma)Q_{13}^{s}(\sigma)~,
\eea 
where $\bfs=(s_1,s_2,s)$, and 
\bea \label{Q12definition}
Q_{13}(\sigma)  = Q_1 Q_2 -q Q_2 Q_1 =(1-q^2) \int_\sigma^{\sigma+2\pi} \dd 
\sigma' \int_\sigma^{\sigma'}\dd \sigma''  V_{be_1}(\sigma') V_{be_2}(\sigma'')~.
\eea
Indeed, 
the relations (\ref{Serre}) can be rewritten as
\bea \label{SerreQ12}
Q_1 Q_{13} &=& q^{-1} Q_{13} Q_1 ~, \qqq Q_2 Q_{13} = q  Q_{13} Q_2~.
\eea 
By making repeated use of these relations together with\footnote{Here we are using 
$[n]_q \equiv \frac{q^n - q^{-n}}{q- q^{-1}} = q^{n-1} + q^{n-3} + \cdots + q^{-n+1}~.$}  
\bea
Q_1 Q_2^n = q^n Q_2^n Q_1 + [n]_q Q_{2}^{n-1} Q_{13}~,
\eea
we can then express an arbitrary product of charges
$Q_1^{s_1}Q_2^{s_2}Q_1^{s_3}Q_2^{s_4}\cdots$ as linear combination of monomials of  the form 
$Q_1^{s_1-s}Q_2^{s_2-s}Q_{13}^{s}$, reducing any screened vertex operator to  
\eqref{N3screenedvertexoperator}.

By using the screened vertex operators $V_{\alpha}^\bfs$ one may represent
more general chiral vertex operators via 
\begin{equation}\label{SVO-CVOa}
V_\alpha^\bfs(\sigma)v_{\beta}=z^{\Delta_{\alpha}}V_s\big({}^{\;\,\alpha}_{\beta'\beta}\big)(z)v_{\beta}~, \qqq \text{with} \quad \beta'=\beta+\alpha+b(s_1e_1+s_2e_2)\,.
\end{equation}
It is clear that we cannot represent the most general 
chiral vertex operators in terms of screened vertex operators 
using only integer values of $s_{1},s_{2}$. It is therefore useful to observe that
$Q_{1}(\sigma)$, $Q_{2}(\sigma)$ and $Q_{13}(\sigma)$ 
are proportional to positive self-adjoint operators, allowing us to consider
also non-integer values for the powers $s_{1}$, $s_2$, $s$ in (\ref{N3screenedvertexoperator}).  
The arguments used in \cite{T01,T03} for the case $N=2$ can easily be adapted to 
establish that $Q_{1}(\sigma)$ and $Q_{2}(\sigma)$ are positive self-adjoint.
The operator 
$Q_{13}(\sigma)$ defined 
in \eqref{Q12definition} can be shown to be proportional to a 
positive operator by  expressing it as an ordered double integral and normal-ordering the
product $V_{be_1}(\sigma') V_{be_2}(\sigma'')$. 
Considering non-integer $s_{1}$, $s_2$  allows us to get arbitrary values for the 
weights $(\beta',\alpha,\beta)$ appearing in (\ref{SVO-CVOa}). 

It is furthermore important to note that the difference between $\beta'$ and $\beta$ is independent of
the parameter 
$s$ in \eqref{N3screenedvertexoperator}. 
By varying $s$ one may therefore define via \eqref{SVO-CVOa} an infinite family of 
chiral  vertex operators
$V_s\big({}^{\;\,\alpha}_{\beta'\beta}\big)(z)$ 
intertwining between the same two representations $\beta'$ and $\beta$. 
The correspondence 
\eqref{CVOvsblocks} between
chiral vertex operators and conformal blocks on the three-punctured sphere associates with each  chiral  vertex operator
$V_s\big({}^{\;\,\alpha}_{\beta'\beta}\big)(z)$ a conformal block 
$\cF_{s}\in{\rm CB}(C_{0,3})$. It seems plausible that the conformal blocks
$\cF_s$ generate a basis for  the physically relevant
subspace of conformal blocks on $\cC_{0,3}$.

To generalize these observations to $N>3$ let us note that it directly follows from (\ref{Serre})
that the space of screened vertex operators forms a module for the nilpotent sub-algebra
of $\cU_q(sl_N)$, with action of the generators $E_i$ represented by right multiplication with $Q_i$.
One may define screening charges $Q_{ij}(\sigma)$ associated with the 
generators $E_{ij}$, $j>i$ of $\cU_q(sl_N)$ recursively via $Q_{i,i+1}(\sigma)\equiv Q_i(\sigma)$  
together with 
\begin{equation}
Q_{ij}(\sigma)\,=\,Q_{ik}(\sigma)Q_{kj}(\sigma)-qQ_{kj}(\sigma)Q_{ik}(\sigma)\,,\quad
1\leq i<k<j\leq N\,.
\end{equation}
More general screened vertex operators can be 
constructed as
\begin{equation}\label{SVO-def}
\begin{aligned}
V_{\alpha}^\bfs(\sigma)=
V_{\alpha}(\sigma) \big[Q_{12}^{s_{12}}(\sigma)&Q_{13}^{s_{13}}(\sigma)\cdots 
Q_{1N}^{s_{1N}}(\sigma)\big]\times\cdots \\
\times\big[&Q_{23}^{s_{23}}(\sigma)\cdots Q_{2N}^{s_{2N}}(\sigma)\big]\times
\cdots \times [Q_{N-1,N}^{s_{N-1,N}}(\sigma)]\,,
\end{aligned}\end{equation}
where $\bfs=\{s_{ij};i<j\}$.
This may be extended to a module for the Borel-subalgebra of $\cU_q(sl_N)$ with generators 
$H_i$ and $E_i$ by identifying $H_i$ with the adjoint action of $-\ii p^i/b$.
For general $\alpha$ one may identify the resulting module with the Verma module 
of $\cU_q(sl_N)$
with weight $\omega_\alpha=- \alpha/b$.
The observations above imply a relation between
screened vertex operators and representations of $\cU_q(sl_N)$ on the level of vector spaces. 
This relation will be strengthened considerably below. 

As before one may use the screened vertex operators $V_{\alpha}^\bfs(\sigma)$ to define
chiral vertex operators via the obvious generalization of  \eqref{SVO-CVOa} to $N>3$. 
Fixing the difference $\beta'-\beta$ defines a subspace in the 
space of parameters $\bfs$ of dimension $\frac{1}{2}(N-1)(N-2)$. This dimension coincides with
half of the dimension of the moduli space of flat connections on $\cC_{0,3}$, which was 
previously found to be equal to the dimension of the space of parameters labeling
inequivalent elements in ${\rm CB}(\cC_{0,3})$. This observation raises our hopes
that the screened
vertex operators $V_{\alpha}^\bfs(\sigma)$ can indeed be used to construct bases for 
${\rm CB}(\cC_{0,3})$.

\paragraph{Degenerate fields:} For special values of $\alpha$ and $\bfs$ one may observe that
the screened vertex operators satisfy certain differential equations relating derivatives
of $V_{\alpha}^\bfs(\sigma)$ to the $W_N$-currents. These fields are called degenerate fields.
The basic example for this phenomenon 
occurs 
in the case $\alpha=-b\omega_1$ when $[p^i,V_{\alpha}^\bfs(\sigma)]=\ii bh_iV_{\alpha}^\bfs(\sigma)$,
where $h_i$, $i=1,\dots,N$ are the weights of the fundamental representation.
We will use the simplified notation $D^i(\sigma)$ for the screened vertex operators
$V_{\alpha}^\bfs(\sigma)$ satisfying these conditions.
They satisfy an $N$\ts{th}-order operator differential equation \cite{Bilal:1988jg}, which is essentially equivalent to the equations expressing the decoupling of null-vectors
in the Verma module $\cV_{-b\omega_1}$ within the framework of \cite{BPZ}, 
see e.g.~\cite{Bowcock:1993wq}\cite{Fateev:2007ab}. State-operator correspondence 
in CFT relates the allowed values of $\beta'$ to the so-called fusion rules, the rules determining the
set of labels of the primary fields that can appear in the
operator product expansion of fields $V_{-b\omega_1}(z_1)V_{\beta}(z_2)$. 

The operator product expansion of degenerate fields generates further degenerate fields.
It follows from the fusion rules that the screened vertex operators $V_{-b\lambda}^\bfs(\sigma)$ 
that can be generated by 
recursively by performing 
operator product expansions of the operators $D^i(\sigma)$  are labeled
by the weights $\lambda$ of finite-dimensional representations\footnote{We identify $M_{i}\equiv M_{\omega_i}$ in our notations, with $\omega_i$ being the
weight of the representation $\wedge^i\square$.} $M_{\lambda}$ of $\cU_q(sl_N)$.
The allowed powers of screening charges 
collected in $\bfs=\{s_{ij},i<j\}$ are constrained by the fusion rules determining  the fields $V_{\beta'}(z_2)$
appearing in the operator  product expansion $V_{-b\lambda}(z_1)V_{\beta}(z_2)$. The $s_{ij}$ 
have to be integers constrained by the condition that
$\frac{1}{b}(\beta-\beta')$ coincides with one of the weights of the vectors in $\cR_{\lambda}$.

\subsection{Braiding and fusion of degenerate fields}
\newcommand{\fv}{\mathfrak{v}}
\newcommand{\fw}{\mathfrak{w}}

It was shown in \cite{Bilal:1988jg} that
the degenerate fields $D^i(\sigma)$ 
satisfy exchange relations of the form 
\begin{equation}\label{braidrel}
D^i(\sigma_2)  D^j(\sigma_1)  = 
\sum_{k,l=1}^N
B^{\rm D}(p)^{ij}_{kl}\,
D^k(\sigma_1)D^l(\sigma_2)~.
\end{equation}
Note that the matrix $B^{\rm D}(p)$ appearing in 
(\ref{braidrel}) is operator-valued in general, being dependent on the zero-mode operator $p=(p^1,\dots,p^{N-1})$.
The matrix $B^{\rm D}(p)$ satisfies a
modified form of the Yang-Baxter equation called dynamical Yang-Baxter equation, and 
$B^{\rm D}(p)$ therefore represents an example of what is called a dynamical R-matrix.
It was furthermore shown in~\cite{Cremmer:1989dj} that there exist linear combinations
$\tilde{D}^i(\sigma)=\sum_{j=1}^N c(p)^i_j\,D^j(\sigma)$ satisfying exchange relations of the form 
\eqref{braidrel} with a matrix $\tilde{B}^{\rm D}(p)$ 
that is $p$-independent, and will therefore be denoted by $\tilde{B}$.
The matrix $\tilde{B}$ is different from the
standard braid-matrix $R$ representing the braiding of two fundamental representations
of the quantum group $\cU\equiv\cU_q(sl_N)$ (see appendix~\ref{BasicQG}
for the relevant background on $\cU_q(sl_N)$). $R$ can be obtained from the universal 
R-matrix $\mathcal{R}$ of $\cU_q(sl_N)$ as
\begin{equation}\label{BfromR}
R=P(\pi_\square^{}\otimes\pi_\square^{})(\mathcal{R})\,,
\end{equation}
where $P$ is the operator permuting the two tensor factors of $\mathbb{C}^N\otimes\mathbb{C}^N$.
The relation between $\tilde{B}$ and $R$ was subsequently clarified in 
\cite{Buffenoir:06}, where is was shown that $\tilde{B}$ can be obtained in a similar way
as \eqref{BfromR} from a universal R-matrix $\tilde{\cR}$ that is related to $\cR$ by a
Drinfeld twist $\mathcal{J}\in\cU\otimes\cU$ such that $\tilde{\cR}={\mathcal{J}_{21}}^{-1}\cR\,\mathcal{J}$.
The twist $\mathcal{J}$ that relates $\cR$ and $\tilde{\cR}$ satisfies the cocycle condition
\begin{equation}\label{cocycle}
(\Delta\otimes {\rm id})(\mathcal{J})\cdot\mathcal{J}_{12}\,=\,({\rm id}\otimes \Delta)(\mathcal{J})\cdot \mathcal{J}_{23}\,.
\end{equation}
As explained in appendix \ref{twistedclebschcompositionS}, one may extend
the action of $\mathcal{J}$ to $m$-fold tensor products, allowing us to define operators
\begin{equation}
J_{f}^{(m)}:=(\pi_{f}\otimes\dots\otimes\pi_{f})(\mathcal{J}^{(m)})~.
\end{equation}
The cocycle condition~\eqref{cocycle} implies  that $J_{f}^{(m)}$ is
independent of the order in which tensor products are taken.

Let us consider the space ${\rm VO}_m$ spanned by the compositions of vertex operators 
$D^{i_m\dots i_1}(\sigma_m,\dots,\sigma_1):=
\tilde{D}^{i_m}(\sigma_m)\tilde{D}^{i_{m-1}}(\sigma_{m-1})\cdots \tilde{D}^{i_1}(\sigma_1)$ with
$i_k\in\{1,\dots,N\}$ for $k=1,\dots,m$. The space ${\rm VO}_m$ carries a representation of the braid group
$\cB_m$ with $m$ strands represented in terms of the braid matrices $\tilde{B}$. It follows from 
the fact that $\cR$ and $\tilde{\cR}$ are related by the Drinfeld twist $\mathcal{J}$ that the linear operator
$J_{f}^{(m)}$  maps the braid group representation on ${\rm VO}_m$ to the standard 
braid group representation on the $m$-fold tensor product of fundamental representations of $\cU_q(sl_N)$.

By repeated use of 
the operator product expansion one may construct other degenerate vertex operators 
$D^\bfs_{\lambda}(\sigma)$ starting from the products $D^{i_m\dots i_1}(\sigma_m,\dots,\sigma_1)$.
The vertex operators $D^\bfs_{\lambda}(\sigma)$ that can be obtained in this way
have weights $\lambda$ associated with the finite-dimensional irreducible representations of 
$\cU_q(sl_N)$.
It easily follows from the results described above that the braid group representation generated 
by products of the vertex operators $D^\bfs_{\lambda}(\sigma)$ is isomorphic to the braid group 
representation on the tensor product of the corresponding representations of $\cU_q(sl_N)$.
This implies, in particular, that the vector space spanned by the  vertex operators $D^\bfs_{\lambda}(\sigma)$
with fixed $\lambda$ is isomorphic to the space on which the finite-dimensional
irreducible representation $M_\lambda$ with highest weight $\lambda$ is realized.


\subsection{Conformal blocks with degenerate fields}

\begin{figure}[t]
\centering
\includegraphics[width= \textwidth]{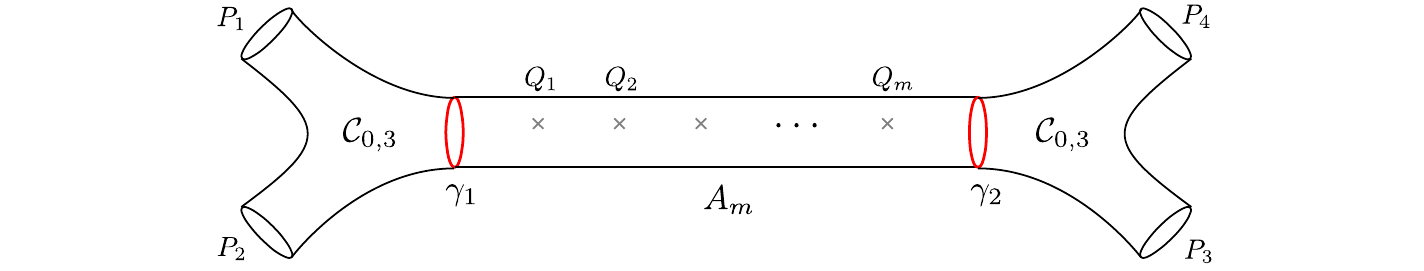}
\caption{Part of a decomposition of $\hat \cC_{g,n+d}$ with $n$ punctures $P_r$ and $d$ marked points $Q_k$ into three-holed spheres $\pants$ and annuli $A_m$ with $m$ marked points.}
\label{annulus}
\end{figure}

Let us
consider conformal blocks associated with
a Riemann surface $\hat{\cC}_{g,n+d}$ with $n+d$ punctures (see figure~\ref{annulus}). 
We assume that fully degenerate representations $\cD_k$ are
associated with the punctures $Q_k$, $k=1,\dots,d$. The remaining $n$ punctures $P_r$, $r=1,\dots, n$
are assumed not to be fully degenerate.  We may alternatively consider $Q_k$,  $k=1,\dots,d$, as a 
collection
of distinguished points on the Riemann surface $\cC_{g,n}$ which has punctures only at $P_r$, $r=1,\dots, n$.
We may start from a  pants decomposition $\sigma=(\gamma,\Gamma)$ of the surface $\cC_{g,n}$. Cutting 
$\cC_{g,n}$ along the simple closed curves contained in $\gamma=\{\gamma_1,\dots,\gamma_{3g-3+n}\}$ 
decomposes the surface into spheres $\cC_{0,3}^v$ 
with three boundary components which can be holes or punctures. The trivalent graph $\Gamma$ on
$\cC_{g,n}$ has 
exactly one vertex within each $\cC_{0,3}^v$; it allows us to distinguish pants decompositions related
by Dehn twists. The pants decomposition specified by $\sigma=(\gamma,\Gamma)$
can always be refined to 
a decomposition into a collection of annuli  $A^e$ and three-holed spheres $T^v$ such that
each of the fully degenerate punctures is contained in one of the annuli $A^e$. This can be done by 
cutting along additional simple closed curves on $\cC_{g,n}$
which do not intersect any of the curves in $\gamma$, do not mutually intersect, and
do not contain any $Q_k$, $k=1,\dots,d$. We furthermore need to cut out discs around the punctures 
$P_r$, $r=1,\dots, n$. 
A chosen orientation of the edges of $\Gamma$ allows us to distinguish 
an incoming and an outgoing boundary component of each annulus $A^e$.

The gluing construction of conformal blocks allows us to construct families of conformal blocks
out of two types of building blocks: The conformal blocks associated with the three-holed spheres, 
and the conformal blocks associated with the annuli $A^e$. Of particular interest
for us will be the latter. Let ${\rm CB}(A_m)$ be the space of conformal blocks 
associated with an annulus with $m$ marked points associated with fully degenerate representations, 
and incoming and outgoing boundary components associated with  non-degenerate
representations. We shall fix the representation associated with the incoming boundary component
to be $\cV_{\beta}$, and we assign fully degenerate representations $\cV_{-b\lambda_1},\dots,
\cV_{-b\lambda_m}$ to the $m$ marked points, respectively. The set of representations associated with the 
outgoing boundary component is then restricted by the fusion rules to a finite set. We will assign to 
the outgoing boundary component the direct sum of all representations allowed by the fusion rules.

It follows from the fusion rules that the resulting space of conformal blocks 
${\rm CB}(A_m)$ is finite-dimensional. A basis for
${\rm CB}(A_m)$ can be constructed explicitly using the degenerate vertex operators defined above,
\begin{equation}\label{discblock}
\cF_{A_m}(v_{m+1}\otimes v_{m}\otimes\cdots\otimes v_1\otimes v_0)=
\big( \,v_{m+1}\,,\,D^{\bfs_m}_{\lambda_m}(v_m|\sigma_m)\;\cdots\; 
D^{\bfs_1}_{\lambda_1}(v_1|\sigma_1)\,v_0\,\big)_{\beta'}\,,
\end{equation}
where $v_0\in\cV_{\beta}$, $v_{m+1}\in\cV_{\beta'}$, and 
$D^{\bfs_k}_{\lambda_k}(v_k|\sigma_k)$ 
are the descendants of the degenerate vertex operators $D^{\bfs_k}_{\lambda_k}(\sigma_k)$
associated with vectors $v_k\in\cV_{-b\lambda_k}$ for  $k=1,\dots,m$. 

It follows from the results discussed in the
previous subsection that the space of conformal blocks ${\rm CB}(A_m)$ is 
naturally isomorphic (as a module for $\cB_m$) to the tensor product of the $m$ finite-dimensional 
representations $M_{\lambda_m},\dots,M_{\lambda_1}$ of the quantum group $\cU_q(sl_N)$,
\begin{equation}\label{W-qgrp}
{\rm CB}(A_m)\,\simeq\,M_{\lambda_m}\otimes\dots\otimes M_{\lambda_1}\,.
\end{equation}
This isomorphism is realized by a linear operators $J^{(m)}_{\lambda_m\dots\lambda_1}(p)$
that is constructed by combining the change of basis removing the $p$-dependence with 
the Drinfeld twist $\mathcal{J}$.

\subsection{Verlinde network operators}

Verlinde network operators are generalizations of Verlinde loop operators \cite{Alday:2009fs,Drukker:2009id}
to Toda CFT of higher rank, and were previously 
studied in \cite{Bullimore:2013xsa}.
In order to define the Verlinde network operators one needs to consider conformal blocks associated with
a Riemann surface $\cC_{g,n+d}$, as above.  
Our definition will be based on certain relations 
between the spaces of conformal blocks with varying number $d$ of fully degenerate insertions.
In order to describe these relations we will employ the set-up introduced in the previous subsection,
in particular 
the decomposition of $\cC_{g,n+d}$ into three-holed spheres $T_v$ and annuli $A^e$,
and the isomorphism  (\ref{W-qgrp}) which can be applied locally for each annulus $A^e$.

This needs to be combined with one further ingredient. We will conjecture that there exist 
exchange relations between fully degenerate chiral vertex operators and generic chiral vertex 
operators, 
\begin{equation}\label{move1}
D^{i} (z_1) V_\alpha^{\bfs}(z_2)   = 
\sum_{j} \sum_{\bfs'}
B^{i,\bfs}_{j, \bfs' }(\alpha) 
V_{\alpha}^{\bfs'}(z_2)D^{j}(z_1)~, 
\end{equation}
where $|z_1|=|z_2|$,
and furthermore operator product expansions of the form 
\begin{equation}\label{move2}
D^{i} (w)\,V_\alpha^{\bfs}(z)  \,=\,
\sum_j
\sum_{\bfs'} F^{i,\bfs}_{j, \bfs'}(\alpha) 
V_{\alpha-b h_j}^{\bfs'}\big(D^j (w-z)v_{\alpha}|z\big)~,
\end{equation}
where $h_s$ are the weights of the fundamental representation with highest weight 
$\omega_1$.
These relations would imply that there exist linear relations between 
the spaces of conformal blocks associated with 
surfaces $\cC_{g,n+d}$ and $\cC_{g,n+d}'$ related by moving a degenerate field from one annulus 
$A^e$ to another one $A^{e'}$. So far we do not have a proof of these relations. We expect that 
such a direct proof should be possible using the free-field representation of the vertex operators. 
One may furthermore note that the relations \eqref{move1}, \eqref{move2} follow from the results 
of \cite{Bullimore:2013xsa} if the vertex operator $V_{\alpha}^\bfs$ is associated with a semi-degenerate
representation~$\cV_\alpha$. Further generalizations of these braid relations were found in \cite{Gomis:2014eya}.
The general 
statement should follow from this special case if the operator product expansion 
of sufficiently many semi-degenerate vertex operators generates vertex operators associated with 
fully non-degenerate representations, as is generally expected.

\begin{figure}[tb]
\centering
\includegraphics[width=\textwidth]{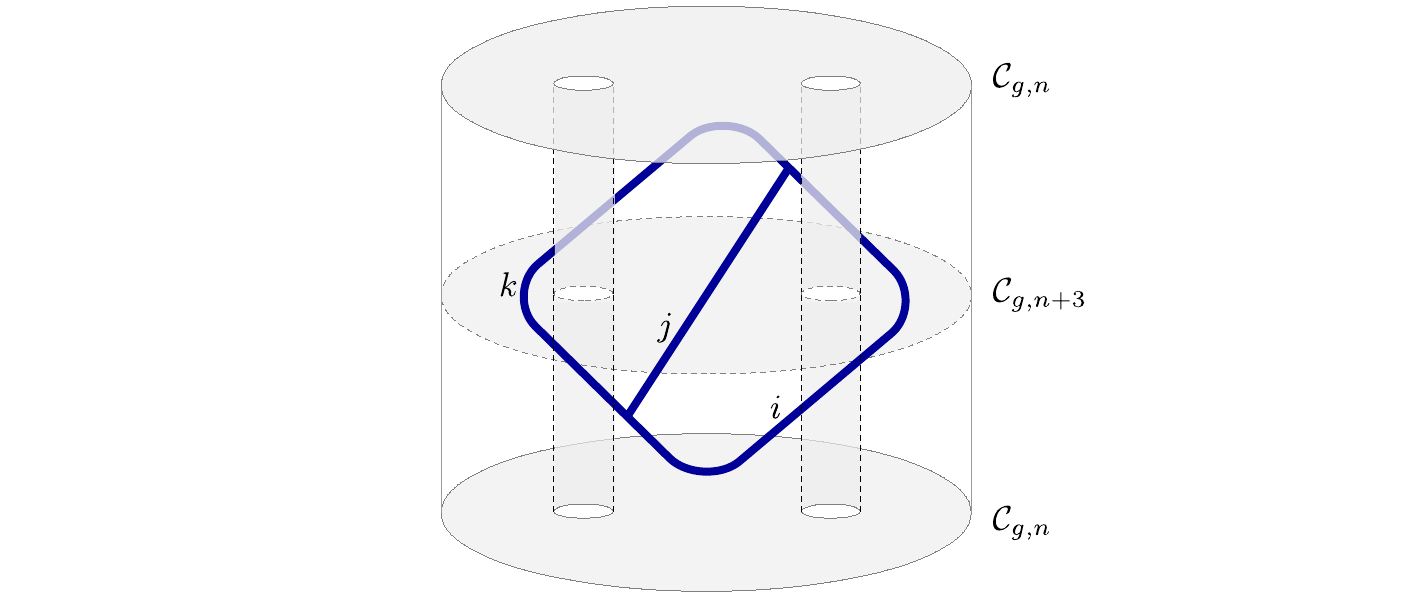}
\caption{Verlinde network operator.}
\label{VerNetOp}
\end{figure}

By combining these ingredients we are now ready to define generalizations of the Verlinde
loop operators as follows (see figure~\ref{VerNetOp}). The isomorphisms (\ref{W-qgrp}) allow us to associate maps between
spaces of conformal blocks with all the maps between tensor products of representations used
in the Reshetikhin-Turaev construction. These maps may be composed with the moves \eqref{move1},
\eqref{move2} describing
the motion of a fully degenerate puncture from one annulus to another. One may in particular
consider compositions of these two types of fusion and braiding operations for degenerate fields 
starting and ending with the 
space of conformal blocks 
associated with a surface $\cC_{g,n}$ with no fully degenerate insertions. It is clear that
such compositions can be labeled by networks  of the same type 
as considered in the previous sections. The construction outlined above associates with 
each network an operator acting on the space ${\rm CB}(\cC_{g,n})$ of conformal blocks 
for a surface $\cC_{g,n}$ with $n$ non-degenerate punctures.

\subsection{Relation to skein algebra}

The Verlinde network operators generate an algebra $\cA_{V}$ realized on the spaces ${\rm CB}(C_{g,n})$.
We are going to argue that this algebra is the same as the algebra of quantized functions
on $\cM^N_{g,n}$.  Recall that the product of two networks is represented in terms of a network with crossings. 
It may be assumed that all crossings are located in annuli $A^e$. In order to describe the resulting 
algebraic relations we may therefore use the isomorphism (\ref{W-qgrp}) allowing us 
to describe the operations
representing the crossing in terms of the intertwining maps used in the Reshetikhin-Turaev construction. 
All skein relations valid in the framework of the Reshetikhin-Turaev construction thereby carry over to 
the Verlinde network operators. In order to verify this claim it suffices to check that the basic
skein relations are preserved by the twist $J_{\lambda_n\dots\lambda_1}$ representing the 
isomorphism (\ref{W-qgrp}). Recall that the basic skein relations~\eqref{generalCrossingMOY} take the form
\bea
\label{genericskeinmoy-rem}
\includegraphics[width=0.93\textwidth]{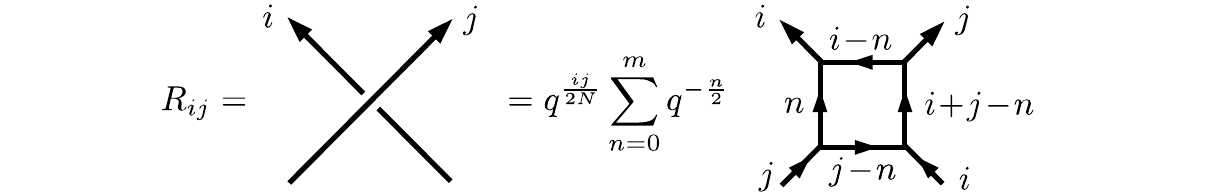}  
\eea
with $m = \text{min}\{i,j,N-i,N-j\}$.
In the framework of Reshetikhin-Turaev one associates to both sides intertwining maps between
the representations
 $M_{j}\otimes M_{i}$ and $M_{i}\otimes M_{j}$ of
the Hopf algebra $\mathcal{U}_q(sl_N)$.
The intertwining map on the left of  \eqref{genericskeinmoy-rem} is represented by the 
operator $R_{ij}$ defined in \eqref{Reval}, 
the operators on the right of \eqref{genericskeinmoy-rem}, in the following denoted 
as $C(n)^{ij}_{ji}$, are 
 compositions of Clebsch-Gordan maps, as introduced in Section \ref{SECquantization}. 
 If the quasitriangular Hopf-algebra $\mathcal{U}=\mathcal{U}_q(sl_N)$ is replaced 
 by a quasi-triangular Hopf-algebra $\tilde{\mathcal{U}}$ related to $\mathcal{U}$ by a 
 Drinfeld twist
 one may construct the intertwining maps $\tilde C(n)^{ij}_{ji}$ between representations 
 of $\tilde{\mathcal{U}}$ from 
 the intertwining maps $C(n)^{ij}_{ji}$
 of ${\mathcal{U}}$ in a natural way, as is explained in Appendix \ref{qGpBackground}. The construction is such that
 we have
 $\tilde C(n)^{ij}_{ji}=J^{-1}_{ij} C(n)^{ij}_{ji} J_{ji}^{}$, provided that $J_{ij}$ is the 
 similarity transformation relating $R_{ij}$ and $\tilde{B}_{ij}$ as 
 $\tilde{B}_{ij}=J_{ij} ^{-1} R_{ij} J^{}_{ji}$. 
 It follows immediately  
 that $\tilde C(n)^{ij}_{ji}$ and $\tilde B_{ij}$ 
 will satisfy a relation of the same form as \eqref{genericskeinmoy-rem}.

One may in particular consider Verlinde network operators associated with networks that are
confined to a disc $\mathbb{D}$ embedded
in an annulus $A^e$. It is easy to see that such network operators will
act on ${\rm CB}(C_{g,n})$ as multiplication by a number. This number represents one of the 
link invariants obtained by using the Reshetikhin-Turaev construction. This construction
is thereby recognized as a special case of the definition of the Verlinde network operators given
above.

\newpage
\section{Spectrum} \label{secSpectrum}

In the previous section we have found strong hints that the construction of loop and network operators using 
Fock-Goncharov coordinates defines an algebra isomorphic to the algebra generated by the 
Verlinde network operators in the Toda CFT. We had furthermore observed in subsection~\ref{comHam} that the resulting 
algebra shares some features with the algebra of operators in a quantum integrable model,
containing large commutative sub-algebras. Simultaneously diagonalizing these sub-algebras 
defines natural representations. The next natural step in our program is to compare 
the representations resulting from the quantization of the Fock-Goncharov coordinates, and 
from Toda conformal field theory. As an outlook, we collect here a few observations and 
conjectures concerning this problem.

\subsection{Spectrum in Toda field theory}

In order to determine the set of representations of the algebra $W_N$ appearing in the spectrum of Toda
field theory one may follow the example of Liouville theory. For studying the spectral problem of Toda 
field theories it is natural to consider canonical quantization on a  
spacetime having the geometry of a two-dimensional cylinder with time coordinate
$t$ and a spatial coordinate $\sigma\sim\sigma+2\pi$. 
At time $t=0$ one may decompose the fields $\phi(\sigma)\equiv \phi(0,\sigma)$ as 
$\phi(\sigma)=\phi_0+\chi(\sigma)$, where the zero mode is 
defined as the average
$\phi_0=\frac{1}{2\pi}\int_0^{2\pi}\dd\sigma\;\phi(\sigma)$. 
One may then observe that the interaction 
term $\int_0^{2\pi}\dd\sigma\;\ex^{b(e_i,\phi(\sigma))}$ in the Hamiltonian is equal to 
$2\pi \,\ex^{b(e_i,\phi_0)}$ up to terms of order $b^2$. 
This means that zero modes and oscillators completely decouple in the limit
$b^2\ra 0$. Canonical quantization of the oscillators yields a 
Hamiltonian of the form
\begin{equation}\label{Hcanq}
H\,=\,H_0+\mathsf{N}+\bar{\mathsf{N}}+\mathcal{O}(b^2)\,,
\qqq \text{with} \quad  H_0=\frac{1}{2}(p,p)+2\pi \mu \ex^{b(e_i,\phi_0)}\,,
\end{equation}
which is defined on the Hilbert space $\cH=L^2(\mathbb{R}^{N-1})\otimes \cF$
by choosing the Schr\"odinger representation on $L^2(\mathbb{R}^{N-1})$
for the operators $p_i$ and $\phi_0^j$ satisfying
$[\phi_0^i,p_j]=\ii\delta_{j}^{i}$, and by realizing two sets of  oscillators $a_n^i$
and $\bar{a}_n^i$, $i=1,\dots,N-1$, $n\in\mathbb{Z}\setminus \{0\}$, on the Fock space $\cF$ in the 
usual way. The operator $\mathsf{N}$ satisfies $[\mathsf{N},a_n^i]=-na_n^i$, and commutes with $\bar{a}_n^i$,
and similarly for $\bar{\mathsf{N}}$.

The problem of determining the
spectrum of the
Hamiltonian $H$ therefore reduces in the limit $b^2\ra 0$ 
to the problem of finding the spectrum of 
the quantum-mechanical 
system with Hamiltonian $H_0$. 
This system is a well-known quantum 
integrable model called the open Toda chain, and its spectrum 
is known exactly \cite{STS,Wa,KK}. 
The spectrum is purely continuous, with generalized
eigenstates in one-to-one correspondence with the orbits of vectors
$\mathfrak{p}\in\mathbb{R}^{N-1}$ under the Weyl group of $SU(N)$.
This suggests that the spectrum of Toda field theory will be purely continuous.

Observing furthermore that  the interaction 
terms $\int_0^{2\pi}\dd\sigma\;\ex^{b(e_i,\phi(\sigma))}$ are closely related to the screening charges of the 
free-field realization of the algebra $W_N$, it becomes straightforward to generalize the arguments 
used in \cite{CT} to determine the spectrum of Liouville theory to Toda field theories with 
$N>2$, leading to the conclusion that the spectrum has the form
\begin{equation}\label{Todaspec}
\cH\,=\,\int_{\mathbb{S}}\dd \beta \;\mathcal{V}_\beta\otimes\bar{\mathcal{V}}_\beta\,,
\end{equation}
where $\mathcal{V}_\beta$ and $\bar{\mathcal{V}}_\beta$ are representations of the algebra
$W_N$ generated by holomorphic currents $W^k(z)$ and their anti-holomorphic counterparts $\bar{W}^k(\bar{z})$,
respectively,
$\mathbb{S}$ denotes the set of Weyl-orbits of vectors of the form
$\beta=Q+\ii \mathfrak{p}$, with $\mathfrak{p}\in\mathbb{R}^{N-1}$ and $Q=\rho(b+b^{-1})$. 
It follows from this form of the spectrum that correlation functions 
such as 
\begin{equation} \label{correlFctalphaVV}
\big\langle\, \alpha_{\infty}\,|\,V_{\alpha_2}(z_2,\bar{z}_2)\,V_{\alpha_1}(z_1,\bar{z}_1)\,
|\,\alpha_0\,\big\rangle\,,
\end{equation}
can be represented in the form 
\begin{equation}\label{holofact}\begin{aligned}
\big\langle\, \alpha_{\infty}\,|\,& V_{\alpha_2}(z_2,\bar{z}_2)\,V_{\alpha_1}(z_1,\bar{z}_1)\,
|\,\alpha_0\,\big\rangle\\
&=\int_{\mathbb{S}}\dd \beta\int_{\mathbb{V}} \dd s_1\dd s_2\dd \bar{s}_1\dd \bar{s_2}\;C_{s_2\bar{s}_2}(\alpha_\infty,\alpha_2,\beta)
C_{s_1\bar{s}_1}(Q-\beta,\alpha_1,\alpha_0)\\ 
&\hspace{4cm}\times
\cF\big[{}_{\alpha_{\infty}}^{\alpha_2}s_2|\beta|s_1{}^{\alpha_1}_{\alpha_0}\big](z_2,z_1)
\bar{\cF}\big[{}_{\alpha_{\infty}}^{\alpha_2}\bar{s}_2|\beta|\bar{s}_1{}^{\alpha_1}_{\alpha_0}\big](\bar{z}_2,\bar{z}_1)\,,
\end{aligned}\end{equation}
where 
\begin{equation}
\cF\big[{}_{\alpha_{\infty}}^{\alpha_2}s_2|\beta|s_1{}^{\alpha_1}_{\alpha_0}\big](z_2,z_1)
=\big(\, v_{\alpha_{\infty}}\,,\, V_{s_2}\big({}^{\;\alpha_2}_{\alpha_\infty\beta}\big)(z_2)\,V_{s_1}\big({}^{\;\alpha_1}_{\beta\alpha_0}\big)(z_1)\,v_{\alpha_0}\,\big)_{\alpha_\infty}\,.
\end{equation}
In order to arrive at  the expansion \eqref{holofact} we have inserted a complete set of intermediate states
from $\cH$ between the two vertex operators $V_{\alpha_2}(z_2,\bar{z}_2)$ and  $V_{\alpha_1}(z_1,\bar{z}_1)$,
and we have furthermore assumed that the resulting matrix elements can be expanded as
\begin{equation}\label{holofact03}\begin{aligned}
\big\langle\, & Q-\alpha_{3}\,|\, V_{\alpha_2}(z,\bar{z})\,
|\,\alpha_1\,\big\rangle\,=\,\\
&\quad =\,\int_{\mathbb{V}} \dd s\,\dd \bar{s}\;C_{s\bar{s}}(\alpha_3,\alpha_2,\alpha_1)\;
\big(\,v_{\alpha_3}\,, V_s\big({}^{\;\,\alpha_2}_{\alpha_3\alpha_1}\big)(z)\,v_{\alpha_1}\,\big)_{\alpha_3}
\big(\,v_{\alpha_3}\,, V_{\bar{s}}\big({}^{\;\,\alpha_2}_{\alpha_3\alpha_1}\big)(\bar{z})\,v_{\alpha_1}\,\big)_{\alpha_3}\,.
\end{aligned}\end{equation}
The assumption \eqref{holofact03} will hold for any set
of chiral vertex operators $V_s\big({}^{\;\,\alpha_2}_{\alpha_3\alpha_1}\big)(z)$ labeled by an
index $s$ taking values in the set ${\mathbb{V}}$ with measure $\dd s$ which may be continuous or discrete
allowing us to define a basis for the space ${\rm CB}(\cC_{0,3})$ via \eqref{CVOvsblocks}.
If the index set ${\mathbb{V}}$ or measure $\dd s$ depends nontrivially on the choice of ${\alpha_i}$,
one will have to modify \eqref{holofact} in an obvious way.

It seems likely that the usual consistency conditions of the conformal bootstrap can only be satisfied 
if the conformal blocks with $\beta\in\mathbb{S}$ form a basis for  the space of conformal blocks ${\rm CB}(\cC_{0,4})$
on the four-punctured sphere $\mathbb{P}^1\setminus\{0,z_1,z_2,\infty\}$, 
as is known to be the case for $N=2$ (Liouville theory) \cite{TV13}.
It is furthermore easy to show that the Verlinde loop operators associated with the curve $\gamma_s$ 
separating 
punctures $z_1$ and $0$ from $z_2$ and $\infty$ act diagonally in this basis, with eigenvalues parameterized in 
terms of $\beta=Q+\ii\mathfrak{p}$. The formulas for the eigenvalues look simplest if one represents the 
vectors $\mathfrak{p}\in\mathbb{R}^{N-1}$ in terms of vectors $\mathfrak{a}\in\mathbb{R}^{N}$ lying on the
plane $\sum_k {a_k}=0$, where $a_k$ are the components of $\mathfrak{a}$ with respect to an
orthonormal basis for $\mathbb{R}^{N}$. The eigenvalues $\lambda_{\omega_1}(a)$
of the Verlinde loop operator associated with the 
curve $\gamma_{s}$ and the degenerate fields $D^i$, for example, can then be represented as
\begin{equation}\label{EVformula}
\lambda_{\omega_1}(a)\,=\,\sum_{i=1}^N \ex^{-2\pi \ii ba_i}\,. 
\end{equation}

The considerations above  motivate us to propose that the Hilbert space of Toda conformal blocks has the form
\bea\label{HCB}
{\rm HCB}(\cC_{g,n}) &=& \int_{\mathbb{S}}\; \prod_{\text{edges $e$}}  \dd \beta_e\, \bigotimes_{\text{vertices $v$}} \,{\rm HCB}(\cC_{0,3}^v) ~.
\eea
Here ${\rm HCB}(\cC_{0,3}^v)$ is the Hilbert space of conformal blocks on the three-punctured spheres $\cC_{0,3}^v$,
with $v=1,\dots,2g-2+n$, that appear in the pants decomposition of $\cC_{g,n}$.
The conformal blocks constructed by gluing will diagonalize the Verlinde loop operators for the cutting curves in the pants decomposition.

It should be noted that our conjecture \eqref{HCB} appears to be necessary for having consistency conditions
of the conformal bootstrap like crossing symmetry, modular invariance, or locality realized
in the usual way. Let us consider, for example, the condition of 
mutual  locality of the vertex operators $V_{\alpha}(z,\bar{z})$ within correlation functions such as
\begin{equation} \label{locality}
\big\langle\, \alpha_{\infty}\,|\,V_{\alpha_2}(z_2,\bar{z}_2)\,V_{\alpha_1}(z_1,\bar{z}_1)\,
|\,\alpha_0\,\big\rangle
=
\big\langle\, \alpha_{\infty}\,|\,V_{\alpha_1}(z_1,\bar{z}_1)\,V_{\alpha_2}(z_2,\bar{z}_2)\,
|\,\alpha_0\,\big\rangle\,.
\end{equation}
Such relations would hold if (i) the conformal blocks 
satisfy braid relations of the form
\begin{equation}\label{braidrel}
\begin{aligned}
&\cF\big[{}_{\alpha_{\infty}}^{\alpha_2}s_2|\beta|s_1{}^{\alpha_1}_{\alpha_0}\big](z_2,z_1)=\\
&\hspace{1.5cm}=\int_{\mathbb S} \dd\beta'
\int_{\mathbb T}\dd t_2\dd t_1\;
B_{\beta\beta'}\big[{}_{\alpha_{\infty}}^{\alpha_2}{}^{\alpha_1}_{\alpha_0}\big]_{s_2s_1}^{t_2t_1}\;
\cF\big[{}_{\alpha_{\infty}}^{\alpha_1}t_1|\beta|t_2{}^{\alpha_2}_{\alpha_0}\big](z_1,z_2)\,,
\end{aligned}
\end{equation}
and (ii) kernels 
$B_{\beta\beta'}\big[{}_{\alpha_{\infty}}^{\alpha_2}{}^{\alpha_1}_{\alpha_0}\big]_{s_2s_1}^{t_2t_1}$
and  structure functions $C_{s\bar{s}}(\alpha_3,\alpha_2,\alpha_1)$ appearing in \eqref{holofact} 
satisfy suitable orthogonality relations.
Equations \eqref{braidrel} may then be interpreted 
as relations between two different bases for the space of conformal blocks ${\rm HCB}(\cC_{0,4})$. 
The spectrum 
\eqref{Todaspec} thereby gets related to the spectrum 
\eqref{HCB} of Verlinde loop operators
on the space of conformal blocks.

\subsection{Spectrum of quantized trace functions}

We have explained in section~\ref{comHam} that a maximal set of commuting Hamiltonians on the symplectic moduli space $\bar \cM_{g,n}^N$ (with fixed holonomies around the punctures) consists of the cutting loops that specify a pants decomposition of $\cgn$ together with the corresponding pants networks.
This implies that all these loop and network operators can be simultaneously diagonalized.

We are going to argue that the spectrum of the Hamiltonians associated with the cutting 
curves coincides with the expected spectrum of the corresponding Verlinde loop operators.
To this aim we are going to observe that some essential features of the spectrum can be anticipated without
solving the eigenvalue problem explicitly. 

\paragraph{Example for $N=2$:}
As a warm-up, let us consider the case of $SL(2)$-connections on $\cC_{1,1}$.
In the polarization $\mathsf{p}=\frac12(\hat B + \hat C)$ and 
$\mathsf{q} = -\frac12(\hat B-\hat C)$ with $[\mathsf{p} ,\mathsf{q}]=-i\hbar$, 
the A-cycle operator~\eqref{AcycOpSL2C11} becomes
\bea\label{hA1PQ}
\hat{A} &=& 2\cosh \mathsf{p} + \ex^{-\mathsf{q}}~,
\eea
which defines a positive self-adjoint, but unbounded operator on $L^2(\mathbb{R})$.
A complete set of
eigenfunctions for this difference operator is known \cite{Ka,FT}. The spectrum is purely
continuous, and supported in the semi-infinite interval $(2,\infty)$.
The (generalized) eigenvalues are non-degenerate and may 
therefore be parameterized in terms of a parameter  $s\in\mathbb{R}^+$ as $2\cosh(s)$. 

Without having solved the spectral problem explicitly one could 
still note that the term  $\ex^{-\mathsf{q}}$ appearing in the definition of $\hat{A}$ decays 
exponentially for $\mathsf{q}\ra \infty$, and that the term $2\cosh \mathsf{p}$ can be seen as 
a ``deformation'' of the usual  term $\mathsf{p}^2$ in Schr\"odinger operators.
One may therefore expect that the spectral problem for $\hat{A}$ 
will in some respects be similar to the spectral problem for the Liouville quantum mechanics with 
Hamiltonian $H_0$~\eqref{Hcanq}. The extend to which this is the case was clarified in \cite{FT}.
In the Schr\"odinger  representation with $\mathsf{q}$ represented on wave-functions $\psi(q)$
as an operator of multiplication with $q$ one may construct two linearly independent Jost 
solutions, eigenfunctions $f_s^\pm(q)$ of $H_0$ that behave for $q\ra\infty$ as
$f_s^\pm(q)= \ex^{\pm\frac{i}{\hbar}sq} +o(1)$ \cite{FT}. 
The Jost solutions grow exponentially
at the opposite end $q\ra -\infty$. One may, however, find a
function $R(s)$ such that the  
linear combination 
$\psi_s(q)= f^+_s(q)+R(s)f^-_s(q)$ decays rapidly for $q\ra -\infty$. Only this linear 
combination  may  appear in the
spectral decomposition of $\hat{A}$.
Note  that $\psi_{-s}(q)$ is related to $\psi_s(q)$ by a relation of the form 
$\psi_{-s}(q)=R(s)\psi_s(q)$, analogous to the case of 
a Schr\"odinger operator with repulsive potential. 

The fact that the spectrum is bounded from below by $2$ follows from 
the observation that $\hat{A}$ is the sum of two positive self-adjoint operators, and its
spectrum therefore bounded from below by the spectrum of $2\cosh \mathsf{p}$.

\paragraph{Example for $N=3$:}

Although we will not be able to find explicit diagonal representations for the A-cycle operators for higher rank, we can  
anticipate features of their spectra by a similar reasoning.
In order to see how the observation above generalize, let us next consider the quantization of the trace
functions associated with $SL(3)$-connections on
$\cC_{1,1}$.

In order to represent the trace functions 
associated with A- and B-cycles in a form that will be convenient for the quantization, let us 
start by fixing the eigenvalues of the holonomy ${\bf M}$ around the puncture 
in terms of constants $\gamma_1 = a_1a_2b_1b_2c_1c_2$ and $\gamma_2 = xy$:
\bea
\left\{ \frac1{a_1 a_2b_1b_2c_1c_2xy}, \frac1{xy}, a_1 a_2b_1b_2c_1c_2x^2y^2\right\} = \left\{ \frac1{\gamma_1\gamma_2}, \frac1{\gamma_2},  \gamma_1\gamma_2^2 \right\}~.
\eea
We must then choose a polarization with  momenta $P_i$ and positions $Q_j$ satisfying $\{P_i, Q_j\} = \delta_{ij}P_i Q_j$.
The nice Poisson bracket~\eqref{betaAlphaPBC11} between the monomials $\alpha_i$ and $\beta_j$ appearing in the A- and B-cycle suggests the following choice:
\bea
P_1 &=& \alpha_1  =  b_1 c_1 x ~,\qqq Q_1 = \prod_i \beta_i^{-\kappa_{1i}^{-1}} = a_1^{-\frac23}a_2^{-\frac13}b_1^{-\frac23}b_2^{-\frac13}x^{-\frac13}y^{-\frac23} ~, \nn
P_2 &=& \alpha_2 = b_2 c_2 y  ~,\qqq Q_2 = \prod_i \beta_i^{-\kappa_{2i}^{-1}} = a_1^{-\frac13}a_2^{-\frac23}b_1^{-\frac13}b_2^{-\frac23}x^{-\frac23}y^{-\frac13} ~,  \nn
P_3 &=& x~, \qqq\qqq   Q_3 = (a_2b_2c_2)^{-1}~.
\eea
The A-cycle trace functions are then expressed as
\bea
{A}_1 &=&  P_1^{\frac13} P_2^{\frac23}+\frac{P_1^{\frac13}}{P_2^{\frac13}} + \frac1{P_1^{\frac23}P_2^{\frac13}}  + \frac{P_1^{\frac13}P_2^{\frac23}P_3}{\gamma_1 \gamma_2^2 Q_1 Q_2} + P_1^{\frac13} P_2^{\frac23} \frac{Q_1 Q_3}{Q_2^2} \left( 1 + \frac1{\gamma_2} \right) + \frac{P_1^{\frac13} Q_2(1+P_3)}{\gamma_1 \gamma_2 P_2^{\frac13} Q_1^2Q_3}   ~, \nn
{A}_2 &=&  P_1^{\frac23} P_2^{\frac13}+\frac{P_2^{\frac13}}{P_1^{\frac13}} + \frac1{P_1^{\frac13}P_2^{\frac23}}  + \frac{ P_2^{\frac13}Q_1Q_3}{P_1^{\frac13}P_3 Q_2^2} +  \frac{P_2^{\frac13} Q_1 Q_3}{P_1^{\frac13}Q_2^2} \left( \frac1{P_3} + \frac1{\gamma_2} \right) + \frac{P_1^{\frac23} P_2^{\frac13}Q_2}{\gamma_1 \gamma_2  Q_1^2Q_3} (1+P_3)   ~. \nonumber
\eea
If we take the limit $Q_1, Q_2\to \infty$ with $Q_1/Q_2$ finite we see that only the first three terms 
survive.  The quantization of these expressions is straightforward following our discussions above,
allowing us to define operators $\hat{A}_i$, $i=1,2$.

By generalizing the arguments used in the case $N=2$ above one finds that the 
eigenvalues $A_i'$ of the operators $\hat{A}_i$ can be parameterized in terms of a vector
$(s_1,s_2)\in\mathbb{R}^2$ as follows
\bea
A_1' =  \xi_1+ \xi_2+ \xi_3~, \qqq A_2' = \frac1{\xi_1} + \frac1{\xi_2 }+ \frac1{\xi_3}~,
\eea
where $\xi_a$ are defined  for $a=1,2,3$ as 
\bea
\big( \xi_1, \xi_2, \xi_3\big) = \big(\ex^{\frac13 (s_1+2s_2)}  ,~ {\ex^{\frac13 (s_1-s_2)} } , ~ \ex^{-\frac13(2s_1+s_2)}  
\big)~.
\eea
The expressions for the eigenvalues $A_i'$ are manifestly invariant under Weyl symmetry permuting the $\xi_a$. 

Furthermore, it seems
likely that the repulsive nature of the dependence on $Q_1,Q_2$ will imply that the spectrum is 
parameterized by the Weyl-orbits of vectors $\mathfrak{s}=(s_1,s_2)\in\mathbb{R}^2$. In 
order to see this, let $\psi_{\mathfrak{s}}(\mathfrak{q})$,  $\mathfrak{q}=(q_1,q_2)$,
be a joint  eigenfunction of 
the operators $\hat{A}_i$, $i=1,2$, in a Schr\"odinger-type
representation where the quantum operators corresponding to
the classical variables $\log Q_i$ are diagonal with eigenvalues $q_i$. 
The wave-function $\psi_{\mathfrak{s}}(\mathfrak{q})$ will be of the form
\begin{equation}
\psi_{\mathfrak{s}}(\mathfrak{q})\,=\,\psi_{\mathfrak{s}}^0(\mathfrak{q})
+\cO\big(\ex^{-\frac{1}{2}(q_1+q_2)}\big)\,,
\end{equation}
where $\psi_{\mathfrak{s}}^0(\mathfrak{q})$ is an eigenfunction of the operators 
\[
\hat{A}_1^0 =  \mathsf{P}_1^{\frac13} \mathsf{P}_2^{\frac23}+ \mathsf{P}_1^{\frac13} \mathsf{P}_2^{-\frac13}  +  {\mathsf{P}_1^{-\frac23}\mathsf{P}_2^{-\frac13}} ~,
\qqq
\hat{A}_2^0 =  \mathsf{P}_1^{\frac23} \mathsf{P}_2^{\frac13}+ \mathsf{P}_1^{-\frac13} \mathsf{P}_2^{\frac13}+  {\mathsf{P}_1^{-\frac13}\mathsf{P}_2^{-\frac23}}     ~,
\]
where $\mathsf{P}_i=\ex^{\frac{\hbar}{\ii}\frac{\partial}{\partial q_i}}$. A joint  eigenfunction of $\hat{A}_i^0 $
is given by $\ex^{\frac{\ii}{\hbar}(s_1q_1+s_2q_2)}\equiv \ex^{\frac{\ii}{\hbar}(\mathfrak{s},\mathfrak{q})}$, 
where $(\mathfrak{s},\mathfrak{q})$ denotes the standard scalar product of 
the vectors $\mathfrak{s}$ and $\mathfrak{q}$ in $\mathbb{R}^2$. The Weyl-invariance of the eigenvalues
of $\hat{A}_i^0$ implies that  other joint eigenfunctions of $\hat{A}_i^0$ are given
by the functions $\ex^{\frac{\ii}{\hbar}(w(\mathfrak{s}),\mathfrak{q})}$, where $w(\mathfrak{s})$ is the vector obtained 
by the action of the element $w$ of the Weyl-group $\mathcal{W}$ of $SL(3)$ on $\mathfrak{s}\in\mathbb{R}^2$. 
It follows that the most general joint eigenfunction of the $\hat{A}_i^0 $ has the form 
\begin{equation}\label{psi0}
\psi_{\mathfrak{s}}^0(\mathfrak{q})\,=\,\sum_{w\in\mathcal{W}} C_w(\mathfrak{s})
\ex^{\frac{\ii}{\hbar}(w(\mathfrak{s}),\mathfrak{q})}\,.
\end{equation}
The exponential growth of the terms in $\hat{A}_i$ depending on $q_i\equiv\log{Q_i}$ will 
imply that the wave-functions $\psi_{\mathfrak{s}}(\mathfrak{q})$ have to decay very rapidly when 
$q_1,q_2\ra\infty$. This will imply that $\psi_{\mathfrak{s}}(\mathfrak{q})$ has to satisfy two reflection
relations that must be compatible with the structure \eqref{psi0} of the $\psi_{\mathfrak{s}}^0(\mathfrak{q})$.
The reflection relations must therefore be of the form 
\begin{equation}\label{refrel}
\psi_{w_i(\mathfrak{s})}(\mathfrak{q})\,=\,R_i(\mathfrak{s})\psi_{\mathfrak{s}}(\mathfrak{q})\,,\qqq i=1,2\,,
\end{equation}
with $w_i$ two different elements of $\mathcal{W}$. By composition of the Weyl reflections 
$w_i$, one may generate reflection relations corresponding to all
elements of the Weyl group $\mathcal{W}$. The resulting relations
will determine the coefficients $C_w(\mathfrak{s})$ in \eqref{psi0} completely up to an overall normalization.
We are thereby lead to the conclusion that the spectrum is indeed parameterized by the Weyl-orbits of 
vectors $\mathfrak{s}\in\mathbb{R}^2$.

A very similar structure is found for the trace functions $B_i$ associated with the 
B-cycles.  The explicit expressions turn out to be of the form
\bea
B_1= Q_2+\frac{Q_1}{Q_2} + \frac1{Q_1} +\cO\big((P_1P_2)^{\frac{1}{6}}\big),\quad
B_2=\frac{1}Q_2+\frac{Q_2}{Q_1} + {Q_1} +\cO\big((P_1P_2)^{\frac{1}{6}}\big),
\eea
suggesting that the eigenvalues can be parameterized in terms of real positive numbers
$\zeta_a$, $a=1,2,3$, satisfying $\zeta_1\zeta_2\zeta_3=1$, as  
\bea
B_1 \to  \zeta_1 + \zeta_2 + \zeta_3~, \qqq B_2 \to \frac1{\zeta_1} + \frac1{\zeta_2}+ \frac1{\zeta_3}~.
\eea
Our observations above suggest that the spectrum of the trace-functions $A_1$ and $B_1$
coincides with the spectrum \eqref{EVformula} of the corresponding Verlinde loop operators.

\paragraph{Higher rank:}
The remarks above seem to generalize to cases with $N>3$. In the case $N=4$, for example, 
one may note that the trace function $A_2$ of the A-cycle holonomy  in the second antisymmetric representation (see~\eqref{AcyclesSL4}) has a similar structure as observed in the case $N=3$ above, 
with the following leading term:
\bea
A_2^0 = 2 \cosh \frac12 (p_1 + p_3) + 2 \cosh \frac12 (p_1 -p_3) + 2 \cosh \frac12 (p_1 +2p_2+ p_3) ~,
\eea
where we defined $p_i\equiv \log P_i$. This further strengthens our confidence that the spectra
of quantized trace functions will coincide with the spectra of the Verlinde loop operators in 
Toda CFT.

\subsection{Concluding remarks}

One of the most important problems is clearly to find useful bases for the space of conformal blocks 
${\rm CB}(\cC_{0,3})$. A natural possibility would be to define such bases by diagonalizing a maximal 
set of commuting Verlinde network operators. Natural as it may be, it seems technically
much harder to analyze the spectrum of network operators along similar lines as described above.
As an example let us consider  the pants network $N_1$ on $\pants$ in the case $N=3$.
We may use  the coordinates
\bea
p_1 = a_1~, \qqq q_1 = y~,
\eea
and denote the eigenvalues of $\bfA$ by $\tilde \alpha_1 = \alpha_1^{1/3}\alpha_2^{2/3}$ and $\tilde \alpha_2 = \alpha_1^{1/3}\alpha_2^{-1/3}$, and similarly for $\bfB$ and $\bfC$.
The pants network function can then be represented as
\small
\begin{align}
& N_1 = \frac{p_1}{{\tilde\beta_1} {\tilde\gamma_1} {\tilde\gamma_2}} \bigg[  1+  \frac{2 {\tilde\beta_1}^2 {\tilde\gamma_1}^2 {\tilde\gamma_2}^3}{{\tilde\alpha_1} p_1^3}+{\tilde\alpha_1} {\tilde\beta_1} {\tilde\gamma_1} &\\
 & + \frac{1}{p_1^2} \left( \frac{{\tilde\beta_1} {\tilde\gamma_1}^2 {\tilde\gamma_2}^2}{{\tilde\alpha_1} {\tilde\alpha_2}}+\frac{{\tilde\beta_1}^2 {\tilde\gamma_1}^2 {\tilde\gamma_2}^3}{{\tilde\alpha_1}}+\frac{{\tilde\beta_1}^2 {\tilde\gamma_1} {\tilde\gamma_2}}{{\tilde\alpha_1}}+{\tilde\alpha_2} {\tilde\beta_1} {\tilde\gamma_1}^2 {\tilde\gamma_2}^2+{\tilde\beta_1}^2 {\tilde\beta_2} {\tilde\gamma_1} {\tilde\gamma_2}^2+\frac{{\tilde\beta_1} {\tilde\gamma_1} {\tilde\gamma_2}^2}{{\tilde\beta_2}} \right) &\nn
&+  q_1 \left( \frac{{\tilde\alpha_2} {\tilde\beta_1}^2 {\tilde\beta_2} {\tilde\gamma_1}^2 {\tilde\gamma_2}^4}{{\tilde\alpha_1} p_1^3}+\frac{{\tilde\alpha_2} {\tilde\beta_1}^2 {\tilde\beta_2} {\tilde\gamma_1} {\tilde\gamma_2}^2}{{\tilde\alpha_1} p_1^2}+\frac{{\tilde\beta_1} {\tilde\beta_2} {\tilde\gamma_1}^2 {\tilde\gamma_2}^3}{{\tilde\alpha_1} p_1^2}+\frac{{\tilde\beta_1} {\tilde\beta_2} {\tilde\gamma_1} {\tilde\gamma_2}}{{\tilde\alpha_1} p_1}+\frac{{\tilde\alpha_2} {\tilde\beta_1} {\tilde\gamma_1} {\tilde\gamma_2}^3}{p_1^2}+\frac{{\tilde\alpha_2} {\tilde\beta_1} {\tilde\gamma_2}}{p_1}+\frac{{\tilde\gamma_1} {\tilde\gamma_2}^2}{p_1}+1 \right) &\nn
 &+ \frac{1}{q_1} \left( \frac{{\tilde\beta_1}^2 {\tilde\gamma_1}^2 {\tilde\gamma_2}^2}{{\tilde\alpha_1} {\tilde\alpha_2} {\tilde\beta_2} p_1^3}+\frac{{\tilde\beta_1}^2 {\tilde\gamma_1}^2 {\tilde\gamma_2}^2}{{\tilde\alpha_1} {\tilde\alpha_2} {\tilde\beta_2} p_1^2}+\frac{{\tilde\alpha_1} {\tilde\beta_1} {\tilde\gamma_1}}{p_1}+{\tilde\alpha_1} {\tilde\beta_1} {\tilde\gamma_1}+\frac{{\tilde\beta_1}^2 {\tilde\gamma_1} {\tilde\gamma_2}}{{\tilde\alpha_2} p_1^2}+\frac{{\tilde\beta_1}^2 {\tilde\gamma_1} {\tilde\gamma_2}}{{\tilde\alpha_2} p_1}+\frac{{\tilde\beta_1} {\tilde\gamma_1}^2 {\tilde\gamma_2}}{{\tilde\beta_2} p_1^2}+\frac{{\tilde\beta_1} {\tilde\gamma_1}^2 {\tilde\gamma_2}}{{\tilde\beta_2} p_1}  \right)\bigg]~. & \nonumber
\end{align} \normalsize
We see that there is no limit in which the terms depending on $q_1$ vanish. This might indicate that the 
spectrum of the corresponding quantum operator is discrete rather than continuous. It would be
very interesting if one could describe the spectrum of this operator more precisely.

One may also note that the free-field construction of chiral vertex operators gave us families of conformal blocks
labeled by a space of parameters which has the same dimension as Lagrangian subspaces of $\cM_{0,3}^{N}$.
This indicates that the conformal blocks that can be obtained in this way may represent a basis for 
${\rm CB}(\cC_{0,3})$.

Let us finally note that the geometric engineering of gauge theories of class $\cS$, combined with 
the topological vertex technique has led to a prediction for the 
structure functions $C_{s\bar{s}}(\alpha_3,\alpha_2,\alpha_1)$ appearing in \eqref{holofact}, see \cite{Mitev:2014isa,Isachenkov:2014eya}.
The labels $s$ for a basis of ${\rm CB}(\cC_{0,3})$ would thereby be identified with geometric
data of the local Calabi-Yau manifold used in the geometric engineering of the class $\cS$ theories
associated with $\cC_{0,3}$. It would be very interesting to identify the meaning of this parameter
within conformal field theory, or within the quantum theory obtained by quantizing $\cM^{N}_{0,3}$.

\acknowledgments

We would like to thank Mathew Bullimore, Tudor Dimofte, Nadav Drukker, Lotte Hollands, and Bruno Le Floch for useful discussions.
The work of MG was supported by the German Science Foundation (DFG) within the Research Training Group 1670 ``Mathematics Inspired by String Theory and QFT'' and by the Swiss National Science Foundation.
The work of IC was supported by the People Programme (Marie Curie Actions) of the European Union's Seventh Framework Programme FP7/2007-2013/ under REA Grant Agreement No. 317089. 
IC acknowledges her student scholarship from the Dinu Patriciu Foundation “Open Horizons”, which supported part of her studies.
Support from the DFG in the framework of the SFB 676 \emph{Particles Strings, and the Early Universe} is gratefully acknowledged.

\appendix

\section{Fock-Goncharov coordinates}\label{AppFG}

In~\cite{FG}, Fock and Goncharov constructed coordinates systems on the moduli space of flat connections over a punctured surface $\cgn$. They assume that the surface is hyperbolic, that is $\chi(\cgn) = 2-2g-n<0$, and has at least one puncture, $n\ge 1$. 
Their approach consists in ``localizing'' flat connections on the triangles of an ideal triangulation of $\cgn$ (in which the vertices of the triangles are at punctures).
However, since a triangle is contractible, any flat connection on it is trivial.
This difficulty can be overcome by considering \emph{framed} flat connections, which means adding on each puncture a flag that is invariant under the holonomy around the puncture.

\subsection{Coordinates associated with $N$-triangulations}

\begin{figure}[t]
\centering
\includegraphics[width=3.4in]{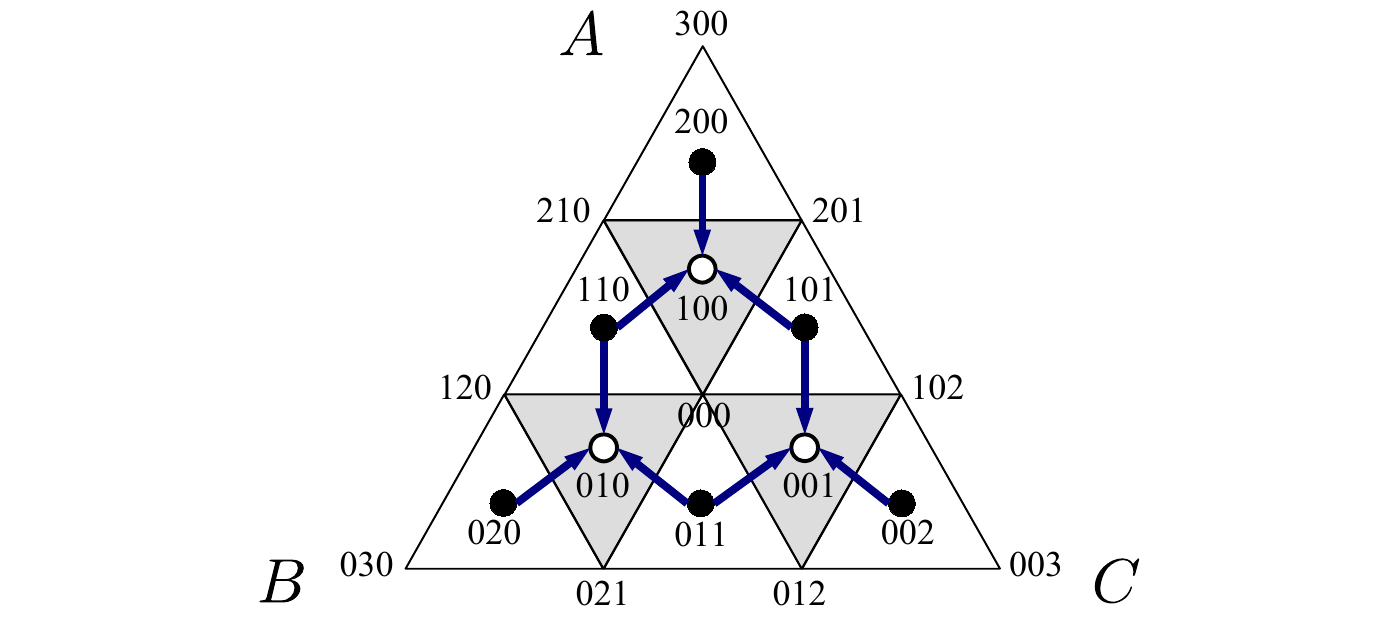}
\caption{$N$-triangulation of a triangle (here for $N=3$). The triples of non-negative integers $a,b,c$ sum to $3$ on lattice points $p_{abc}$, to $2$ on white triangles associated with lines $l_{abc}$ (black dots), and to $1$ on black triangles associated with planes $P_{abc}$ (white dots). The internal lattice point is associated with the 3-space $V_{000}$. Arrows between dots indicate incidence relations.}
\label{3triangle}
\end{figure}

Let us review how to build coordinates on the moduli space of framed flat connections over a punctured surface $\cgn$
(We mostly follow section~4 in~\cite{Dimofte:2013iv}, and appendix A in~\cite{Gaiotto:2012db}).
Given an $SL(N)$-vector bundle with a framed flat connection over $\cgn$, each triangle of an ideal triangulation of $\cgn$ gives rise to a (generic) configuration of three flags $\{A,B,C\}$ associated with the vertices.
A flag $A$ in an $N$-dimensional complex vector space $V_N$ is a collection of nested subspaces
\bea
A: \qquad 0 = A_0 \subset A_1 \subset A_2 \subset \cdots \subset A_N = V_N~, 
\eea
with $\dim[A_a] = a$.
An $N$-triangulation consists in a further decomposition of each triangle into small triangles, whose vertices are at the $N+2 \choose 2$ lattice points $p_{abc}$ with
\bea
a+b+c=N~, \qqq a,b,c \in \bZ_{\ge0}~.
\eea
See figure~\ref{3triangle} for an example with $N=3$.
There are $N^2$ small triangles for each ideal triangle, $N+1 \choose 2$ of which are ``upright'' (white) and labeled by the solutions of $a+b+c=N-1$, while $N \choose 2$ are ``upside-down'' (black) and labeled by the solutions of $a+b+c=N-2$.
To every white triangle we associate a line $l_{abc}$ in $V_N$ arising as the intersection of the corresponding subspaces of the flags $\{A,B,C\}$. More explicitly, writing $A^a \equiv A_{N-a}$ with codim$[A^a] =a$, we have
\bea
l_{abc} = A^a \cap B^b \cap C^c~, \qqq a+b+c = N-1~.
\eea
Similarly, we associate to every black triangle a plane $P_{abc}$, and to every internal lattice point with $a+b+c=N-3$ a 3-space $V_{abc}$:
\bea
P_{abc} = A^a \cap B^b \cap C^c~, \qqq a+b+c = N-2~, \nn
V_{abc} = A^a \cap B^b \cap C^c~, \qqq a+b+c = N-3~.
\eea
The plane $P_{abc}$ on a black triangle contains all three lines $l_{(a+1)bc}$, $l_{a(b+1)c}$, and $l_{ab(c+1)}$ on the adjacent white triangles.
In turn, the 3-space $V_{abc}$ on an internal lattice point contains all three planes on the surrounding black triangles, and thus all six lines on the surrounding white triangles.

This collection of subspaces on the $N$-triangulation allows us to associate coordinates to every lattice point (excluding the vertices at the punctures).
We can define coordinates $x_{abc}$ associated with the internal lattice points $p_{abc}$ with $a+b+c = N-3$ as the triple-ratio of the six surrounding lines contained in $V_{abc}$
(the neighborhood of an internal lattice point looks like the 3-triangulation shown in figure~\ref{3triangle}).
The 3-space $V_{abc}$ contains the three flags
\bea
\tilde A : &&0 \subset l_{(a+2)bc} \subset P_{(a+1)bc} \subset V_{abc}~, \nn
\tilde B : &&0 \subset l_{a(b+2)c} \subset P_{a(b+1)c} \subset V_{abc}~, \nn
\tilde C : &&0 \subset l_{ab(c+2)} \subset P_{ab(c+1)} \subset V_{abc}~.
\eea
Fixing the flags to be $\tilde A = (\mathbf{a_1}, \mathbf{a_1}\wedge \mathbf{a_2})$, with $\bf a_i$ vectors in $V_{abc}$, we can write the triple-ratio as
\bea \label{3ratio}
x_{abc} = \mathbf{\frac{\langle a_1 \wedge a_2 \wedge b_1 \rangle \langle b_1 \wedge b_2 \wedge c_1 \rangle \langle  c_1\wedge c_2 \wedge a_1 \rangle }{\langle a_1 \wedge a_2 \wedge c_1 \rangle \langle b_1 \wedge b_2 \wedge a_1 \rangle \langle  c_1\wedge c_2 \wedge b_1 \rangle} }~.  
\eea
Here the notation $\langle \mathbf{v_1 \wedge v_2\wedge v_3}\rangle$ means the determinant of the matrix expressing the vectors $\mathbf{v_1,v_2,v_3}$ in a unimodular basis for the 3-space containing them (this triple-ratio is the inverse of the one originally defined by Fock and Goncharov, see appendix A in~\cite{Gaiotto:2012db}).

\begin{figure}[t]
\centering
\includegraphics[width=3.4in]{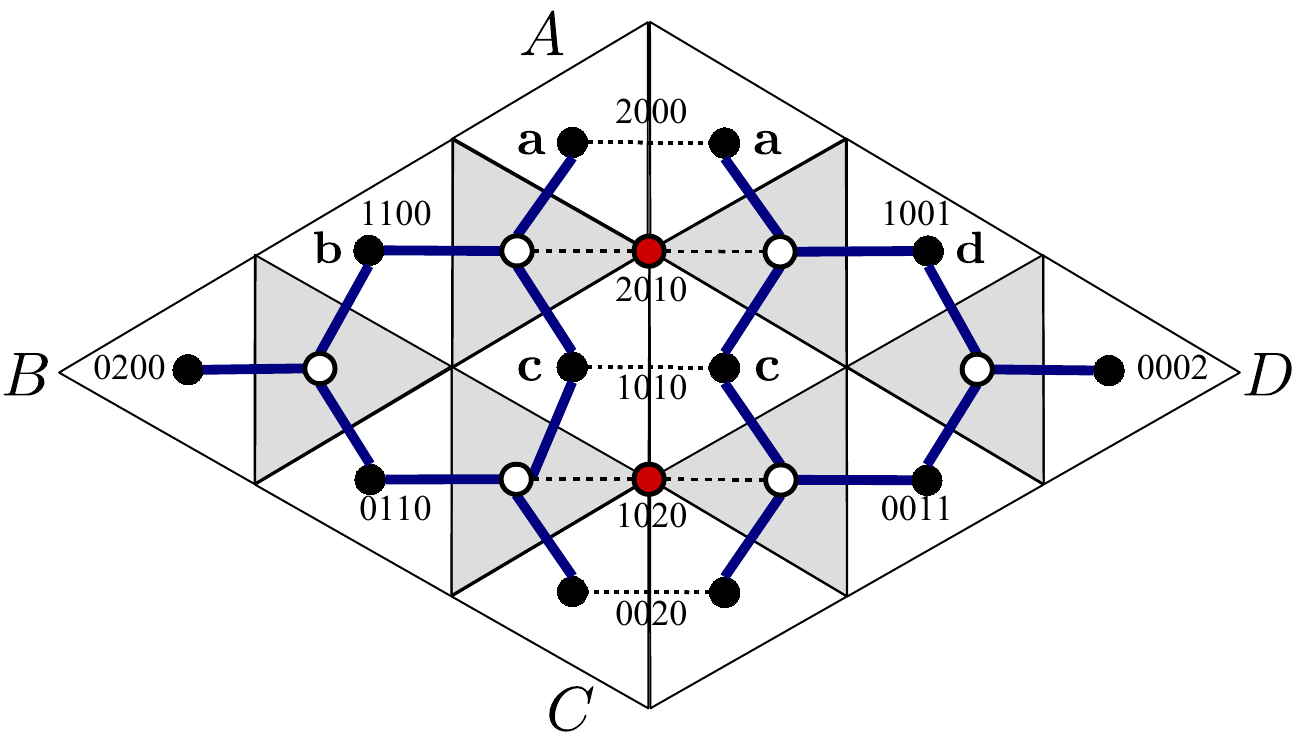}
\caption{Two 3-triangulated triangles $ABC$ and $ACD$ glued along the edge $AC$. The quadruples of non-negative integers $a,b,c,d$ sum to $3$ on lattice points $p_{abcd}$, and to $2$ on white triangles associated with lines $l_{abcd}$. Lines and planes adjacent across the edge are identical. Each edge point is surrounded by four lines.}
\label{crossratio}
\end{figure}

It remains to define coordinates for the lattice points on the edges of the ideal triangulation.
Along the common edge of two glued $N$-triangulated triangles, adjacent white triangles are associated with the same lines, and adjacent black triangles to the same planes (see figure~\ref{crossratio}).
A lattice point on an edge is thus surrounded by four lines, which allows us to define a cross-ratio coordinate.
For example, at the point $p_{2010}$ on figure~\ref{crossratio}, we choose four vectors $\mathbf{a}\in l_{2000}$, $\mathbf{b}\in l_{1100}$, $\mathbf{c}\in l_{1010}$, $\mathbf{d}\in l_{1001}$, all contained in the plane $P_{1000}$, which give the cross-ratio
\bea\label{crossratiox}
x_{2010} =\mathbf{\frac{\langle a \wedge b \rangle \langle c\wedge d\rangle }{\langle a \wedge d \rangle \langle b\wedge c\rangle }}~.
\eea

An ideal triangulation of the surface $\cgn$ has $-2\chi(\cgn)$ triangles and $-3 \chi(\cgn)$ edges.
There are $N-1\choose 2$ internal lattice points in each triangle and $N-1$ lattice points on each edge, so the total number of independent coordinates provided by triple-ratios and cross-ratios is
\bea
\# \{\text{$x$-coordinates} \}= - \chi(\cgn) \dim[SL(N,\bC)]~.
\eea
This agrees with the dimension~\eqref{dimMgnN} of the moduli space of flat $SL(N,\bC)$-connections on $\cgn$.

\subsection{Snakes and projective bases}

\begin{figure}[t]
\centering
\includegraphics[width=3in]{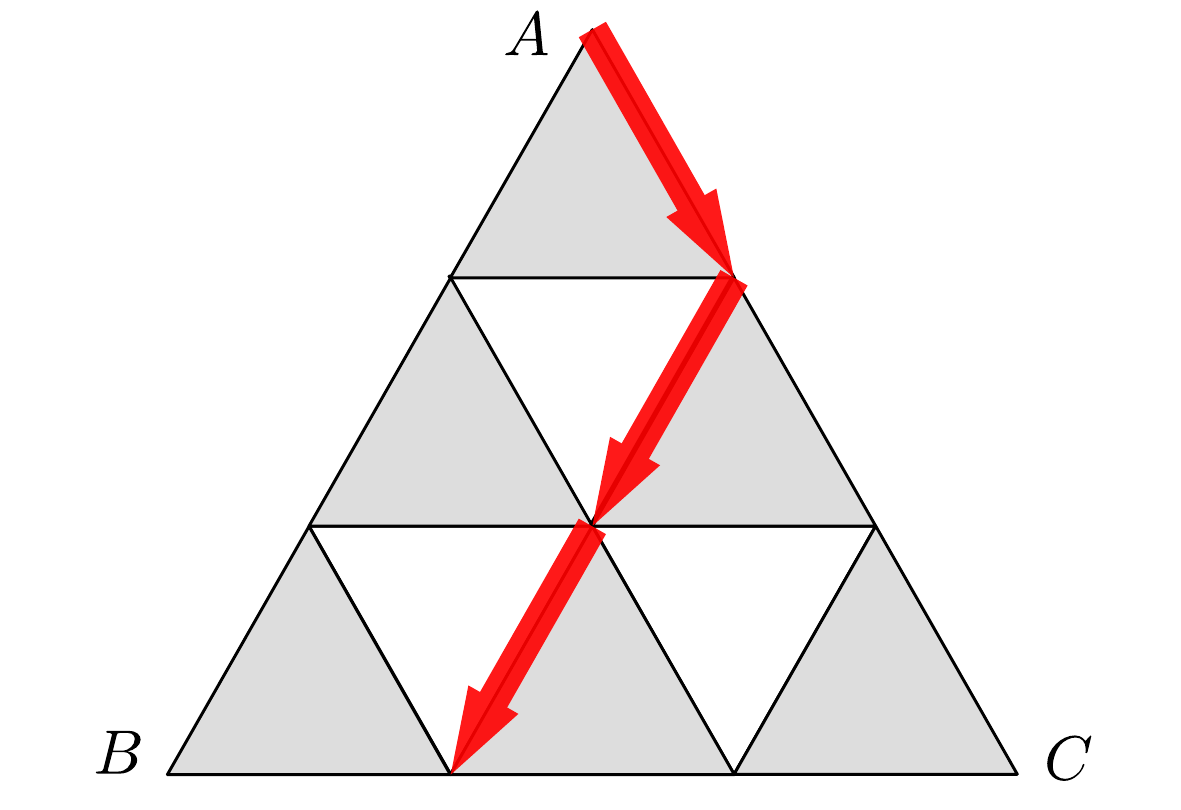}
\caption{Example of a snake on the $(N-1)$-triangulation of a triangle (here for $N=4$).}
\label{snakeExample}
\end{figure}

\begin{figure}[tb]
\centering
\includegraphics[width=\textwidth]{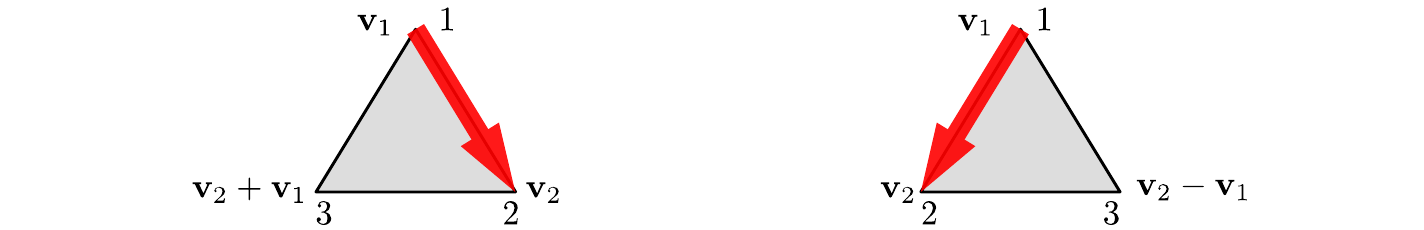}
\caption{Segment of a snake along an edge of a black triangle in an $(N-1)$-triangulation. The vertices of the triangles correspond to 3 coplanar lines $l_1,l_2,l_3$. Picking a vector $\mathbf{v}_1\in l_1$ at the start of the segment determines $\mathbf{v}_2\in l_2$ by the rule that $\mathbf{v}_2 \pm \bfv_1 \in l_3$ depending on whether the segment is oriented clockwise or counterclockwise around the triangle.}
\label{SnakeSegment}
\end{figure}

Fock and Goncharov showed how to construct projective bases in the $N$-dimensional vector space $V_N$ related to three flags $A,B,C$ at the vertices of a triangle. 
Each basis is represented by a \emph{snake}, that is an oriented path on the edges of the $(N-1)$-triangulation from one vertex to the opposite side (see figure~\ref{snakeExample} for an example).
Notice that in the $(N-1)$-triangulation the vertices correspond to lines, and the upright (black) triangles to planes.
Each segment of the snake goes from a line $l_1$ to a line $l_2$, and the plane they define also contains a third line $l_3$ at the third vertex of the black triangles.
Given a vector $\mathbf{v}_1\in l_1$, a vector $\mathbf{v}_2\in l_2$ is determined by the rule that $\mathbf{v}_2+\bfv_1 \in l_3$ if the segment is oriented clockwise around the black triangle, and $\mathbf{v}_2-\bfv_1 \in l_3$ if it is oriented counterclockwise (see figure~\ref{SnakeSegment}).
To construct the projective basis $\{\bfv_1,\bfv_2,  \cdots, \bfv_N\}$ defined by a snake, we can choose any vector $\mathbf{v_1}$ at the first vertex of the snake and use this rule to determine all the subsequent vectors.

A projective basis can be transformed into another one by a sequence of simple moves.
\begin{figure}[tb]
\centering
\includegraphics[width=\textwidth]{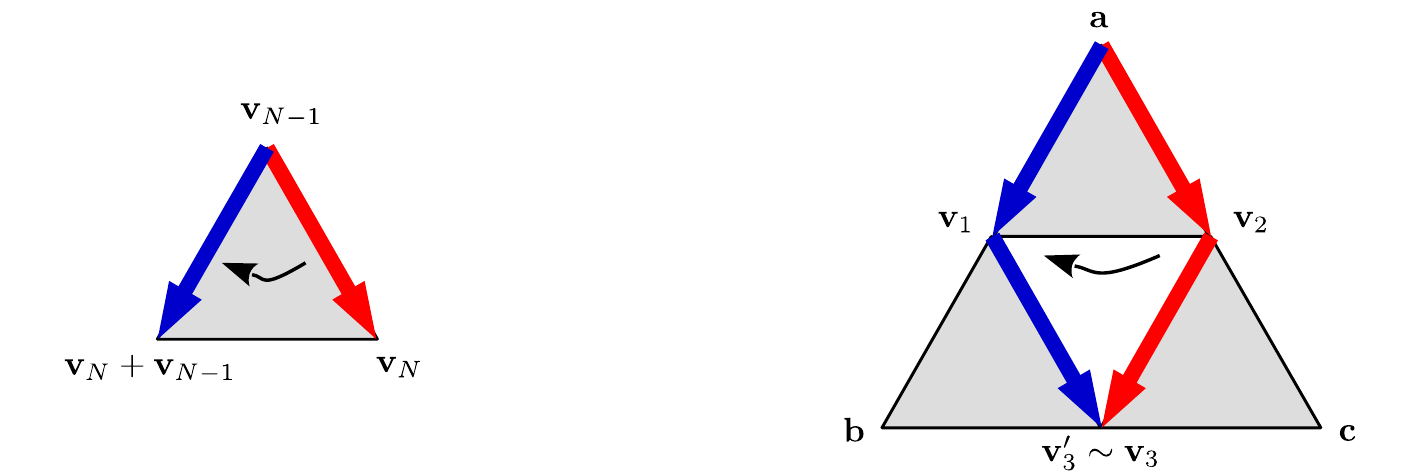}
\caption{Elementary moves for segments of a snake. \emph{Left}: Move I can only occur on the last segment of a snake. \emph{Right}: Move II can occur on any two consecutive segments of a snake.}
\label{SnakeMoves}
\end{figure}
The moves I and II shown in figure~\ref{SnakeMoves} can be expressed in terms of two types of elementary $GL(N)$ matrices.
Let $\varphi_i : SL(2) \to GL(N)$ be the canonical embedding corresponding to the $i^\text{th}$ root $\lambda_i - \lambda_{i+1}$.
Then the two types of elementary matrices can be written as 
\bea
F_i = \varphi_i \begin{pmatrix} 1&0\\1&1\end{pmatrix}~, \qqq
H_i (x) = \text{diag} (\underbrace{1, \cdots,1}_{i \text{ times}}, x, \cdots, x)~.
\eea
For example, for $N=3$ we have the following elementary matrices:
\bea
& F_1 = \begin{pmatrix} 1&0&0\\1&1&0\\0&0&1\end{pmatrix}~, \quad F_2 = \begin{pmatrix} 1&0&0\\0&1&0\\0&1&1\end{pmatrix}~,& \nn
& H_1(x) = \begin{pmatrix} 1&0&0\\0&x&0\\0&0&x\end{pmatrix}~, \quad H_2(x) = \begin{pmatrix} 1&0&0\\0&1&0\\0&0&x\end{pmatrix}~.&
\eea
Note that $F_i$ and $H_j(x)$ commute unless $i=j$. 

Move I flips the last segment of a snake across a black triangle, which according to the rule in figure~\ref{SnakeSegment} transforms the projective basis as
\bea
\begin{pmatrix} \mathbf{v}_1\\ \vdots \\ \mathbf{v}_{N-1} \\ \mathbf{v}_N \end{pmatrix}  \mapsto \begin{pmatrix} \mathbf{v}_1\\ \vdots \\ \mathbf{v}_{N-1} \\ \mathbf{v}_N + \mathbf{v}_{N-1} \end{pmatrix}  =  F_{N-1}  \begin{pmatrix} \mathbf{v}_1\\ \vdots \\ \mathbf{v}_{N-1} \\ \mathbf{v}_N \end{pmatrix}  ~.
\eea
Move II takes any two consecutive segments of a snake across a pair of adjacent black and white triangles, see figure~\ref{SnakeMoves}.
For a choice of vector $\bfa$ at their first common vertex, the initial and final snakes define two vectors $\mathbf{v}_3$ and $\mathbf{v}_3'$ spanning the line at the vertex where the snakes meet again.
The proportionality function is precisely the coordinate $x_{abc}$ for the internal lattice point associated with the white triangle
\bea
\mathbf{v}_3' = x_{abc} \mathbf{v}_3~.
\eea
This can be checked by writing down the triple-ratio~\eqref{3ratio} for the choice of flags $A=(\mathbf{a}, \mathbf{v}_1)$, $B=(\mathbf{b}, \mathbf{v}_3)$, $C=(\mathbf{c}, \mathbf{v}_2)$ and using the relations provided by the black triangles as in figure~\ref{SnakeSegment} (that is $\mathbf{v}_1 = \mathbf{v}_2 + \mathbf{a}$, $\mathbf{b} = \mathbf{v}_3'+\mathbf{v}_1$, $\mathbf{c} = \mathbf{v}_3 - \mathbf{v}_2$).
The transformation between the bases associated with the two snakes in figure~\ref{SnakeMoves} acts as
\bea
\begin{pmatrix} \mathbf{a}\\ \mathbf{v}_2 \\ \mathbf{v}_3 \end{pmatrix} \mapsto \begin{pmatrix} \mathbf{a}\\ \mathbf{v}_1 \\ \mathbf{v}_3' \end{pmatrix} 
=  F_1 H_2(x_{abc})  \begin{pmatrix} \mathbf{a}\\ \mathbf{v}_2 \\ \mathbf{v}_3 \end{pmatrix} ~.
\eea

\begin{figure}[tb]
\centering
\includegraphics[width=3in]{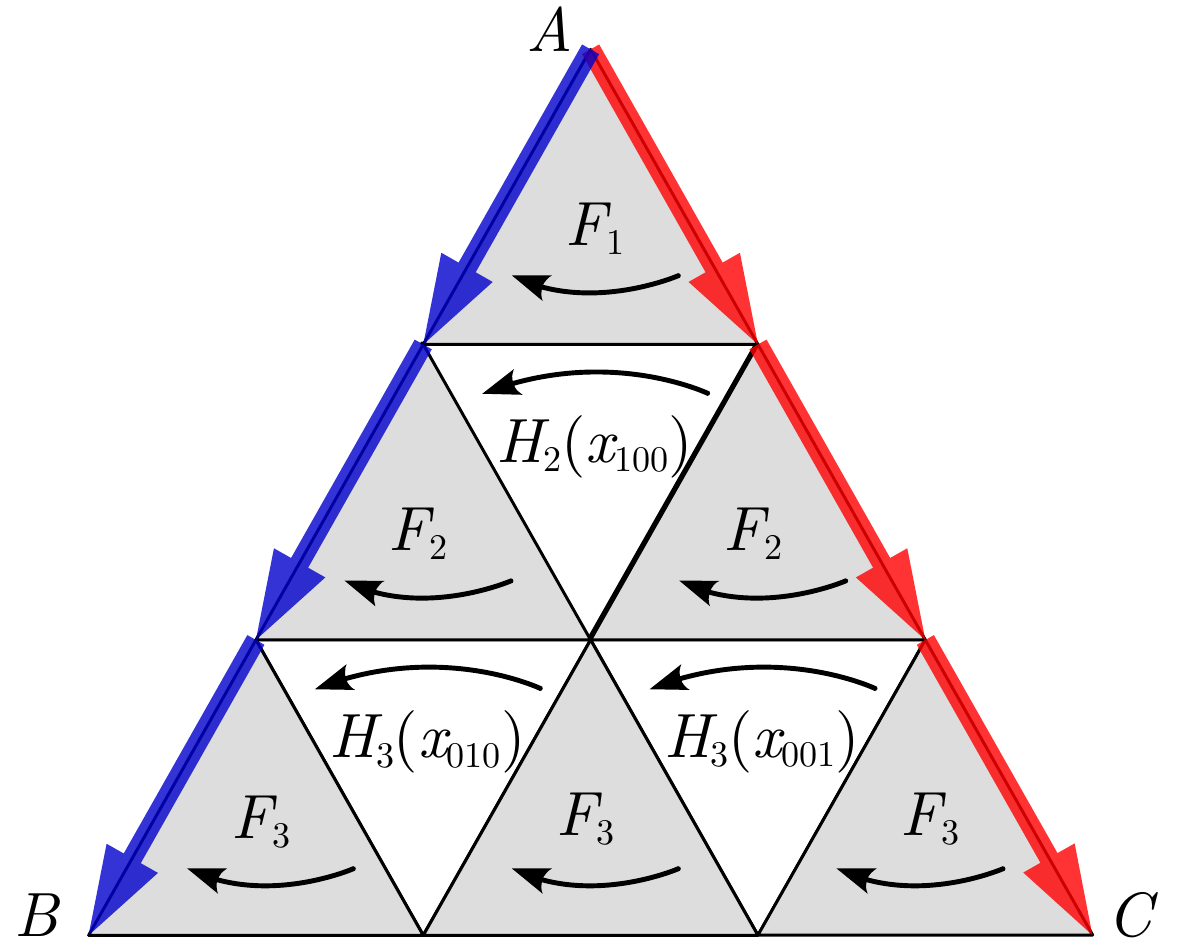}
\caption{Elementary matrices $F_i$ and $H_i(x)$ in an $(N-1)$-triangulation (here for $N=4$). The transformation $\cF$ from the snake along the edge $AC$ to the snake along $AB$ can be decomposed into a sequence of moves across the black and white triangles, corresponding to matrices $F_i$ and $H_i(x)$.}
\label{FHmatrices}
\end{figure}

More generally, we can see that moving the $i^\text{th}$ segment of snake across a black triangle corresponds to acting with a matrix $F_i$, while moving it across a white triangle corresponds to acting with a matrix $H_i(x)$, where $x$ is the face coordinate associated with the white triangle.
An important snake transformation consists in moving a snake from one edge of a triangle across the entire face to the next edge, rotating clockwise around the initial vertex (see figure~\ref{FHmatrices}).
For $N=2$ this gives simply $\cF  = F_1$. For $N=3$, this gives 
\bea
\cF  = F_2F_1H_2(x_{000})F_2 = \begin{pmatrix} 1&0&0\\ 1&1&0\\ 1&1+x_{000} & x_{000}\end{pmatrix}~.
\eea
For $N=4$ as in figure~\ref{FHmatrices} this gives
\bea
\cF  = F_3F_2H_3(x_{010})F_3F_1H_2(x_{100})F_2H_3(x_{001})F_3~.
\eea

\begin{figure}[t]
\centering
\includegraphics[width=\textwidth]{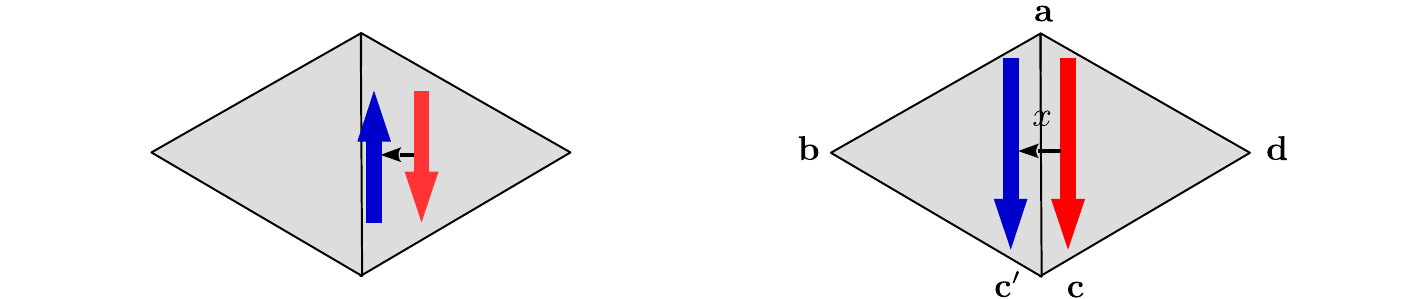}
\caption{\emph{Left}: reversal of the orientation of the snake along an edge of a triangle.
\emph{Right}: snake crossing an edge of a triangle with a cross-ratio coordinate $x$.}
\label{SmoveFlip}
\end{figure}

Another type of move is the reversal of the orientation of a snake along an edge of a triangle, see figure~\ref{SmoveFlip}.
This reverses the order of the basis vectors, and changes the sign of every even-numbered vector, thus multiplying by an anti-diagonal matrix
\bea \label{Smatrix}
S = \begin{pmatrix} \cdots & 0&0&1\\ \cdots & 0&-1&0 \\ \cdots & 1&0&0  \vspace{-.17cm} \\  \iddots& \vdots & \vdots& \vdots \end{pmatrix}~.
\eea

Finally, we can move a snake from one triangle to another across their common edge.
If the coordinates along the edge followed by the snake are $\{ x_1, x_2, \ldots, x_{N-1} \}$ (in this order), this transformation acts as 
\bea \label{edgematrix}
H_1(x_1) H_2(x_2) \cdots H_{N-1}(x_{N-1}) = \text{diag}(1, x_1, x_1 x_2, \ldots, x_1x_2 \cdots x_{N-1} )~.
\eea
This can again be checked by writing down the cross-ratio~\eqref{crossratiox} for the edge coordinate $x$ in figure~\ref{SmoveFlip} and using the relations provided by the black triangles as in figure~\ref{SnakeSegment}.

\section{Background on $sl(N)$ and $\cU_q(sl_N)$}
\label{qGpBackground}

\subsection{Lie algebra conventions}\label{Lieconv}

The vectors $e_i$ with $i=1,\ldots , N-1$ are the simple roots of the Lie algebra $sl(N)$ 
and $(\cdot,\cdot)$ denotes the scalar product on the Cartan subalgebra of $sl(N)$, with 
the Cartan matrix given by $\kappa_{ij}= (e_i, e_j)$. It is often useful to represent the root/weight
subspace of $sl(N)$ as a hyperplane in $\mathbb{R}^N$, allowing us to 
represent the simple roots as $e_i = u_i - u_{i+1}$ with $u_i$ a unit vector of $\bR^N$.
The simple roots are dual to the fundamental weights $\omega_i$ in the sense that 
$(e_i, \omega_j) = \delta_{ij}$. In general they can be written as 
$\omega_i = \sum_{j=1}^{i} u_j - \frac{i}{N}\sum_{k=1}^N u_k$. 
Their inner product is given by the inverse Cartan matrix 
$(\omega_i,\omega_j)=\kappa_{ij}^{-1}$ and their sum is the Weyl vector $\rho=\sum_i \omega_i$.
The weights of the fundamental representation of $sl(N)$ with highest weight 
$\omega_1$ are $h_i = \omega_1 - \sum_{j=1}^{i-1}e_j$, with $i = 1, \ldots, N$ and
$(h_i, h_j) = \delta_{ij} - \frac1N $.

\subsection{Basic definitions}\label{BasicQG}

We review some basic notions about quantum groups that are relevant here. 
For a complete background and set of axioms we refer the reader to some of the 
standard references \cite{CP:1994}.

Abstractly, a {\it Hopf algebra} is a collection $(\cU,m,\eta,\Delta,\epsilon,S)$, 
with unit $\eta:\bC(q)\rightarrow\cU$, product $m:\cU\otimes\cU\rightarrow\cU$, 
coproduct $\Delta:\cU\rightarrow\cU\otimes\cU$, counit $\epsilon:\cU\rightarrow\bC(q)$ 
and an algebra antiautomorphism called antipode $S:\cU\rightarrow\cU$. The triple 
$(\cU,m,\eta)$ is a unital associative algebra and $(\cU,\Delta,\epsilon)$ is a 
counital associative coalgebra. $\Delta, \epsilon$ are unital algebra homomorphisms 
and satisfy the following relations
\bea
(\Delta\otimes\tid)\circ\Delta &=& (\tid\otimes\Delta)\circ\Delta ~
\qquad\text{(co-associativity)} \nn 
(\epsilon\otimes\tid)\circ\Delta &=& (\tid\otimes\epsilon)\circ\Delta=\tid \nn
m\circ(S\otimes\tid)\circ\Delta &=& \eta\circ\epsilon=m\circ(\tid\otimes S)
\circ\Delta ~\qquad\text{(pentagon)}~.
\eea
The quantum group $\cU=\cU_q(sl_N)$ is an example of a Hopf algebra, generated 
as an associative algebra over $\bC(q)$ by $E_i,~F_i,~K_i^{\pm 1}$ for 
$i=1,\ldots ,N-1$ satisfying 
\bea
K_iK_i^{-1}=1=K_i^{-1}K_i ~,~ K_iE_j=q^{(i,j)}E_jK_i ~,~ 
K_iF_j=q^{-(i,j)}F_jK_i  ~~~\text{for}~~~ i,j=1,\ldots ,N-1~, \nnm
\eea
where $(i,j)\in\{2,-1,0\}$ when $|i-j|=0,1$ or $\geq2$, and
\bea
[E_i,F_j]=\delta_{ij}\frac{K_i-K_i^{-1}}{q-q^{-1}}~,~~i,j=1,\ldots ,N-1~,
\qquad [E_i,E_j]=0 ~,~ [F_i,F_j]=0 ~, ~~ |i-j|\geq 2 ~.\nnm
\eea
For $|i-j|=1$, the generators satisfy the quantum Serre relations
\bea
E_i^2E_j-(q+q^{-1})E_iE_jE_i + E_jE_i^2=0 \quad , \quad 
F_i^2F_j-(q+q^{-1})F_iF_jF_i + F_jF_i^2=0 ~. \nnm
\eea
The Hopf algebra structure is given by 
$\epsilon(K_i)=1,~\epsilon(E_i)=0=\epsilon(F_i)$, 
\bea
\Delta(K_i)&=&K_i\otimes K_i  ~~~,~~~\Delta(E_i)=E_i\otimes K_i+1\otimes E_i ~~~,~~~ 
\Delta(F_i)=F_i\otimes 1+K_i^{-1}\otimes F_i ~\nnm
\eea
and
\bea
S(K_i) &=& K_i^{-1} ~~~,~~~ S(E_i)=- E_i K_i^{-1} ~~~,~~~ S(F_i)=- K_i F_i ~. \nnm  
\eea
A {\it quasitriangular} Hopf algebra $\cU$ additionally has an invertible element 
$\cR\in \cU\otimes\cU$, the universal $R$-matrix of $\cU$, which satisfies a set of 
relations. For example, $\cR\Delta(a)=\Delta'(a)\cR$ for any $a\in\cU$, where 
$\Delta'(a)=P\circ\Delta(a)$ and $P$ is the permutation homomorphism mapping 
$P(a\otimes b)=(b\otimes a)$. 

With each algebra $\cU$, we can associate the category $\Rep\cU$ of its finite 
dimensional linear representations, with objects $V,~W,\ldots$ that are finitely 
generated left $\cU$-modules and with morphisms that are $\cU$-linear homomorphisms. 
The action of $\cU$ on an $\cU$-module $V$ induces a representation 
$\pi_V:\cU\rightarrow\End~V$.

The coproduct in a quasitriangular Hopf algebra $\cU$ induces the tensor product in 
$\Rep\cU$, where for $\cU$-modules $U,~V$ and $\forall a\in\cU$, 
$\pi_{U\otimes V}(a)=(\pi_U\otimes\pi_V)(\Delta(a))$. The unit object in $\Rep\cU$ is 
the ground field $\bC(q)$ equipped with the action of $\cU$ by means of the counit 
$\epsilon$. Furthermore, there exist associativity and left and right action 
homomorphisms that make $\Rep\cU$ a monoidal category (see for example \cite{RT1}). 
Moreover, for any $\cU$-modules $V,~W$ there is a mapping called braiding 
\bea
R_{VW}=P\circ(\pi_V\otimes\pi_W)(\cR):V\otimes W\rightarrow W\otimes V ~,
\eea 
which is a morphism of $\Rep\cU$ and turns this into a braided monoidal category.

For triples of $\cU$-modules $U,~V,~W$, where $U\subset V\otimes W$, a Clebsch-Gordan 
map 
$C^U_{VW}:V\otimes W\rightarrow U$ 
intertwines the actions on $V\otimes W$ and $U$,
\bea 
\label{intertwiningCG}
C_{VW}^U (\pi_V \otimes \pi_W)\Delta(x) = \pi_U(x) C_{VW}^U, 
\qquad x\in\cU ~.
\eea
Clebsch-Gordan maps are represented graphically as trivalent vertices in 
figure~\ref{GluingNetworks}. 
We note that the representations $V$, $W$ and $U$ may be reducible, in general. 
We may, in particular, consider cases that $V$, $W$ and $U$ can be tensor
products of representations such as $V=V_1\otimes\dots \otimes V_m$, for example. 
Such Clebsch-Gordan maps $C_{VW}^U$ can be constructed as compositions of 
Clebsch-Gordan maps intertwining the respective representations on the tensor factors.
Avoiding cases where $q$ is a root of unity, one may find any irreducible finite-dimensional 
representation in a sufficiently high tensor power  of the fundamental representation.

The conjugate Clebsch-Gordan 
maps $C_U^{VW}: U \rightarrow V\otimes W$ satisfy the property
\bea
\label{intertwiningCG3}
C_U^{VW} \pi_U(x) = (\pi_V \otimes \pi_W)\Delta(x) C_U^{VW} ~.
\eea
A special subset of intertwiners are the {\it cap} and {\it cup} maps intertwining between a 
tensor product $V\otimes \bar{V}$ and the trivial representation $\mathbf{1}\simeq \bC(q)$, 
respectively, where $\bar{V}$ is the conjugate representation to  $V$.
In the Reshetikhin-Turaev construction, one considers
general morphisms $M$ in $\Rep\cU$ between 
tensor products of modules $V_1\otimes\ldots\otimes V_m$ and $W_1\otimes\ldots\otimes W_n$.

\subsection{Twisted (compositions of) Clebsch-Gordan maps}
\label{twistedclebschcompositionS}
  
 Two quasitriangular Hopf algebras $\mathcal{U}$ and $\tilde{\mathcal{U}}$ are related by a 
 Drinfeld twist $\cJ\in\cU\otimes\cU$ if the co-products and $R$-matrices are related by
 \begin{equation}
  \tilde\Delta(\cdot)=\cJ^{-1}\Delta(\cdot)\cJ,\qqq
  \tilde \cR_{12}^{}=\cJ^{-1}_{21}\cR_{12}\cJ_{12}^{}\,.
 \end{equation}
Here we are using leg-numbering notation where the subscripts refer to tensor factors.
The twist will preserve co-associativity if 
\begin{equation}\label{cocycle-2}
(\Delta\otimes {\rm id})(\cJ)\cdot \cJ_{12}\,=\,({\rm id}\otimes \Delta)(\cJ)\cdot \cJ_{23}\,,
\end{equation}
The action of $\cJ$ may be extended to $m$-fold tensor products by defining recursively
\begin{equation}
\cJ^{(k+1)}:=({\rm id}^{\otimes{(k-1)}}\otimes\Delta)(\cJ^{(k)})\cdot
 ({\rm id}^{\otimes{(k-1)}}\otimes \cJ)\,,
\qquad
\cJ^{(2)}:=\cJ\,.
\end{equation}
The condition (\ref{cocycle-2}) ensures that the operator
\begin{equation}
J_{V_m\otimes\dots\otimes V_1}^{(m)}:=(\pi_{V_m}\otimes\dots\otimes\pi_{V_1})(\cJ^{(m)})\,,
\end{equation}
is independent of the order in which tensor products are taken.

Given a Clebsch-Gordan map $C_{VW}^U:V\otimes W \rightarrow U$ 
one may define 
\bea
\label{twistedClebschdef}
\tilde{C}_{VW}^U := C_{VW}^U \cdot J_{VW}^{}~ \,,\qquad J_{VW}:=(\pi_V\otimes\pi_W)(\cJ)\,.
\eea
It is easy to see that $\tilde{C}_{VW}^U$ is a Clebsch-Gordan map intertwining the representation 
defined on $V\otimes W$ using the twisted coproduct $\tilde{\Delta}$ with the representation $U$.
Similarly, by defining $\tilde{C}_U^{VW} := J^{-1}_{VW} C_U^{VW}$ one can show that 
$\tilde C_U^{VW} \pi_U(x) = (\pi_V \otimes \pi_W)\tilde\Delta (x) \tilde C_U^{VW}$. 

Given a general intertwining map $C:V_1\otimes\ldots\otimes V_m\ra W_1\otimes\ldots\otimes W_n$
it seems natural to define
\begin{equation}\label{twistedClebschdef-gen}
\tilde{C}:=(J^{(n)}_{W_1\otimes\ldots\otimes W_n})^{-1}\cdot M\cdot J^{(m)}_{V_1\otimes\ldots\otimes V_m}\,.
\end{equation}
If $M$ is represented as the composition of "more elementary" Clebsch-Gordan maps
one needs to verify the consistency of definition \eqref{twistedClebschdef-gen} with \eqref{twistedClebschdef}.
This boils down to the consideration of two cases, as will be discussed in the following. 

\begin{figure}[t]
\centering
\includegraphics[width=0.9\textwidth]{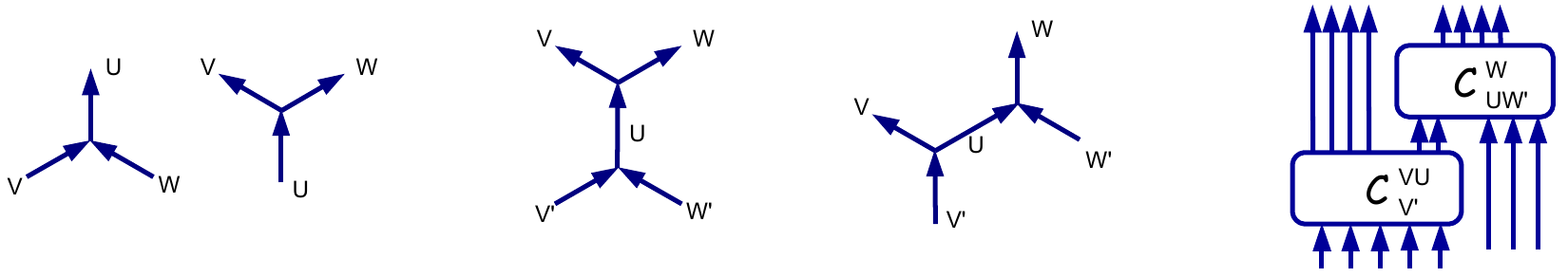}
\caption{({\it Left}): Clebsch-Gordan maps. ({\it Middle}): Gluing of 
vertices. ({\it Right}): Generic network.}
\label{GluingNetworks}
\end{figure}

\paragraph{Twisting composite intertwiners (I):} 

Let  $C_{T}^{UVW} : T \rightarrow U \otimes V \otimes W$ 
be constructed as $C_S^{UV}C_T^{SW}$.  
This composition may be re-expressed in terms of $\tilde{C}_S^{UV}\tilde{C}_T^{SW}$
by using 
\bea
C_{S}^{UV}C_T^{SW}=C_{S}^{UV}J_{SW}^{}J^{-1}_{SW}C_T^{SW} = 
C_{S}^{UV}J_{SW}\tilde C_T^{SW}~.
\eea
Decomposing $J_{SW}=\sum_k \pi_{S}(x_{k})\otimes \pi_W(x^k)$, one may use the 
intertwining property satisfied by $C_{S}^{UV}$ to calculate
\begin{equation}\begin{aligned}
C_{S}^{UV}\,J_{SW}&= C_{S}^{UV} \cdot \sum_k \pi_{S}(x_{k})\otimes \pi_W(x^k) = \\
&=\sum_k(\pi_U\otimes\pi_V)(\Delta(x_{k}))\otimes \pi_W(x^k) \cdot C_{S}^{UV} 
= J_{(U\otimes V)\otimes W}\,C_{S}^{UV}~. \nnm
\end{aligned}\end{equation}
Noting that $C_{S}^{UV}=J_{UV}\tilde{C}_{S}^{UV}$, we arrive at
\bea 
\label{pullingclebsches}
C_{S}^{UV}C_T^{SW}=J_{(U\otimes V)\otimes W}J_{UV}\tilde C_{S}^{UV}\tilde C_T^{SW}~.
\eea
In a similar way one may verify that 
$C_S^{VW}C_T^{US}=J_{U\otimes(V\otimes W)}J_{VW}\tilde C_{S}^{VW}\tilde C_T^{US}$.

\paragraph{Twisted compositions (II):} Clebsch-Gordan maps of the type $C_U^{VW}$ 
and $C^U_{VW}$ can also be composed in the following ways
\bea
\label{genericnetworks}
 C^{VW}_{V'W'} := C_U^{VW} C_{V'W'}^U \quad , 
\quad C^{VW}_{V'W'} := C_{UW'}^W C_{V'}^{VU} ~.
\eea
The second type of composition is diagrammatically represented 
in figure \ref{GluingNetworks}.

Recall that the representations $\pi_V$, $\pi_U,\ldots$ are not necessarily 
irreducible. As such, the blocks in figure~\ref{GluingNetworks} should 
themselves be thought of as networks with a general number of incoming and 
outgoing legs and iterated compositions of Clebsch-Gordan maps inside each block. 
Compositions of the first type in equation \eqref{genericnetworks} are twisted trough 
\bea
\label{twistednetwork}
\tilde N^{VW}_{V'W'} = \tilde C_U^{VW} \tilde C_{V'W'}^U = 
J^{-1}_{VW} C_U^{VW} C_{V'W'}^U J_{V'W'} ~.
\eea 
Compositions of the second type 
\bea
\label{flippednetwork}
\hat N^{VW}_{V'W'} = C_{UW'}^W C_{V'}^{VU} = \tilde C_{UW'}^W J^{-1}_{UW'} 
J_{VU} \tilde C_{V'}^{VU} ~
\eea
also transform by conjugation with the $J$-factors of equation \eqref{twistednetwork}. 
To show that $\hat N^{VW}_{V'W'}$ is twisted by acting on its free legs as  
\bea
\hat N^{VW}_{V'W',\text{twist}} = J^{-1}_{VW} \hat N^{VW}_{V'W'} J_{V'W'} ~,
\eea
we insert the identity in \eqref{flippednetwork} 
\bea
\hat N^{VW}_{V'W',\text{twist}} 
= J^{-1}_{VW}  C_{UW'}^W J_{V(U\otimes W')} J_{V(U\otimes W')}^{-1} 
J_{(V\otimes U) W'} J_{(V\otimes U) W'}^{-1} C_{V'}^{VU} J_{V'W'} ~.\nnm
\eea  
$J_{V(U\otimes W')}$ and $J_{(V\otimes U) W'}^{-1}$ can be shifted past $C_{UW'}^W$ 
and $C_{V'}^{VU}$ to cancel $J^{-1}_{VW}$ and $J_{V'W'}$ 
\bea 
C_{UW'}^W J_{V(U\otimes W')}=C_{UW'}^W \sum_k \pi_V(x_k) \otimes 
(\pi_U\otimes\pi_{W'})\Delta(x^k) = \sum_k \pi_V(x_k) \otimes\pi_W(x^k) 
C_{UW'}^W ~,\nnm
\eea
where $\sum_k\pi_V(x_k)\otimes\pi_W(x^k)C_{UW'}^W = J_{VW}C_{UW'}^W$ 
and similatly for $J_{(V\otimes U) W'}^{-1}$. So we find
\bea
\hat N^{VW}_{V'W',\text{twist}} &=&  C_{UW'}^W J_{V(U\otimes W')}^{-1} 
J_{(V\otimes U) W'} C_{V'}^{VU} = 
\tilde C_{UW'}^W J^{-1}_{UW'} J_{V(U\otimes W')}^{-1} J_{(V\otimes U) W'} 
J_{VU}\tilde C_{V'}^{VU} ~.\nnm
\eea 
By the cocycle condition~\eqref{cocycle-2} 
$J^{-1}_{UW'} J_{V(U\otimes W')}^{-1} J_{(V\otimes U) W'} J_{VU}$ 
reduces to the identity evaluated on the tensor product of modules 
$V\otimes U\otimes W'$ and 
\bea
\hat N^{VW}_{V'W',\text{twist}} = \tilde C_{UW'}^W \tilde C_{V'}^{VU}~
\eea
as claimed. Therefore general networks $\hat N^{VW}_{V'W'}$,  
constructed as iterated compositions of Clebsch-Gordan maps,  
are twisted by $J$-factors which act on their external free legs.


\providecommand{\href}[2]{#2}\begingroup\raggedright\endgroup

\end{document}